\documentclass[twoside,12pt]{book}
\usepackage[english]{babel}
\usepackage{amsmath,amsthm}
\usepackage{amsfonts}
\usepackage{dsfont}
\usepackage{graphicx}
\usepackage{dsfont}
\usepackage{epstopdf}

\begin{document}
%%%%% frontmatter  $$$$$$
    \frontmatter

\title{Modified Theories of Gravity: Traversable Wormholes}

\author{Miguel A. Oliveira\footnote{email:miguel@cosmo.fis.fc.ul.pt} \\Centro de Astronomia
e Astrof\'{\i}sica da Universidade de Lisboa,\\ Campo Grande, Ed. C8
1749-016 Lisboa, Portugal}

\maketitle

\begin{center}
{\bf Abstract}
\end{center} 
This MSc thesis is divided in to two parts. The first, covers the foundations of theories of gravitation, and, the second incorporates original work on the subject of the existence of traversable wormholes in $f(R)$ modified theories of gravity.
 
A short incursion in the field of scalar-tensor theories had to be made, owing to an apparent inconsistency in the result previously found.

    \newpage{\pagestyle{empty} \cleardoublepage}

    %\documentclass[a4paper]{scrbook}
%\usepackage{vmargin}
%\setpapersize{A4}
%\usepackage{doublespace}

%\begin{document}
%\begin{titlepage}

\begin{center}
UNIVERSIDADE DE LISBOA\\
FACULDADE DE CI\^ENCIAS\\
DEPARTAMENTO DE F\'ISICA
\vskip .6 cm
\begin{figure}[h]
\begin{center}
%\mbox{\epsfig{file=logo.eps,width=.2\linewidth}}
\includegraphics[width=5cm]{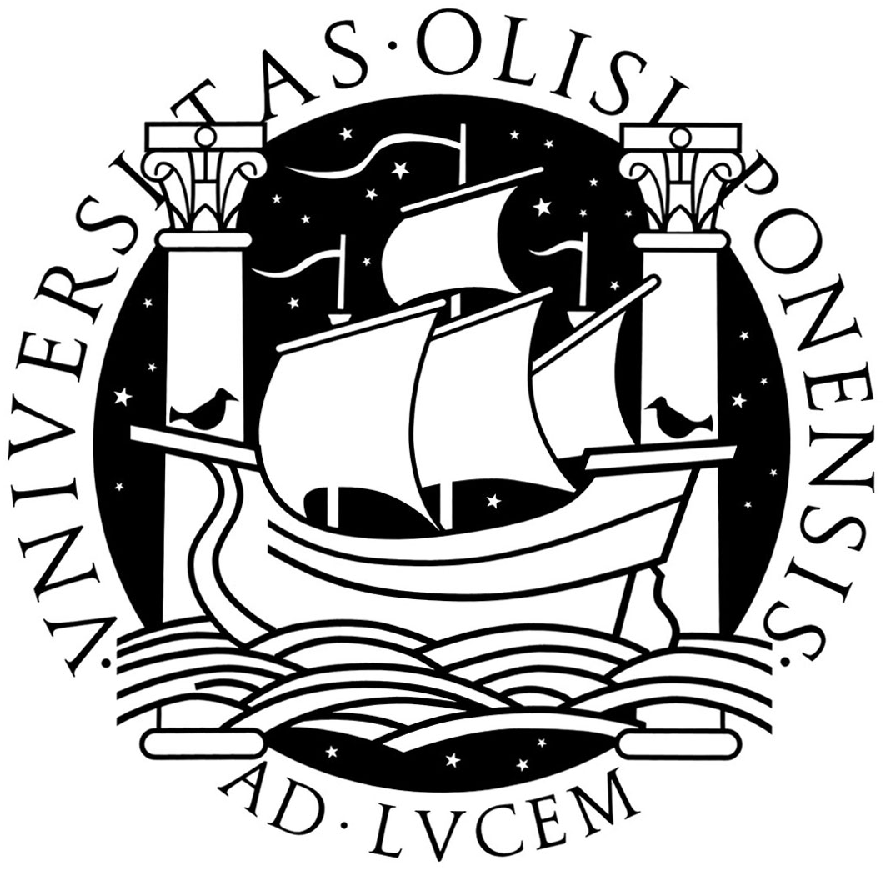}
\end{center}
\end{figure}
\vskip .25 cm

{\Huge \bf {Modified Theories of Gravity:\\ Traversable Wormholes}
\vskip .25 cm
}
%\vskip .25 cm
\vskip 1. cm
{\Large{\bf Miguel \^Angelo Oliveira}}
\vskip 1.35 cm
MESTRADO EM ASTRONOMIA E ASTROF\'ISICA
\vskip 1.6 cm
2011
\end{center}
%\end{titlepage}
%\end{document}

    \newpage{\pagestyle{empty} \cleardoublepage}

    %\documentclass[a4paper]{scrbook}
%\usepackage{vmargin}
%\setpapersize{A4}
%\usepackage{doublespace}

%\begin{document}
%\begin{titlepage}

\begin{center}
UNIVERSIDADE DE LISBOA\\
FACULDADE DE CI\^ENCIAS\\
DEPARTAMENTO DE F\'ISICA
\vskip .6 cm
\begin{figure}[h]
\begin{center}
%\mbox{\epsfig{file=logo.eps,width=.2\linewidth}}
\includegraphics[width=5cm]{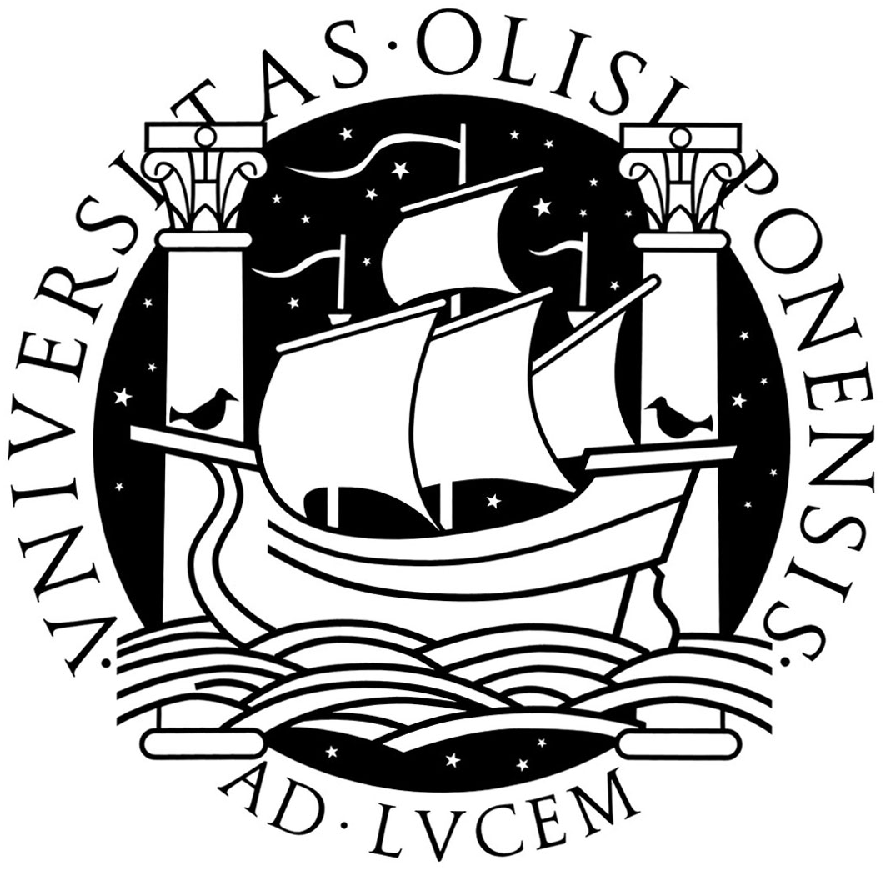}
\end{center}
\end{figure}
\vskip .25 cm

{\Huge \bf {Modified Theories of Gravity:\\ Traversable Wormholes}
\vskip .25 cm
}
%\vskip .25 cm
\vskip 1. cm
{\Large{\bf Miguel \^Angelo Oliveira}}
\vskip 1.35 cm
{\large Thesis supervised by:\\ Doctor Francisco S. N. Lobo and Professor Paulo Crawford}
\vskip 1.35 cm
MESTRADO EM ASTRONOMIA E ASTROF\'ISICA
\vskip 1.6 cm
2011
\end{center}
%\end{titlepage}
%\end{document}

    \newpage{\pagestyle{empty} \cleardoublepage}

\thispagestyle{empty}

\begin{flushright}
    \vspace*{10cm}
    In the beginning, God created the heavens and the earth;\\
    In Him we live and move and have our being.\\ \vspace*{5mm}
    The Bible
\end{flushright}

    \newpage{\pagestyle{empty} \cleardoublepage}

\thispagestyle{empty}

\begin{center}
\Large
Modified Theories of Gravity: Traversable Wormholes.

Miguel \^{A}ngelo Oliveira \emph{MSc Thesis}
\end{center}
\vspace{1 cm}

\begin{center}
    \bf Abstract
\end{center}

In recent years, the overwhelming amount of observational data supporting the \emph{late-time accelerated expansion of the Universe} has thrusted cosmology in particular and gravitation in general to the forefront of scientific research. In fact, the problem of the origin of the so called \emph{dark energy}, stands as one of the most tantalizing scientific issues of today's theoretical investigations, both in the field of fundamental physics and in that of astrophysics.

In this context, note must be given to the  appearance of modifications of Einstein's General Relativity, especially designed to deal with this problem, although simultaneously addressing other inconsistencies of the standard view of gravity, such as the dark matter problem and inflation. This is the case of the scalar-tensor theories of gravity and the $f(R)$ modified theories of gravity.
In this work, we consider these modifications, but focus on the existence of a specific type of exact solution: traversable wormholes.
These are hypothetical tunnels in space-time, and are primarily
useful as ``gedanken-experiments" and as a theoreticians probe of the
foundations of general relativity, although their existence as a solution
of the field equations may be regarded as a viability condition of the
theory.

We begin by analyzing the possibility of the existence of these wormholes in $f(R)$ modified theories of gravity. We conclude that this particular class of modifications does indeed posses these hypothetical space-time tunnels as exact solutions. However, given the well known correspondence between $f(R)$ gravity and Brans-Dicke theories, we present new exact vacuum wormhole geometries that generalize the well-known solutions in the literature.

\vspace{1 cm}

{\bf Keywords:} Modified Theories of Gravity, General Relativity , Equivalence Principle, Exact Solutions, Traversable Wormholes.

    \newpage{\pagestyle{empty} \cleardoublepage}

\thispagestyle{empty}

\begin{center}
\Large
Teorias Modificadas da Grvidade: Wormholes Transit\'{a}veis.

Miguel \^{A}ngelo Oliveira \emph{Tese de Mestrado}
\end{center}
\vspace{0.5 cm}

\begin{center}
    \bf Resumo
\end{center}

Recentemente, a enorme quantidade de dados observacionais suportando a \emph{expans\~{a}o do universo tardio}, propulsionou a cosmologia em particular e a gravita\c{c}\~{a}o em geral para a linha da frente da investigação cient\'{i}fica.  De facto, o problema da origem da chamada \emph{energia escura},  constitui um dos problemas mais empolgantes da investiga\c{c}\~{a}o actual, tanto no campo da f\'{i}sica fundamental como no da astrof\'{i}sica.
Neste contexto, \'{e} de notar o aparecimento de modifica\c{c}\~{o}es da Relatividade Geral de Einstein especialmente concebido para lidar com este problema, mas que simultaneamente tentam dar resposta a outras inconsist\^{e}ncias do modelo padr\~{a}o da gravidade, como são por exemplo o problema da mat\'{e}ria escura e o da infla\c{c}\~{a}o. Destas destacamos, as teorias escalares tensoriais e, as teorias $f(R)$ da gravita\c{c}\~{a}o.

Neste trabalho, consideramos estas modifica\c{c}\~{o}es mas, focamos em particular a exist\^{e}ncia de um conjunto de solu\c{c}\~{o}es exactas: os wormholes transit\'{a}veis.

Estes s\~{a}o hipot\'{e}ticos  t\'uneis no espa\c co-tempo e, s\~{a}o primariamente \'uteis como ``experi\^encias de pensamento'' e, como sondagens te\'oricas  dos fundamentos da Relatividade Geral. No entanto, a sua existencia como solu\c{c}\~{o}es exactas pode ser entendida como um crit\'{e}rio de viabilidade de uma dada teoria.

Come\c{c}amos por analisar a possibilidade da existencia destes wormholes no contexto das teorias $f(R)$ da gravita\c{c}\~{a}o. Concluimos que esta classe particular de modifica\c{c}\~{o}es da Relatividade Geral, admite de facto a existencia destas solu\c{c}\~{o}es exactas que se constituem com o tuneis hipot\'{e}ticos no espaço-tempo. No entanto, dada a bem conhecida correspondencia  entre as teorias $f(R)$ da gravidade e as teorias de Brans-Dicke, apresentamos novas geometrias t\'uneis em vacuo que generalizam as j\'{a} bem conhecidas solu\c{c}\~{o}es aprsentadas na literatura.

\vspace{ 0.5 cm}

{\bf Palavras-chave:} Teorias Modificadas da Gravidade, Relatividade Geral, Princ\'{i}pio da Equival\^encia, Solu\c{c}\~{o}es Exactas, T\' uneis Tansit\'{a}veis.

    \newpage{\pagestyle{empty} \cleardoublepage}

    \tableofcontents

    \newpage{\pagestyle{empty} \cleardoublepage}
    \newpage{\pagestyle{empty} \cleardoublepage}
    \listoffigures

    \newpage{\pagestyle{empty} \cleardoublepage}

    %%%%%%%%%%%%%%%%%%%%%%%%%%%%%%%%%%%%%%%%%%%%%%%%%%%%%%%
%%%%%%%%       Preface      %%%%%%%%%%%%%%%%%%%%%%%%%%%
%%%%%%%%%%%%%%%%%%%%%%%%%%%%%%%%%%%%%%%%%%%%%%%%%%%%%

\chapter{Preface}

This thesis is a required part of an MSc program in Astronomia e Astrof\'{i}sica,  taught at the Faculdade de Ci\^{e}ncias da Universidade de Lisboa, and the work herein was carried out at Centro de Astronomia e Astrof\'{i}sica da Universidade de Lisboa (CAAUL). I declare that this thesis is not substantially the same as any that I have submitted for a degree or diploma or other qualification at any other university and that no part of it has already been or is being concurrently for any such degree or diploma or any other qualification.

Chapters 3 and 4 are the result of a collaboration with Francisco S. N. Lobo, and resulted in publications listed below: \begin{itemize}
  \item Francisco S. N. Lobo and Miguel A. Oliveira, ``Wormhole geometries in $f(R)$ modified theories of gravity'', Phys. Rev. D {\bf 80}, 104012 (2009)
  \item Francisco S. N. and Lobo Miguel A. Oliveira, ``General class of vacuum Brans-Dicke wormholes'', Phys. Rev. D {\bf81}, 067501 (2010).
\end{itemize}

\subsection*{Plan of the Thesis}

This thesis contains five chapters. The first, is a general introduction, with some very brief historical remarks, to the subject of modified theories of gravity. The second chapter, Foundations, is meant as a non-axiomatic account of the principles that underlie (in the author's choice), gravitational theories in general, also in this chapter there is a summary of some well known exact solutions. The next two chapters are the analyses themselves. Chapter 3 is devoted to the existence of wormholes in $f(R)$ modified theories of gravity and chapter 4 is a proposal of a new class of general Brans-Dicke wormholes. This is followed by chapter 5 which contains the conclusions and discussions.

    % ----------------------------------------------------------------
%  ************************************************
% **** -----------------------------------------------------------
%\documentclass[12pt]{report}
%\usepackage{dsfont}
%\usepackage{amssymb}
%\usepackage{graphicx}
% ----------------------------------------------------------------
%\vfuzz2pt % Don't report over-full v-boxes if over-edge is small
%\hfuzz2pt % Don't report over-full h-boxes if over-edge is small

% ----------------------------------------------------------------
%\begin{document}

% ----------------------------------------------------------------

\vspace{2 cm}
\chapter{Acknowledgements}
Completing a MSc Thesis was part of a life-long goal, which is now fulfilled. However, because \emph{no man is an island, entire of itself;  [and] every man is a piece of the continent, a part of the main,}\footnote{John Donne, ``Devotions upon Emergent Occasions'' (1623), XVII.} a time comes when we must, not only give credit where it's due, but also recognize that our work is only made possible by the commitment of many others.

Among these, I would especially like to thank my thesis supervisors: Professor Paulo Crawford for all his support, advice, encouragement and particularly  because he was the first to  believe in me; and I am also deeply grateful to Francisco Lobo, for his friendship and, for helping me to publish two  chapters of this work: with him, I have learned more than I could have imagined.

I would also like to thank Centro de Astronomia e Astrof\'{i}sica da Universidade de Lisboa (CAAUL), for the use of the physical resources available. Equally, I am grateful to Funda\c{c}\~{a}o para a Ci\^{encia} e Tecnologia (FCT), for financial support.

And last, but by all means not least, I'd like to thank S\'{o}nia Queir\'{o}s: thank you, because with your shining presence, you always managed to light the way!
%\end{document}
% ----------------------------------------------------------------

%%%%%%% mainmatter %%%%%%%%%%%%%%%%%%%

    \mainmatter

    \part{Foundations}

    % ----------------------------------------------------------------
%%************************************************
% **** -----------------------------------------------------------
%\documentclass[12pt]{report}
%\usepackage{graphicx}
%\usepackage{dsfont}
%\usepackage{amssymb}
%\usepackage{amsmath}
% ----------------------------------------------------------------
%\vfuzz2pt % Don't report over-full v-boxes if over-edge is small
%\hfuzz2pt % Don't report over-full h-boxes if over-edge is small

% ----------------------------------------------------------------
%\begin{document}

%\date{}%
%\dedicatory{}%
%\commby{}%
% ----------------------------------------------------------------
% ----------------------------------------------------------------
\chapter{Introduction}\label{intro}

\section{Brief outline}

 General Relativity is a very successful, well tested, and predictive theory. It was first set forth by Einstein in 1915, and, not only has it passed all experimental tests that have been devised, but it has also become the basis for the description of virtually all gravitational phenomena known to date, thus establishing itself as a true paradigm, in the field of Gravitation in particular, and in Physics in general.

 However, despite this astonishing success, the mere fact that GR is a scientific theory makes it provisional, tentative or probational! Inevitably, provided scientific research follows it's normal course, some piece of information (experimental or otherwise) will come along, that doesn't easily fit within the framework of any given theory, regardless of how well constructed it may be. For General Relativity, the acceleration of the universe, --- prompting the introduction of dark energy ---, the rotation curves of galaxies and the mass discrepancy of clusters of galaxies --- supporting the existence of dark matter (see \cite{dark_side} for a review), and finally, on a more fundamental plane, GR's obstinate resistance to all attempts at it's quantization \cite{how_far}, constitute a set of instances that strongly motivate the introduction of Modified Theories of Gravitation (MTG's).

 From a slightly different point of view, modifying something is perhaps the best way to gain insight into it. If we have a set of ideas --- which we take to be an \emph{ad hoc} and, extremely loose definition of a theory ---, we may ask such questions as: what are the foundational principles upon which the theory rests? How important is each of them and, can we build a different consistent theory that leaves out, any number and combination of them? We could also be interested in looking at this theory from a broader perspective and investigating wether it can be viewed as a part of a more general class of theories. So, wether we are dropping principles from the theory, or looking at it from a higher viewpoint,\emph{``the plan, the great idea, is this: abstraction and generalization.''}\footnote{R.P. Feynman, {\it The Feynman Lectures on Physics}, 22-3.}

 Restricting ourselves, to the field of Gravity, we can take GR --- or more precisely, it's foundational principles ---, as a starting point, and, analyze simple theories that could be used as tools to assess how far (and in what direction), we can deviate, from this theory. These theories, are to be considered toy or straw-man theories, they must be simple and easy to handle, each of them deviating from GR in only one aspect. We can then discuss the viability of such toy theories, and, use them to gain a greater and deeper understanding of the problems involved in the conception of MTG's. This is, therefore, a trial-and-error approach, to modified gravity.

 The program we have just outlined, although in itself limited --- for a criticism of such a program see \cite{sot-PhD}, chapter 7 ---,  is still too vast and ambitions, to be undertaken in it's entirety by the author in the context of the present MSc Thesis. Therefore, the choice was made to restrict the number of theories, and the number of aspects of these theories to be considered. This restriction cut the immense forest of possibilities, down to mainly two theories and, one aspect of these theories. These are, respectively, $f(R)$ and Brans-Dicke modified theories of gravitation and, the existence of static traversable wormholes, respectively. Many other theories, will of course be mentioned, as will be numerous other features of these same theories, but the work revolves around these two theories and the existence of wormholes.

 This thesis, is outlined as follows: the remaining part of this chapter, is devoted to a sketch of the field of MTG's. The purpose of this section is to highlight the `cognate' nature of GR and modified gravity. Also,  we emphasize GR's role as a basis for the construction of modified theories of gravity.
Chapter 2, deals mainly with the foundations of Einstein theory, since this is to be the starting point for MTGs. In this chapter, we review the principles upon which theories of gravity in general rest. These principles may be, either relaxed or relinquished altogether, in order to produce alternative theories.

Chapter 3, is based on a work that is the outcome of a scientific collaboration with Doctor Francisco Lobo, concerning wormhole geometries in $f(R)$ modified theories of gravity \cite{mod-grav}.

Chapter 4, is also based upon a collaboration involving the author, that emerged to clarify some technical issues, raised in the context of the work just mentioned. These concern the equivalence between $f(R)$ and Brans-Dicke theories one the one hand, and the existence of wormhole geometries on the other \cite{brans-wh}.

Finally in chapter 5, we present some conclusions and future perspectives of this work.

\section{The emergence of GR and Modified Gravity}

Once Einstein finished the Special Theory of Relativity (SR), the next logical step was, of course, to handle --- perhaps in a covariant way, although there was no consensus about this at the time ---, the problem of Gravitation. After all, the displacement of the perihelion of Mercury, had cast doubts about the Newtonian treatment of this problem in general. About this situation we can quote \cite{Crawford}:
\begin{quote}
    Summarizing, Einstein's approach was embodied in heuristic principles that guided his search from the beginning in 1907. The first and more lasting one was the `Equivalence Principle" which states that gravitation and inertia are essentially the same.
This insight implies that the class of global inertial frames singled out in the special relativity can have no place in a relativistic theory of gravitation. In other words, Einstein was led to generalize the principle of relativity by requiring that the covariance group of his new theory of gravitation be larger than the Lorentz group. This
will lead him trough a long journey and in his first step, already in his review of 1907, Einstein formulated the assumption of complete physical equivalence between a uniformly accelerated reference frame and a constant homogeneous gravitational field. That is, the principle of equivalence extends the covariance of special relativity
beyond Lorentz covariance but not as far as general covariance. Only later Einstein formulates a "Generalized Principle of Relativity" which would be satisfied if the field equation of the new theory could be shown to possess general covariance. But Einstein's story appealing to this mathematical property, general covariance, is full
of ups and downs.
\end{quote}

The first attempts, followed the most logical route: to incorporate gravity, into SR. It soon became clear, however,  --- to Einstein at least ---, that gravity could not be described, through the use of a scalar field alone. Einstein had tried using a field $c$ corresponding to the speed of light, and this attempt was not only unsuccessful, but also, enough to convince him of the futility of a `scalar gravity'\footnote{The story is no so simple, of course! But this sketch only attempt at highlighting the fact that Einstein eventually abandoned scalar gravity. However see \cite{Crawford,Pais}, for a detailed account.}. Einstein sought, from then on, to use a tensor formalism to attack this problem. Others, like Nordstr\"{o}m, in his first theory, insisted in the use of a scalar field, (the field was $m$, the mass, in that theory), and some others, like Abraham, opposed the use of special relativity (of Lorentz invariance that is), and built theories that were not even covariant! (See section 12b and, Chapter 13 of \cite{Pais, Weyl}). This somewhat messy state of affairs, was eventually resolved in favor of Einstein's tensor approach, and of the Theory of General Relativity, rendering Nordstr\"{o}m's two theories (and other scalar theories of gravity), obsolete and fossilized remnants of the emergence of the geometric view of space-time.

The question, however, about the generality of the principles  of GR, remained well alive. As early as 1919, attempts were made to find a higher order theory of gravity, i.e., a theory in which the field equations are, unlike GR's, of order greater then second. These attempts, due to A. Eddington \cite{Eddington} and H. Weyl, wore motivated, mostly by theoretical completeness.  Also, around the same time (1920), there were discussions about whether it is the metric or the connection, that should be considered the principal field related to gravity. In 1924, Eddington presented a purely affine version of GR in vacuum. Later Schr\"{o}dinger, generalized Eddington's theory to include a non-symmetric metric \cite{Schrodinger}. These are vacuum theories and, great difficulties are encountered when any attempt is made to include matter into them. It is worth mentioning here, one other approach to this question, that is, to have both a metric and a connection that are at least to some extent independent. A good example is the Einstein-Cartan theory, that uses a non symmetric connection and, Riemann-Cartan spaces. This theory allows the existence of torsion and relates it to the presence of spin.

A different class of MTG's, is made of those theories in which, unlike GR, gravity is not represented only by a tensor quantity. The simplest modification, in this context, consists in the addition of a scalar field, --- after all this is the simplest type of field there is under transformations --- that along with the tensor one, is part of the description of gravity. Such a field, would have to involve (in a Lagrangian formulation) non-minimal couplings, otherwise it could simply be considered as a matter field. Jordan in 1955, and later, Brans and Dicke in 1961, developed the (Jordan-)Brans-Dicke theory and, subsequent generalizations of it are called Scalar-Tensor theories of gravity (see \cite{sot-PhD} sec. 2.4 and chapter 4).

If instead, we consider vector fields, we could have a theory in which gravity is described solely by such a field. This was sketched,  in 1908, by H. Minkowski. We could also have, in addition to the tensor field two others, namely, a scalar and a vector one, which is the case of Bekenstein's Tensor-Vector-Scalar (TeVeS) theory proposed in 2004  \cite{Bekenstein}. It has a curious motivation: to account for the anomalous rotation curves of galaxies, Milgrom proposed to avoid introducing dark matter by changing Newton's laws  \cite{Milgrom}, this is called Modified Newtonian Dynamics (MOND); since this theory is not relativistic, TeVeS was crafted to be the relativistic extension of MOND.
Einstein-Aether theory is another theory of this kind. In this one, a dynamical vector (but not a scalar) field is added. The aether is a preferred frame, (whose role is played precisely by this vector field), that would have to be determined on the basis of some yet unknown physics. It is interesting to note this frame may lead to Lorentz invariance violations.

There remains the possibility of adding a second tensor field to the description of gravity. In most theories produced by this method, the second tensor field plays the role of a metric. A good example is Rosen's bimetric theory, in which, in addition to the dynamical space-time metric, there is a second, non-dynamical flat metric. Since this flat non-dynamical metric, indicates the existence of a prior geometry, Rosen's theory is not background-independent. Nowadays, most of the interest in these theories, comes from variable speed of light cosmology. This is because, we can use a metric in Maxwell's equations different from the space-time metric. While the space-time metric defines the geometry, this new metric governs the propagation of light. The result is a theory in which, the propagation of light is different from what we would expect, based on the space-time metric. This is used, as an alternative to address the problems usually treated within the paradigm of inflation.

More recently, based on motivations stemming from the representation of gravitational fields near curvature singularities and the creation of a viable first order approximation to a quantum theory of gravity, among others (see \cite{dark_side}, and references therein), attempts have been made to generalize the Einstein-Hilbert Lagrangian. The latter, as is well known is linear in $R$, and the feeling was, that there is no \emph{a priori} reason for this very special form. Therefore, some of these, involve ``quadratic Lagrangians'', including terms of the form $R^{\,2},\ R_{\mu\nu}R^{\mu\nu},\ R_{\alpha\beta\mu\nu}R^{\alpha\beta\mu\nu},\ \varepsilon^{\alpha\beta\mu\nu}R_{\alpha\beta\gamma\delta}R^{\gamma\delta}_{\phantom{a}\phantom{a}\mu\nu},\  C_{\alpha\beta\mu\nu}C^{\alpha\beta\mu\nu}$. In this context, a more general form of Lagrangian, involving an arbitrary function of the curvature scalar $f(R)$ was later analyzed \cite{Bu70}. While the first $f(R)$ models, were motivated by inflationary scenarios --- a good example being  Starobinsky's model, were $f(R)=R-\Lambda+\alpha R^{\,2}$ ---, the more recent ones, are directed at the solution of the problem of the late-time acceleration of the universe. We will refrain from exposing $f(R)$ theories in detail here, since chapter 3, will be devoted to work involving them, and will therefore contain a somewhat more detailed introduction to this class of theories.

The search for quantum gravity, has produced a theory, known as \emph{string theory}, (which we will simply describe as a perturbative, and hence background-dependent theory, that uses objects known as strings as the fundamental building blocks for interactions), this is believed to be a viable theory unifying all (four) physical interactions. One of the predictions of this theory is the existence of extra spatial dimensions. Moreover, recent developments in string theory, have motivated the introduction of the brane-world scenario, in which the 3-dimensional observed universe, is imbedded in a higher-dimensional space-time. Although, most brane-word scenarios --- notably the Randall-Sundrum type --- produce ultra-violet modifications to General Relativity, i.e., extra-dimensional gravity dominates at high energies, there are those that lead to infra-red
modifications, i.e., those in which extra-dimensional
gravity dominates at low energies.

A different class of these brane-word scenarios, exhibits an interesting characteristic, that consists in the presence of the late-time cosmic acceleration, even when there is no dark-energy field. This exciting feature is called ``self-acceleration'', and the class of models where it arises, is named the Dvali-Gabadadze-Porrati (DGP) models. It is important to note, however, that these DGP models offer a paradigm for nature, that is fundamentally different from dark energy models of cosmic acceleration, even those with same expansion history \cite{dark_side}.

We have mentioned above, some quadratic curvature invariants, that have been tentatively added to the Einstein-Hilbert Lagrangian. One other noteworthy possibility, is the Gauss-Bonnet invariant, namely:
\begin{equation}\label{GBI}
    \mathcal{G}=R^{\,2}-4R_{\mu\nu}R^{\mu\nu}+4R_{\alpha\beta\mu\nu}
    R^{\alpha\beta\mu\nu}\,.
\end{equation}
This quantity, $\mathcal{G}$, has some very interesting properties: it is a generally covariant scalar; the variation of the scalar density $\sqrt{-g}\mathcal{G}$, with respect to the metric, is (\emph{via} Gauss's theorem) a total divergence; and it is a topological invariant (see \cite{sot-PhD} section 3.5.1). The second property mentioned, implies immediately that we can safely add $\mathcal{G}$ to the Einstein-Hilbert action, without changing the gravitational field equations.

The Gauss-Bonnet invariant is also important because recent developments in String/M-Theory suggest that unusual gravity-matter couplings may become important at the present low curvature universe. Among these possible couplings, there is a scalar field-$\mathcal{G}$ coupling, known as  Gauss-Bonnet Gravity. The action for this theory is:
\begin{equation}
S=\int d^4x
\sqrt{-g}\left[\frac{R}{2\kappa}-\frac{\lambda}{2}\partial_\mu
\phi \partial^\mu\phi-V(\phi)+f(\phi){\cal
G}\right]+S_M(g^{\mu\nu},\psi)\,,
    \label{GBaction}
\end{equation}
where $\lambda=+1$, for a canonical scalar field and, $\lambda=-1$, for a phantom field, respectively.\footnote{A canonical scalar field, has the action $S=\int {\rm d}^4x \sqrt{-g} \left[ -\frac12
(\nabla \phi)^2-V(\phi) \right]$, leading to $\omega=\frac{p}{\rho}>-1$, whereas a phantom (or ghost) field has a negative kinetic term, $S=\int {\rm d}^4x \sqrt{-g} \left[ \frac12
(\nabla \phi)^2-V(\phi) \right]$, and consequently $\omega<-1$.} Note that in the matter action there is a minimal coupling of the matter to the metric and it is not at all coupled to the scalar field, this makes Gauss-Bonnet gravity a metric theory (see, sec \ref{metric-sec} for a definition of metric theories). There are also generalizations of this theory and, in particular,  we mention Modified Gauss-Bonnet Gravity, with an action including a function of $\mathcal{G}$, that is:
\begin{equation}
S=\int d^4x \sqrt{-g}\left[\frac{R}{2\kappa}+f({\cal
G})\right]+S_M(g^{\mu\nu},\psi)\,.
   \label{modGBaction}
\end{equation}

Finally, since we mentioned string theory, we feel obligated to refer, if only briefly, it's main alternative, that is, Loop Quantum Gravity (LQG). LQG, unlike string theory, is only (and this is already a lot), a quantum theory of gravity, it is not a unified theory of physics. It also does not predict the existence of extra spacial dimensions. LQG is an example of the canonical quantization approach to the construction of a quantum theory of gravity. It is a fully background-independent and non-perturbative quantum theory of gravity \cite{how_far}. There are no experimental data whatsoever, supporting or disproving, any of these two theories. For the moment, all we have to compare them are consistency checks and aesthetic principles.

Many other modified theories exist, but the purpose of this section is only to give a rough picture of the field, and not to be exhaustive. We will need,  in order to construct MTGs, to modify the foundations of GR. The next chapter is a description of these foundations.
%%%%%%%%%%%%%%%%%%%%%%%%%%%%%%%%%%%%%%%%%%%%%%%%%%%%%%%%%%%%%%%%%%%%%%%%%%%%%%%%%%%%%%%%%%%%
% --------------------------------------chain--------------------------
%\include{referenc}
%\end{document}
% ----------------------------------------------------------------\left(

    % ----------------------------------------------------------------
% Chapter on Foundations of General Rlalivity ************************************************
% **** -----------------------------------------------------------
%\documentclass[12pt]{report}
%\usepackage{graphicx}
%\usepackage{dsfont}
%\usepackage{amssymb}
%%\usepackage{amsmath}
% ----------------------------------------------------------------
%\vfuzz2pt % Don't report over-full v-boxes if over-edge is small
%\hfuzz2pt % Don't report over-full h-boxes if over-edge is small

% ----------------------------------------------------------------
%\begin{document}

%\date{}%
%\dedicatory{}%
%\commby{}%
% ----------------------------------------------------------------
% ----------------------------------------------------------------
\chapter{Foundations}

General Relativity, as was already mentioned (Chapter \ref{intro}), is the most widely accepted and the best verified theory of Gravitation. So much so, that to some extent, Modified Theories of Gravity (MTG) are modifications of Einstein's generalized theory of relativity.

In this chapter, we review some of the fundamental principles upon which GR is based. Evidently, any MTG must also rest upon at least some of these principles. However, this chapter is neither an axiomatic formulation of GR, nor of gravitation in general. This is mainly because no claim is made about the independence of all principles stated here, although from a purely logical standpoint, it would be preferable to start off with a minimum of assumptions and derive all others from these, our purpose is to clarify what lies at the very heart of this theory.
We will also not discus those characteristics of a theory that it must fulfil in order to be a theory. Things such as completeness or self-consistency, important as they may be, will simply be assumed.

It has been pointed out \cite{sot-PhD}, that the so called, \emph{experimental tests of General Relativity} (namely, the deflection of light; the shift in the perihelion of Mercury; and, the gravitational redshift of distant light sources eg. galaxies), are in effect, tests of the underlying principles, rather then tests of the theory itself. This is one more reason that motivates the idea that we can find theories tailored to satisfy certain experimental tests. In the following sections we follow \cite{sot-PhD} very closely.

\section{A Theory of Principles}\label{D-frame}

The idea stated above, that the tests are tests of the principles rather than tests of the field equations, suggests the construction of a framework that would be used to analyze theories. This framework would provide the starting point for the construction of viable theories of gravity. General Relativity, as a perfectly viable one, would follow closely such a framework. Dicke proposed that the two following basic assumptions should be included\cite{Will}:
\begin{enumerate}
  \item The set of all physical events is a 4-dimensional manifold, called Spacetime;
  \item The equations are independent of the coordinates used -- Principle of Covariance.
\end{enumerate}
This is called the Dicke Framework. About the first assumption, we make clear that it (by itself) does not presuppose the existence of any other structure (like a metric or an affine connection) in Spacetime. The second, merits close attention, since some authors, like Wald \cite{wald} give a different definition of covariance and distinguish between:
\begin{itemize}
  \item \emph{General Covariance}: there are no preferred vector fields or no preferred basis of vector fields pertaining only to the structure of space which appear in any law of physics.
  \item \emph{Special Covariance}: if $\mathcal{O}$ is a family of observers and, $\mathcal{O'}$ is a second family obtained from the first by ``acting'' on it with an isometry, then if $\mathcal{O}$ makes a measurement on a physical field, then $\mathcal{O'}$ must also be able to make that same measurement. The set of physical measurements are the same for the two families.
\end{itemize}
Dicke added two more requirements, that gravity should be associated with one or more fields of tensorial nature (scalar, vector or tensor), and that the field equations should be derivable from an invariant action via a stationary action principle.

\subsection{The principle of equivalence}

One of the most important principles in GR in particular, and in gravitational theories in general, is the Principle of Equivalence. It reportedly guided Einstein in the construction of the generalized theory \cite{Pais} and has since been the subject of heated discussions.

First of all, we point out that a version of the principle of equivalence was already present in Newtonian Mechanics. It was even mentioned in the first paragraph of Newton's \emph{Principia} (see, \cite{Will} page 13 and Figure 2.1 there).  In that context, this principle states that the inertial and passive gravitational masses are equal, $m_I=m_p$,\footnote{The inertial mass is that present in Newton's third law $\vec{F}=m_I \vec{a}\,.$ The passive gravitational mass is the one in Newton's gravitational law $\vec{F_g}=m_p\vec{g}\,.$}. In this form the principle can also be equivalently stated in the following way: all bodies fall (when in free-fall) with the same acceleration, independently of their composition and mass.

Several current forms (and several different formulations throughout the literature) of the equivalence principle may be distinguished \cite{sot-PhD}:
\begin{itemize}
  \item Weak Equivalence Principle (WEP): the trajectory of an uncharged test particle\footnote{For the relevant definitions, see \cite{Will}.} (for all possible initial conditions ) is independent of it's structure and composition;
  \item Einstein Equivalence Principle (EEP): this assumes that the WEP is valid and, that non-gravitational test experiments have outcomes that are independent of both velocity (Local Lorentz Invariance - LLI), and position in space-time (Local Position Invariance - LPI);
  \item Strong Equivalence Principle (SEP): this assumes that the WEP is valid both for test particles and self-gravitating bodies, and also  assumes LLI and LPI for any local test experiment.
\end{itemize}
We can add that the Schiff conjecture states that: \emph{any complete, self consistent theory of gravity that embodies WEP also necessarily embodies EEP}(see \cite{Will}, Section 2.5).

We stress, once more, that this is not the only formulation of the equivalence principle. A different but equivalent statement, emphasizes the fact that there is an accelerated reference frame where (locally), ``insofar as their mechanical motions are concerned the bodies will behave as if gravity were absent''\footnote{Except for possible tidal effects due to inhomogeneities in the gravitation field.} \cite{Will} for an example see \cite{Pauli}. Also as the terminology is not uniform, what we called here the WEP, is sometimes called Galileo's principle \cite{Rindler}.

The principle of equivalence is a very useful one, since physical consequences may be drawn from it. The WEP, for example, implies the existence of a preferred family of curves, these are the trajectories of test bodies in free-fall. These curves, however, have no \emph{a priori} relation to the geodesics of a metric. So that, even if we postulated the existence of a metric, it's geodesics would not necessarily be the above mentioned family of curves.

The EEP, on the other hand, implies, (since it adds to the requirements of the WEP, LLI and LPI) the existence of a set of frames for which the theory reduces to Special Relativity. These are also the free-fall frames. Since in these frames all bodies are unaccelerated, the LLI guarantees that two such frames located at the same event, with different velocities, will yield the same physical predictions. The LPI, will then extend this to all spacetime events.

There is, therefore, in the case of the EEP a second rank tensor which in the local free-fall frame reduces to a metric conformal to the Minkowski one. We could however, still think of coupling each matter field with a different conformally related second rank tensor. But, even in this case, the EEP demands that the different coupling constants differ by at most a multiplicative constant. So that that the EEP allows us, (after rescaling coupling constants and porforming a conformal transformation) to find a metric $g_{\mu\nu}$ that locally reduces to the Minkowski flat spacetime metric $\eta_{\mu\nu}$.

All of these conformally related metrics $\phi g_{\mu\nu}$, can be used to write down the equations of motion. The metric $g_{\mu\nu}$ can however, be in a sense singled out since, at each spacetime point $\mathcal{P}$, given the existence of \emph{local Lorentz frames} (LLF), a set of coordinates can be found such that:
\begin{equation}
    g_{\mu\nu}=\eta_{\mu\nu}+\mathcal{O}\left(\sum_\alpha|x^\alpha-x^\alpha(\mathcal{P})|^2\right) \qquad \mbox{and} \qquad \frac{\partial g_{\mu\nu}}{\partial x^\alpha}=0\,,
\end{equation}
are valid.

Since, the geodesics of $g_{\mu\nu}$, in LLF are straight lines. Also since, the trajectories of free-falling bodies are straight lines in a local free fall frame. If we identify the LLF and the local free-fall frames, we can conclude that: \emph{the geodesics of $g_{\mu\nu}$, are the trajectories of free falling bodies}.

As for the SEP, it additionally enforces: the extension of the validity of WEP to self-gravitating bodies; and the validity of the EEP (that is, the LLI and LPI, of course), to local gravitational experiments. The only known theory that satisfies the SEP is General Relativity.

There are, however, limitations to the use of these equivalence principles as guidelines for the construction of Modified Theories of Gravity. We mention only three of these: the first is the aforementioned connection between the SEP and GR (it has been claimed that SEP is only valid for GR, although no proof of such statement has ever been offered); the second is related to the definition of `test particles', i.e., how small must a particle be so that we can neglect it's gravitational field, (note that the answer will probably be theory-dependent); the third, linked to the fact that some theories may appear to either satisfy or not a given EP depending on the variables used to describe the theory, --- notably, this is the case of Scalar-Tensor Theories and the Jordan and Einstein frames, see Sec. \ref{Sec:II}.

\subsection{The metric postulates}\label{metric-sec}

The above are some of the foundational postulates that a theory must be built upon, but suppose we are given such a theory (action or field equations), and we want to verify whether or not it satisfies some principle, what mathematical form do the principles take? Conversely, given a principle, what mathematical form does it take and how does it constraint our theory?

The so-called metric postulates, are not principles in themselves, they are otherwise mathematically formulated propositions with which a theory must comply in order to be compatible with the principles.

Thorne and Will proposed the following metric postulates:
\begin{enumerate}
  \item Defined in space-time, there is, a second rank non-degenerate tensor, called a metric, $g_{\mu\nu}\,$\footnote{More precisely a pseudo-metric since in GR it will turn out not to be positive-definite. };
  \item If $T_{\mu\nu}$ is the stress-energy tensor, associated with non-gravitational mater fields, and if $\nabla_{\mu}$ is a covariant derivative derived from the Levi Civita connection associated with the metric above, then $\nabla_{\mu}T^{\mu\nu}=0\,$.
\end{enumerate}
Theories that satisfy the metric postulates are called \emph{metric theories}. We note two things about the metric postulates: first, that geodesic motion can be derived from the second metric postulate \cite{sot-PhD,Fock}; second, that the definition of $T^{\mu\nu}$ is somewhat vague and imprecise, as is the notion of non-gravitational fields \cite{meta}.

Before we close this section, we must once again emphasize that, the principles stated here are nothing but one choice out of many possible ones. For example, \cite{dinverno}, has a somewhat different list of principles.  That list contains, Mach's principle, (which we will discuss briefly later) a sort of philosophical conjecture about inertia and the matter distribution of the Universe; the principle of equivalence; the principle of covariance; the principle of minimal gravitational coupling, which is stated as saying that ``no terms explicitly containing the curvature tensor\footnote{This will be defined in the next section.} should be added in making the transition from the special to the general theory''; and, the correspondence principle, that in the case of gravitation means that GR must, in the limit of weak gravitational fields reduce to Newton's gravitation.

At this point, geometry `sneaks in', --- it was there all along, since the first assumption of the Dicke framework, because the definition of manifold is geometric in nature. The metric, however, is a manifestly geometric quantity, it is obviously used to calculate distances, as it's name suggests. So for the rest of this chapter, we will be discussing geometry alongside General Relativity!

\section{Geometry as the description of space-time}

The basic concept in differential geometry is that of a \emph{manifold}. This is, intuitively, just a set that locally `looks like' $\mathds{R}^n$. For a rigorous definition of manifold and, detailed discussion of differential geometry see \cite{wald, dinverno, HawEll, Gravitation}.

In the context of theories of gravitation, as was already stated, spacetime is a 4-dimensional manifold, where we define a symmetric non-degenerate metric $g_{\mu\nu}$, and a quantity related to parallel transport called a connection (see \cite{dinverno} for an introduction, and \cite{HawEll} for a more advanced treatment), $\Gamma^{\lambda}_{\phantom{a}\mu\nu}$. This connection, by it's relation with parallel transport leads naturally to a definition of derivative adapted to curved manifolds, this is the covariant derivative denoted $\tilde{\nabla}$, in general. It's definition is:
\begin{equation}
    \tilde{\nabla}_\mu T^\nu_{\phantom{a}\sigma}=\partial_\mu T^\nu_{\phantom{a}\sigma}+\Gamma^{\nu}_{\phantom{a}\mu\alpha}T^\alpha_{\phantom{a}\sigma}-\Gamma^{\alpha}_{\phantom{a} \mu\sigma}T^\nu_{\phantom{a}\alpha}\,.
\end{equation}
It is important to note, that we have made no association of the connection $\Gamma^{\nu}_{\ \mu\nu}$ with the metric $g_{\mu\nu}$. This will be an extra assumption and, there will be a connection related to the metric called the Levi-Civita connection. We will use the symbol $\tilde{\nabla}$ to denote this general covariant derivative, and $\nabla$ denotes the one obtained from the Levi-Civita connection (\cite{wald} contains a derivation of the Levi-Civita connection and \cite{Schrodinger} elaborates on the use of more general ones).

The notion of curvature of a manifold is given by the \emph{Riemann Tensor}, which can be constructed from this generic connection as follows:
\begin{equation}
    {\cal R}^\mu_{\phantom{a}\nu\sigma\lambda}=
    -\partial_\lambda\Gamma^\mu_{\phantom{a}\nu\sigma}
    +\partial_\sigma\Gamma^\mu_{\phantom{a}\nu\lambda}
    +\Gamma^\mu_{\phantom{a}\alpha\sigma}
    \Gamma^\alpha_{\phantom{a}\nu\lambda}
    -\Gamma^\mu_{\phantom{a}\alpha\lambda}
    \Gamma^\alpha_{\phantom{a}\nu\sigma}\,,
\end{equation}
which does not depend on the metric and is antisymmetric in it's last indices.

To describe the relation between the connection and the metric, we introduce the non-metricity tensor:
\begin{equation}
    Q_{\mu\nu\lambda}=-\tilde{\nabla}_\mu g_{\nu\lambda}
\end{equation}
and, the Weyl vector:
\begin{equation}
    Q_\mu=\frac{1}{4}Q_{\mu\nu}^{\phantom{a}\phantom{b}\nu}
\end{equation}
which is just the trace of the non-metricity tensor in it's last two indices.

Moreover, the antisymmetric part of the connection is the Cartan Torsion Tensor:
\begin{equation}
    S_{\mu\nu}^{\phantom{a}\phantom{a}\lambda}=\Gamma^\lambda_{\phantom{a}[\mu\nu]}\,.
\end{equation}
One of the traces of the Riemann tensor is called the Ricci Tensor. Now, there are two possibilities for this contraction, either the first and the second indices or, the first and the third are contracted (due to the antisymmetry of the Riemann tensor, a contraction of the first and the fourth indices is equal to a contraction of the first and the third, with an additional minus sign), that is:
\begin{equation}
    {\cal R}_{\mu\nu}\equiv {\cal R}^\sigma_{\phantom{a}\mu\sigma\nu}=-{\cal R}^\sigma_{\phantom{a}\mu\nu\sigma} \qquad\mbox{or,}\qquad {\cal R}'_{\mu\nu}\equiv {\cal R}^\sigma_{\phantom{a}\sigma\mu\nu}\,.
\end{equation}
We will thus obtain that the second tensor ${\cal R_{\mu\nu}}'$ will be the antisymmetric part of the first ${\cal R_{\mu\nu}}$ for a symmetric connection. The tensor quantity ${\cal R_{\mu\nu}}$ is, of course, nothing but the usual Ricci tensor:
\begin{eqnarray}
               {\cal R}_{\mu\nu}&=&{\cal R}^\lambda_{\phantom{a}\mu\lambda\nu}=\partial_\lambda \Gamma^\lambda_{\phantom{a}\mu\nu}-\partial_\nu \Gamma^\lambda_{\phantom{a}\mu\lambda}+\Gamma^\lambda_{\phantom{a}\sigma\lambda}\Gamma^\sigma_{\phantom{a}\mu\nu}-\Gamma^\lambda_{\phantom{a}\sigma\nu}\Gamma^{\sigma}_{\phantom{a}\mu\lambda}\\
               {\cal R}'_{\mu\nu}&=&-\partial_\nu \Gamma^\alpha_{\phantom{a}\alpha\mu}+\partial_\mu \Gamma^\alpha_{\phantom{a}\alpha\nu}\,
\end{eqnarray}
Using the metric, --- up to now the tensors have been independent from it ---, to contract ${\cal R}_{\mu\nu}$, we may obtain the usual Ricci scalar, whereas through  the use of ${\cal R}_{\mu\nu}'$ we get a null tensor, since the metric is symmetric and ${\cal R}_{\mu\nu}'$ is antisymmetric. That is:
\begin{equation}
    {\cal R}=g^{\mu\nu}{\cal R}_{\mu\nu} \qquad \mbox{and}, \qquad g^{\mu\nu}{\cal R}_{\mu\nu}'\equiv0\,.
\end{equation}

\section{General Relativity}\label{gen-rel}

The principles stated above, are too general if we want to restrict ourselves only to GR. If we are to obtain Einstein's theory, we will have to make further assumptions. This section will explore these assumptions.

We state them here briefly for future reference.
\begin{enumerate}
  \item Torsion dos not play any fundamental role in GR: $S_{\mu\nu}^{\phantom{a}\phantom{a}\lambda}=0$;
  \item The metric is covariantly conserved: $Q_{\mu\nu\lambda}=0$;
  \item Gravity is associated with a second rank tensor field, the metric, and no other fields are involved in the interaction;
  \item The field equations should be second order partial differential equations;
  \item The field equations should be covariant.
\end{enumerate}
One of the features of the above discussion was the independence of the connection relative to the metric, this was an attempt to get a general set of characteristics that a theory must obey. However, the second metric postulate calls upon a notion of covariant derivative --- and consequently of connection ---, that is linked to the metric. This choice of connection --- the \emph{Levi-Civita connection} ---, is one of the most fundamental assumptions of GR. To fulfill this, it turns out that we need two things: firstly, the symmetry of the connection with respect to it's two lower indices, that is:
\begin{equation}\label{sym-conn}
    \Gamma^\alpha_{\phantom{a}\mu\nu}=\Gamma^\alpha_{\phantom{a}\nu\mu}\qquad \Leftrightarrow \quad S^\lambda_{\phantom{a}\mu\nu}=0\,.
\end{equation}
Secondly, the metric must be conserved by the covariant derivative --- or, covariantly conserved:
\begin{equation}\label{cov-con}
    \tilde{\nabla}_\lambda g_{\mu\nu}=0\qquad \Leftrightarrow \qquad Q_{\lambda\mu\nu}=0\,.
\end{equation}
The assumption~(\ref{sym-conn}) means that space-time is torsion-less, while~(\ref{cov-con}) implies that the non-metricity is null.
With these choices, the connection takes the Levi-Civita form:
\begin{equation}\label{Levi-Civita}
    \left\{^\alpha_{\phantom{a}\mu\nu}\right\}=\frac{1}{2}g^{\alpha\beta}\Big[\partial_\mu g_{\nu\beta}+\partial_\nu g_{\mu\beta}-\partial_\beta g_{\mu\nu}\Big]
\end{equation}
These assumptions, increase the symmetry of the Riemann tensor. It now becomes anti-symmetric with respect to the first two indices and, symmetric with respect to the exchange of the two consecutive pairs of indices.

In Newtonian Gravity, the equation that describes the dynamics of the gravitational potential,  is Poisson's equation, $\nabla^2\varphi=4\pi\rho$. Einstein, in his original derivation of the field equations of GR, relied on a close analogy with this equation. In fact, the equations of GR in empty space are simply $R_{\mu\nu}=0$, where this Ricci tensor has been constructed not from the most general connection but, out of the Levi-Civita affinity~(\ref{Levi-Civita}). This is in good analogy with Laplace's equation $\nabla^2\varphi=0$, since the Ricci tensor is a second order differential expression on the components of the connection.

However, to extend this analogy to the case where we have matter, some extra assumptions must be made. The choice of the field(s) is the first assumption: in GR the only field of the theory (the only one whose dynamics we want to describe), is \emph{the metric}. All other fields, are considered `matter fields', i.e., sources of the `gravitational field'. Therefore, we impose (only in GR) that gravity is associated to no field other than the second rank tensor field that represents the metric. This also means of course, that the field equations should have a left side depending only on the metric and, a right side containing the dependence on all other fields, the `matter fields'. As it will turn out, the object that generalizes the distribution of the `matter', and hence plays the role of the source of the field is, the Stress-Energy Tensor $T^{\mu\nu}$ see \cite{Gravitation} for a detailed discussion, and \cite{sot-PhD,meta} for some problems related to the definition of this quantity.

Second, if we are to have an analogy with Poisson's equation, then our field equation must be a second order differential equation.
As for the last requirement above, it stems from the second point in the Dicke framework, mentioned in section~(\ref{D-frame}).

With these assumptions, if we follow the original derivation by Einstein see \cite{wald,Gravitation,Schutz}, we will obtain the field equations for GR:
\begin{equation}\label{GR-eqs}
    G_{\mu\nu}=R_{\mu\nu}-\frac{1}{2}g_{\mu\nu}=kT_{\mu\nu}\,.
\end{equation}
Where $k$ is a constant that depends on the system of units used. In conventional units we have $k=\frac{8\pi G}{c^4}$. We will use however a system in which $c=1$, and therefore, in this system $k={8\pi G}$

\section{The Lagrangian formulation of General Relativity}

The General Theory of Relativity is a classical\footnote{That is non-quantum.} theory, therefore all dynamical content of the theory is enclosed in the Einstein field equations~(\ref{GR-eqs}). Despite this, there are reasons for the use of a lagrangian formulation of GR. Firstly, it is an elegant formulation and has a very great aesthetic appeal. To this, we can add that it is much easier to modify GR and compare different modified theories of gravitation through the use of the lagrangian formulation. And lastly --- although we will not use this criterion here ---, at the quantum level the action does have a physical significance. We will therefore briefly develop this Lagrangian formulation.

\subsection{The variational principle in Classical Mechanics}\label{prob,ger}

In the context of Classical Mechanics, the motion of system of particles, cam be described by a set of $n$ quantities called \emph{generalized coordinates}, $q_i\quad i=1...n$. These generalized coordinates form a n-dimensional Cartesian space called \emph{configuration space}, the motion of the system is a curve in this space and, time is the parameter of this curve.

For monogenic\footnote{The systems for which all forces can be derived from a single scalar potential.} systems, see \cite{Goldstein} for details, an integral principle can be formulated that describes the motion of the system between two instants of time $t_1$ and $t_2$. \emph{Hamilton's Principle} states that, \emph{the variation of a line integral $S$ called the action, for fixed $t_1$ and $t_2$, is zero.}

The action is given by:
\begin{equation}\label{S}
    S=\int_{t_1}^{t_2}L(q_i(t),\dot{q}_i(t),t)dt\,.
\end{equation}
In the above integral, $L$ is the Lagrangian function, and the principle can be stated mathematically in the following form:
\begin{equation}\label{deltaS}
    \delta S=\delta \int_{t_1}^{t_2}L(q_i(t),\dot{q_i}(t),t)dt =0\,.
\end{equation}
If the constraints of the system are holonomic\footnote{Holonomic constraints are those describable by an equation of the form $f(q_i,t)=0$.}, Hamilton's principle is equivalent to Lagrange's Equations\footnote{For non-holonomic constraint we obtain the same equations with an extra term, namely
\begin{eqnarray}
      \frac{d}{dt}\left(\frac{\partial L}{\partial \dot{q}_k}\right)-\frac{\partial L}{\partial q_k}=\sum_l \lambda_l a_{lk}\nonumber
\end{eqnarray}
but now the equations of constraints are needed to complete the solution of the problem.}:
\begin{equation}\label{Lagrange}
    \frac{d}{dt}\left(\frac{\partial L}{\partial \dot{q}_i}\right)-\frac{\partial L}{\partial q_i}=0\qquad i=1...n\,.
\end{equation}

\subsection{Lagrangian Formulation of a Field Theory}
 If we have a field instead of a system of particles, we need a slightly different form of this formalism. Let $M$ be a manifold, $\psi$ a set of fields associated with the theory and, $S[\psi]$ a linear functional\footnote{A real (complex) functional is a map of a function to real (complex) numbers.}.

If we consider one-parameter families of field configurations  $\psi_{\lambda}$ and, if there is a $\frac{dS}{d\lambda}$ at  $\lambda=0$ for all such families starting from $\psi_0$ then the following equation is the definition of $\chi$:
\begin{equation}\label{defchi}
\frac{dS}{d\lambda}=\int_M \chi \delta \psi\,,
\end{equation}
where, $\frac{d\psi_\lambda}{d\lambda}\bigg|_{\lambda=0}=\delta\psi$. The tensor $\chi$, is a dual tensor to $\psi$. It is termed \emph{the functional derivative} of $S$ and, it can be written as:
\begin{equation}\label{Snot}
    \chi=\frac{\delta S}{\delta \psi}\,.
\end{equation}
Taking now, a functional of the form:
\begin{equation}\label{funcL}
    S[\psi]=\int_M\mathcal{L}[\psi]\,,
\end{equation}
where $\mathcal{L}$ is a function of $\psi$ and a finite number of it's derivatives, viz:
\begin{equation}\label{punds}
    \mathcal{L}|_x=\mathcal{L}(\psi(x),\nabla\psi(x),\ldots,\nabla^k\psi(x))\,.
\end{equation}
If $S$ is functionally differentiable and, if  the solutions of the field equations are the field configurations $\psi$ which extremize $S$:
\begin{equation}\label{extr}
    \frac{\delta S}{\delta \psi}\bigg|_{\psi}=0\,,
\end{equation}
then, $S$ is the action of the theory and, $\mathcal{L}$ is the \emph{lagrangian density}. The lagrangian formulation of a field theory --- in close analogy to the particle mechanics case ---, is the definition of this density $\mathcal{L}$.

\subsection{The Einstein-Hilbert action}\label{sec-einhill}

General Relativity is a classical field theory, therefore, the formalism just described is applicable with the metric tensor field $g_{\mu\nu}$ playing the role of the dynamical variable. However, since in the particular case of GR, the connection is not a generic one, but the Levi-Civita connection, as was discussed in section~(\ref{gen-rel}), the question of which of the two we should use arises. This is, of course, absolutely equivalent to whether or not we should take the connections as a dynamical variable.

In the case of the so called \emph{metric formalism}, the only field describing gravity is the metric, this is considered to be covariantly conserved, eq. (\ref{cov-con}), and therefore the connection is the Levi-Civita one. If, however, we want to consider the metric and a general independent symmetric connection both as variables, if moreover we want to derive the properties of the connection from the field equation and, arrive at GR in the end, we have the \emph{Palatini formalism}.

If we adopt the metric formalism, we must make some assumptions about the gravitational action. The first is an imposition of the principle of (general) covariance (see Section \ref{D-frame}): the action must be a generally covariant scalar. This implies that, since the volume element is a tensor density of weight $-1$ (see \cite{dinverno} for a definition), we must have a multiplicative factor of $\sqrt{-g}$, where $g$ is the determinant of the metric tensor $g_{\mu\nu}$. The second, is the assumption that the action depends only on the metric and it's first derivatives. This is assumed so that we can arrive at second order partial differential equations.

In spite of this second assumption, the simplest scalar quantity we can construct with the metric is the Ricci scalar ${R}$, which not only depends on the first derivatives, but, also on the second! There is, in fact, no scalar quantity constructed only with the metric and it's first derivatives. These are just not covariant objects, and therefore, there is no combination of them that would turn out to be covariant!
In these conditions the gravitational action (Einstein-Hilbert action) is defined to be:
\begin{equation}\label{EinsHil}
    S_{EH} =\frac{1}{16\pi G}\int_U \sqrt{-g} R\, d^4x\,,
\end{equation}
The variational principle, in this case, takes the form:
\begin{equation}\label{calc1}
    \delta S_{EH} = \delta \int_U  \sqrt{-g}R\,d^4x = 0
\end{equation}
This can be grouped in three terms:
\begin{equation}\label{calc2}
    \underbrace{\int_U \delta \sqrt {-g} g_{\mu\nu}R^{\mu\nu} d^4x\ +\ \int_U  \sqrt {-g} \delta g_{\mu\nu}R^{\mu\nu} d^4x}_{(I)+(II)}\ +\ \underbrace{\int_U\sqrt {-g} g_{\mu\nu}\delta R^{\mu\nu} d^4x}_{(III)}
\end{equation}
The first two terms can be calculated if we consider the inverse of the metric tensor. This is defined as:
\begin{equation}\label{definv}
    \delta^{\mu}_{\phantom{b}\nu}=g^{\mu\alpha}g_{\alpha\nu}\,,
\end{equation}
taking the variation of both sides of this last equation we obtain,
\begin{equation}\label{invcalc}
    0=\delta (g^{\mu\alpha}g_{\alpha\nu})=\delta g^{\mu\alpha}g_{\alpha\nu}+g^{\mu\alpha}\delta g_{\alpha\nu}
\end{equation}
hence:
\begin{equation}\label{invresult}
    \delta g^{\mu\nu}=-g^{\mu\alpha}g^{\nu\beta} \delta g_{\alpha\beta}\,.
\end{equation}
We also need another important result, obtained by considering the definition of the inverse of a generic matrix:
\begin{equation}\label{matrix1}
    [a_{ij}]^{-1}=\frac{1}{a}\ adj[a_{ij}]\,,
\end{equation}
where, $a$ and $adj(a_{ij})$, are respectively the determinant, and the adjoint (which is itself a matrix ), of the matrix $[a_{ij}]$.

If we calculate this determinant, using Laplace's rule we get:
\begin{equation}\label{determinant}
    a= \sum_{j=1}^{n}a_{ij}\ adj[a_{ij}]\,,
\end{equation}
were we use row $i$, to develop the determinant. We then have the following partial derivative:
\begin{equation}\label{dterresult}
    \frac{\partial a}{\partial a_{ij}}=\ adj[a_{ij}]\,.
\end{equation}
Considering the composite function $a(a_{ij}(x^{k}))$, we have for it's derivative:
\begin{equation}\label{matrixpartial}
    \frac{\partial a }{\partial x^{k}}=\frac{\partial a }{\partial a_{ij}}\frac {\partial a_{ij }}{\partial x^{k}}\,,
\end{equation}
this equation, together with equations~(\ref{matrix1}) and~(\ref{dterresult}) give:
\begin{equation}\label{matrixresult}
\frac{\partial a }{\partial x^{k}}=a[a_{ij}]^{-1}\frac {\partial a_{ij }}{\partial x^{k}}\,.
\end{equation}
If we apply this result to the metric tensor $g_{ab}$, and it's inverse~eq.~(\ref{definv}), leads to:
\begin{equation}\label{metricresult}
    \partial_\mu g =gg^{\alpha\beta}\partial_\mu g_{\alpha\beta}\,.
\end{equation}
Finally, if we use a similar deduction but, with $\delta$, replacing $\partial$ we arrive at:
\begin{equation}\label{metricvariation}
    \delta g =g g^{\mu\nu}\delta g_{\mu\nu}\,.
\end{equation}
We can moreover, based on the above formula, obtain the following result:
\begin{equation}\label{sqrtresult}
    \delta\sqrt{(-g)} = \frac{1}{2}\sqrt{(-g)}g^{\mu\nu}\delta g_{\mu\nu}\,.
\end{equation}
It is now possible, to calculate the term~$(I)+(II)$ in equation~(\ref{calc2}). For this we use the results~(\ref{invresult}) and,~(\ref{sqrtresult}). After some algebraic manipulations this term yields:
\begin{equation}\label{result}
    -\int_U \sqrt{(-g)}\bigg[R^{\mu\nu}-\frac{1}{2}g^{\mu\nu}R\bigg]\delta g_{\mu\nu}d^4x=0\,,
\end{equation}
and, the fact that the integral is equal to zero --- combined with the fact that $\delta
g_{\mu\nu}\neq 0$, since we are considering this as the independent variation --- implies that the Einstein Tensor $G^{\mu\nu}$, (defined below) is null, that is:
\begin{equation}\label{eqEinstein}
    G^{\mu\nu}\equiv R^{\mu\nu}-\frac{1}{2}g^{\mu\nu}R=0\,.
\end{equation}
These are, Einstein's  equations in the absence of matter. However, eq~(\ref{calc2}) has one extra term, which we labeled $(III)$, and which should consequently be null. This term can be rewritten as a surface term, namely:
\begin{equation}\label{EH-Surface}
    \int_U\sqrt {-g} g_{\mu\nu}\delta R^{\mu\nu} d^4x=-2\int_{\delta U} \sqrt{|h|}\delta K d^3x\,,
\end{equation}
where, $\delta U$ is the boundary of $U$, $h$ is the determinant of the 3-metric induced on the $\delta U$ and, $K$ is the trace of the extrinsic curvature\footnote{For a definition of extrinsic curvature see \cite{wald}}.

We can not, despite this, simply make this last term equal to zero, since this would imply that we would have to fix (on the boundary $\delta U$), not only the metric, but also it's derivatives but, we don't have enough degrees of freedom to do this and so the boundary term given by eq. (\ref{EH-Surface}) can't be null.

The solution is to correct the Einstein-Hilbert action so that there is no boundary term.  This is done by adding a term in the following way:
\begin{equation}\label{EH-action-corr}
    S_{EH}'=S_{EH}+\frac{1}{8\pi G}\int_{\delta U} \sqrt{|h|}\delta K d^3x\,.
\end{equation}
For this action, variation with respect to the metric, does lead to Einstein's equations \emph{in vacuo}, ie. $G_{\mu \nu}=0$. There are however different ways to derive this equation, see for example \cite{ohanian}.

\section{The equations in the presence of ``matter fields''}

The action we have been considering up to this point is only associated to the gravitational part of the field theory. Consideration of the sources of the field --- to use a classical language ---, leads us to add another sector to the action, this is the so called mater sector that is associated to the ``mater fields''.

The matter action is defined in the following way:
\begin{equation}\label{matter-action}
    S_M=\int_U \sqrt{-g}\mathcal{L}_M(g_{\mu \nu},\psi)d^4x\,,
\end{equation}
where, $\mathcal{L}_M$, is the matter lagrangian. Variation of this matter action with respect to the metric leads to the stress-energy tensor defined as:
\begin{equation}\label{Tdef}
    T_{\mu\nu}=-\frac{2}{\sqrt{-g}}\frac{\delta \mathcal{L}_M}{\delta
    g^{\mu\nu}}\,.
\end{equation}
If, we finally consider the variation of $S=S_{EH}'+S_M$, with respect to the metric we obtain the full Einstein field equations:
\begin{equation}\label{full-Einstin}
    G_{\mu \nu}=8\pi GT_{\mu \nu}\,.
\end{equation}
These are the equations that govern the dynamics of GR, and any modification of this theory is, at the same time and to some extent, a modification of this action and a modification of these equations.

\section{Exact solutions}

Some of the exact solutions to Einstein's equations are so important, that they merit investigations about wether or not they are still solutions to a general MTG. That is to say, for example: we know of the existence of a static, spherically symmetric, (possibly asymptotically flat), solution to Einstein's equations, but is there such a solution (and, is it the same one) in, say, scalar-tensor theories or, in $f(R)$ modifications of gravity? (For just on example see \cite{Multamaki}.) Similarly, are there any spatially homogeneous and isotropic, constant curvature solutions to a generic MTG? The two solutions just mentioned are, of course, Schwarzschild's solution and Friedman-Lemaître-Robertson-Walker type solutions respectively.

We devote this section, therefore, to a brief exposition of these two solutions, along with a third type of exact solution called wormholes. This is a rather different type of solution, since no known astrophysical objects are described by this type of solution, that is, there are no wormholes that we know of. Wormholes are just a theoretician's probe of the foundations of a gravitational theory, they are ``gedanken-experiments'', and we consider them as such.

\subsection{Schwarzschild's solution}

The description of an empty space-time outside of a static (non-rotating in this case), spherically symmetric, distribution of matter, will be a solution of the Einstein vacuum equations:
\begin{equation}\label{vac-eq}
    R_{\mu \nu}=0\,.
\end{equation}
We suppose the solution to have the same spherical symmetry, present in the source of the field, hence we expect a spherically symmetric space-time (this is an assumption on the spacetime, not a proven fact). We also expect that, far away from the matter, we are in a very good approximation to flat space-time, and so our solution must reduce at spatial infinity to Minkowski's space-time, this is called \emph{asymptotic flatness}.

Starting from the general metric, (see \cite{wald,Rindler}, for details):
\begin{equation}\label{generic}
    ds^2=-e^{A(r)}dt^2+e^{B(r)}dr^2+r^2\left(d\theta^2+\sin^2\theta\, d\varphi^2\right)
\end{equation}
we can determine, the two unknown functions $A(r)$ and $B(r)$, by the use of (\ref{vac-eq}). They turn out to be functions such that:
\begin{equation}\label{coefi}
    \alpha(r)=e^{A(r)}=e^{-B(r)}=\left(1-\frac{2GM}{r}\right)\,.
\end{equation}
Where we have introduced explicitly the gravitational constant $G$,  --- and will continue to do so in this section. Schwarzschild's solution then becomes:
\begin{equation}\label{SCHW}
    ds^2=-\left(1-\frac{2GM}{r}\right)dt^2+\left(1-\frac{2GM}{r}\right)^{-1}dr^2+r^2\left(d\theta^2+\sin^2\theta\, d\varphi^2\right)\,.
\end{equation}
In this metric, the mass $M$ can be identified, by correspondence with Newtonian theory for large spacial distances, with the mass of the central body that we considered to be the source of the field. Therefore, we can immediately conclude, that this metric, in the limit of vanishing mass $M\rightarrow0$, reduces to Minkowski's flat space-time metric, since both $g_{00}\rightarrow 1$, and $g_{11}\rightarrow 1$. We can equally see that, if $r\rightarrow\infty$, we once more recover flat space-time, and therefore we indeed have asymptotic flatness.

One other striking feature of the Schwarzschild solution, is the existence of an
$r$-value for which the metric has a coordinate singularity: $r\equiv r_s=2GM$. This is a consequence of the coordinates used and, at this ``Schwarzschild radius'', only a particle traveling at $v=c$ could remain at rest. Since the only particles capable of this velocity are photons (or zero mass particles in general), no particle can be motionless at $r= r_s$. (See \cite{Gravitation, Rindler}, for a discussion of this subject).

\subsubsection{Birkhoff's theorem}

We mention briefly that, Birkhoff proved a very important extension of the above reasoning. \emph{The unique vacuum solution, with spherical symmetry, to Einstein's equations, is Schwarzschild's solution, even if we relax the assumption about staticity}.

There are important aspects of this theorem in modified theories of gravity, see for example \cite{Birkhoff}.

\subsection{Cosmological solutions}\label{cosmology}

\begin{figure}[c]
 %\begin{center}
  \begin{tabular}{ll}
    %\hline
    \includegraphics[width=2.6in]{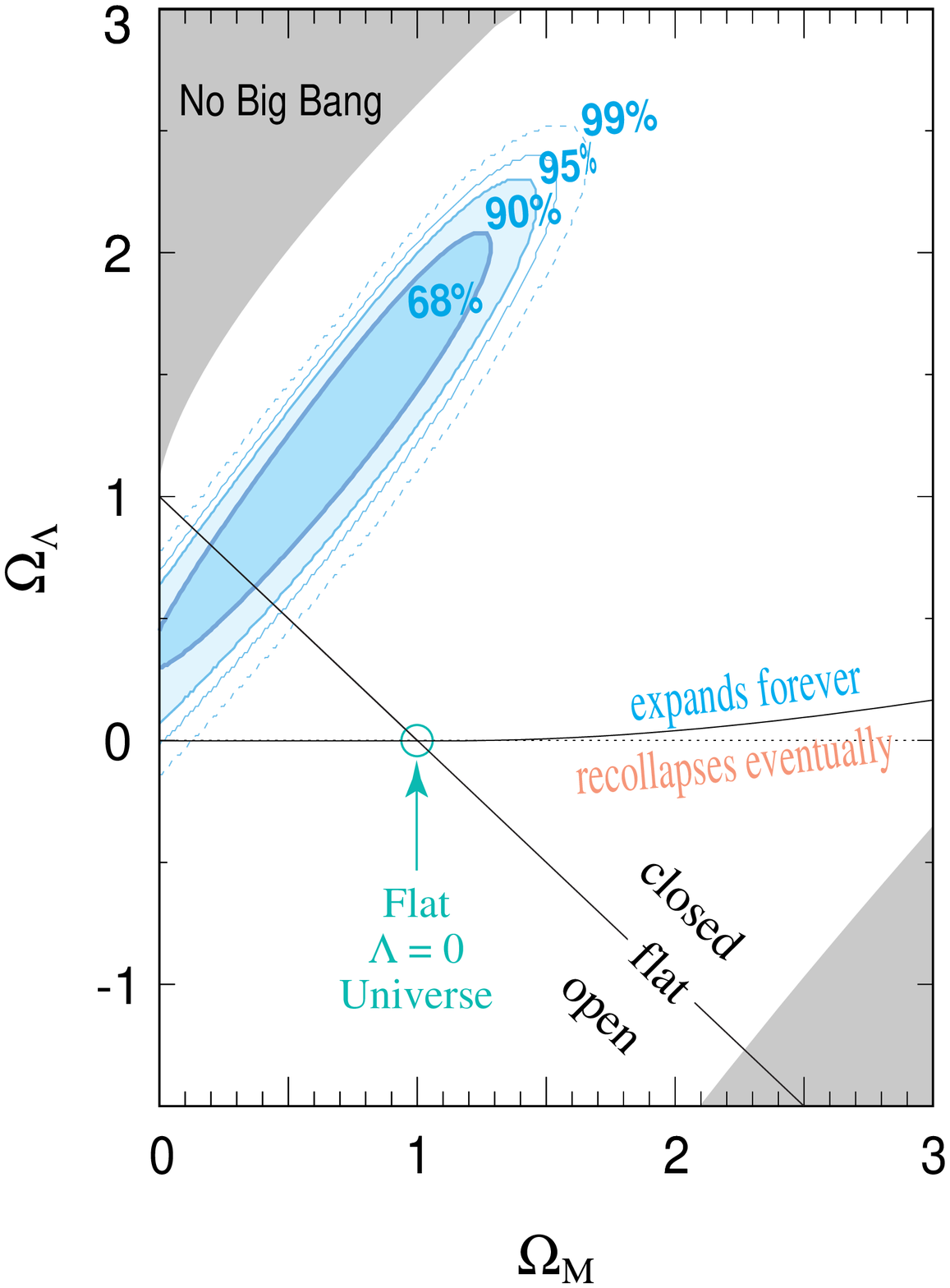} & \includegraphics[width=2.6in]{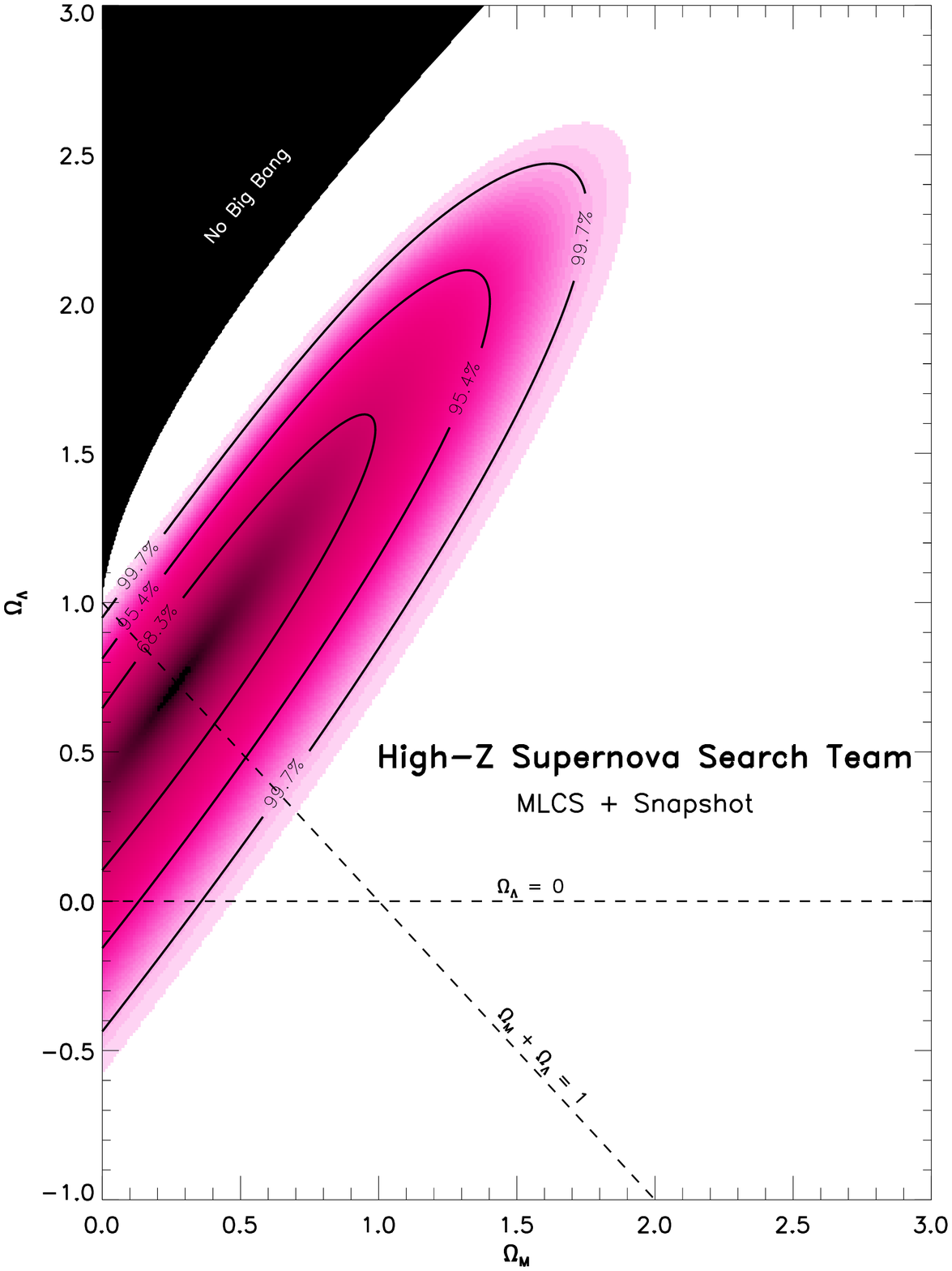} \\
    \includegraphics[width=2.6in]{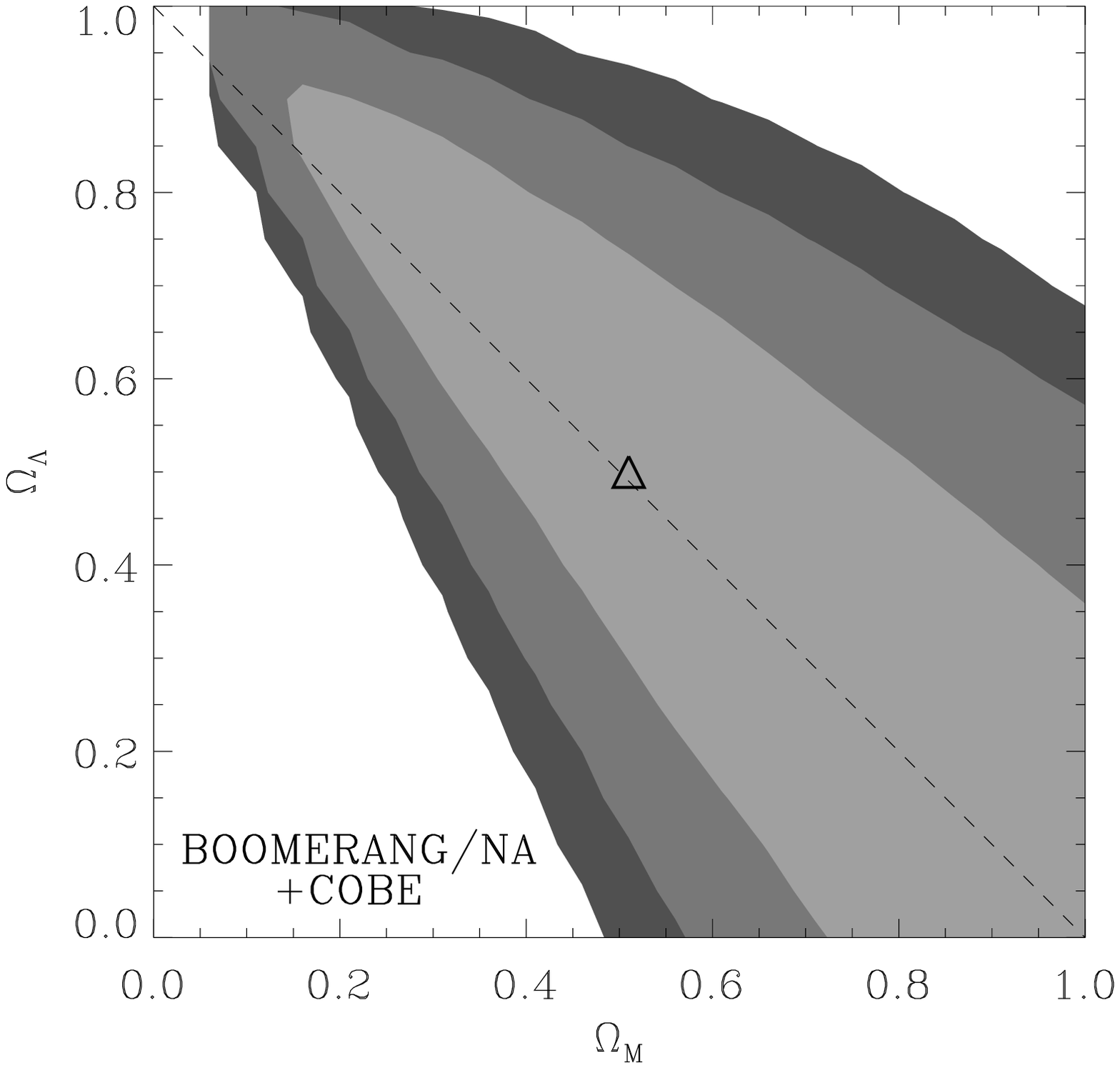} &
      %\hline
  \end{tabular}
  \caption{Constraints in the $\Omega_{\rm M}$-$\Omega_\Lambda$
  plane from: {\bf (top right)} the Supernova Cosmology Project; {\bf (top left)} the High-Z Supernova Team; {\bf (bottom)}the North American flight of the BOOMERANG microwave
  background balloon experiment.
  }\label{Fig:dark}
  %\end{center}
  \label{Fig:dark}
\end{figure}
Cosmology rests on two very well established assumptions: gravity is described by GR; and, the universe is spatially homogeneous and isotropic. We devote, a few brief remarks to the latter of these hypotheses, which is called the \emph{Cosmological Principle}. Homogeneity, which we can roughly consider to mean that the universe `looks the same when we move to another point', is supported by Galaxy redshift surveys, whereas isotropy, taken loosely to mean that there are no preferred directions, is very firmly founded on the properties of Cosmic Microwave Background Radiation (CMB) (For observational data see \cite{CMB}). Before proceeding to a mathematical formulation of these concepts, we add one very important experimentally measured property of the universe, that is Hubble's law: `` the universe is expanding, in such a way that each point is receding from every other, with a velocity proportional to distance''. That is:
\begin{equation}\label{Hubble}
    v_r=Hd\,.
\end{equation}
In this equation $v_r$, is the recessional velocity of a certain point\footnote{This ` point' as we call it, must in fact be, in general, a cluster of galaxies. This is because a single galaxy will move with the Hubble flow, with velocity $v_r$, and will have in addition a peculiar (local) velocity $v_p$. For large enough groups of galaxies the peculiar velocities will average to zero.}, $d$ is the proper distance and, $H$ is Hubble's `constant' that in fact depends on cosmological time.

A space is said to be spatially \emph{homogeneous}, if there exists a one-parameter family of hypersurfaces $\Sigma_t$, foliating the spacetime, such that for any $t$ and $p,\,q \in \Sigma_t$, there exists an isometry of the spacetime metric $g_{\mu\nu}$ which takes $p$ into $q$.

With respect to isotropy, it is very important to note that not all observers can see the universe as isotropic. For example, if we are moving in a spaceship, we will see matter flowing preferably along our direction of motion, and thus the universe will not appear isotropic. This means that, only an observer at rest relative to the mean matter of the universe will consider it to be isotropic. More specifically a spacetime is said to be spatially \emph{isotropic} at each point, if there exists a congruence of timelike curves (or observers), with tangents $u^\mu$, such that it is impossible to construct a geometrically preferred tangent vector orthogonal to $u^\mu$. This is equivalent to the following: given any point $p$, and any two tangent vectors $s^\mu_1,s^\mu_2\in V_P$, there exists an isometry of the metric $g_{\mu\nu}$, that takes $s_1^\mu$ into $s_2^\mu$, (while leaving $p$ and $u^\mu$ at $p$ fixed).

We have as a final corollary that, if the spacetime is both homogeneous and isotropic, then, the homogeneity hypersurfaces $\Sigma_t$, must be orthogonal to the tangents $u^\mu$ to the world lines of the isotropic observers.

The requirements of isotropy and homogeneity, are central since they can be used to restrict the form of $^{(3)}{R}^\alpha_{\phantom{a}\beta\gamma\delta}$, the three tensor on $\Sigma_t$ \cite{wald},
\begin{equation}\label{3-Riemman}
    ^{(3)}R_{\alpha\beta\gamma\delta}=Kh_{\gamma[\alpha}h_{\beta]\delta}\,,
\end{equation}
where $h_{\mu\nu}$ is the restriction of $g_{\mu\nu}$ at $p\in \Sigma_t$ to vector tangent to $\Sigma_t$, and, K is a constant.

In these conditions the metric can be shown to be:
\begin{equation}\label{FLRW}
    ds^2=-dt^2+a^2(t)\left[d\chi^2+\Psi^2_k(\chi)\left(d\vartheta^2+\sin^2\vartheta\ d\phi^2\right)\right]\,,
\end{equation}
where:
\begin{equation}
    \Psi_k(\chi)=
    \left\{ \begin{array}{ll}
              \sin \chi  & \mbox{if $k=+1$}\\
              \chi   & \mbox{if $k=0$}\\
              \sinh \chi & \mbox{if $k=-1$}
            \end{array}
    \right.\,.
\end{equation}
In this equation, the index $k$, is related to the geometry of the 3-surfaces of the universe: $k=0$, describes a universe with a flat Euclidian space geometry; the solution associated with $k=1$, is a positive curvature, hyperspherical universe; and, finally the $k=-1$, is a negative curvature hyperbolic universe.

To solve Einstein's equations~(\ref{full-Einstin}), we need to specify the form of the stress-energy tensor $T_{\mu\nu}$. It turns out that, the most general tensor compatible with the cosmological principle is the following:
\begin{equation}\label{cosmoT}
    T_{\mu\nu}=\rho u_\mu u_\nu+p(g_{\mu\nu}+u_\mu u_\nu)\,,
\end{equation}
where, $\rho$, and $p$ have the usual meaning, that is, the energy density and the isotropic pressure, respectively, of a fluid.

For the metric~(\ref{FLRW}), and the matter content given by~(\ref{cosmoT}), GR yields the following two equations (see, \cite{copeland} for a review):
\begin{eqnarray}
 \label{H-2}
&& H^2 \equiv \left(\frac{\dot{a}}{a}\right)^2
=\frac {8\pi G \rho}{3}-\frac {k}{a^2}\,, \\
\label{dotHeq}
&&\dot{H}=-4\pi G(p+\rho)+\frac{k}{a^2}\,.
\end{eqnarray}
In these equations, $a(t)$ is an unknown function termed the  \emph{scale factor}, it describes the rate of expansion of all distances as a function of time, and,  $H(t)\equiv \left(\frac{\dot{a}}{a}\right)$, is Hubble's constant, the  same constant relating the recessional velocity to distance in~(\ref{Hubble}).

From the conservation of the stress-energy tensor we can derive the following continuity equation:
\begin{equation}
\dot{\rho}+3 H(\rho+p)=0\,.
\label{conteq}
\end{equation}
We can use this to eliminate the term $k/a^2$ from~(\ref{H-2})~and~(\ref{dotHeq}) and obtain:
\begin{eqnarray}
\label{acceleq}
\frac{\ddot{a}}{a}=
-\frac {4 \pi G}{3} \left(\rho+3p\right)\,.
\end{eqnarray}
From this last equation we see that, the expansion of the universe will only be accelerated for $\rho+3p<0$.

Equation (\ref{H-2}), is a balance equation, and it can therefore be written in the form:
\begin{equation}
\Omega(t)-1=
\frac{K}{(aH)^2}\,,
\end{equation}
where,
\begin{equation}
    \left\{ \begin{array}{ll}
               \Omega >1 &\mbox{ or }\rho>\rho_c \mbox{ if $k=+1$}\\
               \Omega =1  &\mbox{ or }\rho=\rho_c \mbox{ if $k=0$}\\
               \Omega <1&\mbox{ or }\rho<\rho_c \mbox{ if $k=-1$}
            \end{array}
    \right.\,.
\end{equation}
Despite the natural expectation, that the expansion of the universe is decelerating, all experimental evidence points startlingly to contrary. That is, the universe, is accelerating! The direct confirmation of this fact, comes from the observation of the luminosity distances of high redshift supernovae.\cite{observations} Further support, comes from CMB anisotropy observations, and, Large Scale Structure (LSS), observations \cite{copeland}. The FLRW model is usually characterized by the following quantities:
\begin{equation}\label{omegas}
    \begin{array}{cc}
    \Omega_M=\frac{\rho_M}{\rho_c}=\frac{8\pi G}{3H^2}\rho_{0M} &\,, \Omega_\Lambda=\frac{\rho_\Lambda}{\rho_c}=\frac{\Lambda}{3H^2}\,.\\
    \end{array}
\end{equation}
$\Omega_\Lambda$ is the density of vacuum energy. The evidence suggests the preferred model is $(\Omega_{0M},\Omega_\Lambda)=(0.3,0.7)$, see fig. (\ref{Fig:dark}).
To solve this problem two main avenues (there are other alternatives, see \cite{copeland}) present themselves before us. We can, in a diversity of ways, modify the matter action, by introducing some new source field, (or a number of source fields, having one among various possible couplings), that fulfils the conditions for accelerated expansion (namely, the already mentioned $\rho+3p>0$ condition). Among these, we find the numerous forms of the so called dark energy, which include the abundant use of one, or more, scalar fields to account for the dynamical aspects of the problem. \cite{copeland} We can, alternatively, modify the gravitational part of the action, and in this way assume that gravity behaves unlike GR at long distances, and thus account for the observed properties of the universe\cite{sot-PhD}.

In this work, we will focus more closely on this second avenue. We however, center our attention not so much on the characteristics of these modifications of gravity that account for this late acceleration of the universe, but more on the existence of certain type of solutions that we find to be probes of the consistency of certain theories.

\subsection{Wormhole solutions}\label{wormholes}

The general approach to the solution of Einstein's equation in the presence of matter (\ref{full-Einstin}), is to specify the sources of the fields embodied in the stress-energy tensor $T^{\mu\nu}$, solve the differential equations and thus, obtain the metric $g_{\mu\nu}$, which we consider to be the solution. The inverse process, is nonetheless also possible. We consider a solution, that is a metric tensor $g_{\mu\nu}$, and  solve the equations for the stress-energy tensor components $T^{\mu\nu}$.

However, despite the validity of this process, not all stress-energy tensors, thus obtained, are acceptable. Certain conditions, called \emph{energy conditions}, must be satisfied by these tensors, so that we can correctly consider them sources of the gravitational field.

We write these conditions for a generic stress-energy tensor $T^{\mu\nu}$, and, we also enumerate the same conditions for a tensor of the form:
\begin{equation}\label{fluidT}
    T^{\mu\nu}=\left[\begin{array}{cccc}
                   \rho & 0 & 0 & 0 \\
                   0 & p_1 & 0 & 0 \\
                   0 & 0 & p_2 & 0 \\
                   0 & 0 & 0 & p_3 \\
                 \end{array}
               \right]\,.
\end{equation}
The pointwise energy conditions are \cite{Lobo-PhD}:
\begin{description}
  \item[Null Energy Condition (NEC):] for any null vector $k^\mu$
  \begin{equation}\label{NEC-gen}
    T_{\mu\nu}k^\mu k^\nu\geq 0\,.
  \end{equation}
  Or, for a stress-energy tensor of the form (\ref{fluidT}):
  \begin{equation}\label{NEC}
    \forall i,\quad\rho+p_i\geq0
  \end{equation}
  \item[Weak Energy Condition (WEC):] for any timelike vector $U^\mu$
  \begin{equation}\label{WEC-gen}
    T_{\mu\nu}U^\mu U^\nu\geq 0\,.
  \end{equation}
  Physically $T_{\mu\nu}U^\mu U^\nu$ is the energy density measured by any timelike observer with four-velocity $U^\mu$. In terms of the above stress-energy tensor we have,
  \begin{equation}\label{WEC}
    \rho\geq0\quad\mbox{and}\quad\forall i,\quad\rho+p_i\geq0
  \end{equation}
  By continuity the WEC implies the NEC.
  \item[Strong Energy Condition (SEC):] for any timelike vector $U^\mu$, the following inequality holds,
  \begin{equation}\label{SEC-gen}
    \bigg(T_{\mu\nu}-\frac{T}{2}g_{\mu\nu}\bigg)U^\mu U^\nu\geq0\,.
  \end{equation}
  Where $T$ is the trace of the stress-energy tensor.  And, in terms of the our special stress-energy tensor,
  \begin{equation}\label{SEC}
    \forall i,\quad\rho+p_i\geq0\qquad\rho+\sum_ip_i\geq0\,.
  \end{equation}
  The SEC implies the NEC but no necessarily the WEC.
  \item[Dominant Energy Condition (DEC):] for any timelike vector $U^\mu$,
  \begin{equation}\label{DEC-gen}
   T_{\mu\nu}U^\mu U^\nu\geq 0\quad\mbox{and}\quad T_{\mu\nu}U^\mu \quad\mbox{is not spacelike}\,.
  \end{equation}
  The DEC, implies that the local energy density be positive and, that the energy flux should be timelike or null. The DEC also implies the WEC and therefore the NEC, but not necessarily the SEC. For the case of the $diag\left\{\rho,p_1,p_2,p_3\right\}$ stress-energy tensor, we have,
  \begin{equation}\label{DEC}
    \rho\geq0\quad\mbox{and}\quad\forall i, \quad p_i\in [-\rho,\rho]\,.
  \end{equation}
\end{description}
For further details and, for the definition the \emph{Averaged Energy Conditions} see, \cite{Lobo-PhD}.
Wormholes are a type of solutions that violate all the pointwise energy conditions. Therefore as was already stated, we only consider this type of solutions as thought experiments or, as a tool for teaching General  Relativity \cite{mor-thor}.

A wormhole solution is characterized by the following space-time metric:
\begin{equation}\label{WH}
    ds^2=-e^{2\Phi(r)}dt^2+\frac{dr^2}{1-b(r)/r}+r^2\left(d\theta^2+\sin^2\theta\, d\varphi^2\right)\,,
\end{equation}
where $\Phi(r)\mbox{ and, }b(r)$, are arbitrary functions of the coordinate $r$. The function $\Phi(r)$, is related to the gravitational redshift and, is therefore called \emph{redshift function}; as for $b(r)$, it is termed the \emph{shape function}, since it determines the shape of the wormhole throat, as will be shortly seen.

The coordinate $r$, is non-monotonic, that is, it decreases form $+\infty$ to a minimum value $r_0$, representing the wormhole throat, where $b(r_0)=r_0$, and them increases from $r_0$ to $+\infty$.

Despite the fact that $g_{rr}$ diverges at the wormhole throat, the proper distance,
\begin{equation}\label{proper-d}
    l(r)=\pm\int_{r_0}^r\frac{dr}{\sqrt{1-b(r)/r}}\,,
\end{equation}
must be finite everywhere. Also the circumference of a circle of radius $r$ is $2\pi r$. Using $l(r)$, the line element (\ref{WH}), can be set in the form:
\begin{equation}\label{WH-proper}
    ds^2=-e^{2\Phi(r)}dt^2+dl^2+r(l)^2\left(d\theta^2+\sin^2\theta\, d\varphi^2\right)\,.
\end{equation}
Consider an equatorial slice $\theta=\pi/2$, of a fixed time $dt=0$ version of metric~(\ref{WH}):
\begin{equation}\label{fixed-t}
    ds^2=\frac{dr^2}{1-b(r)/r}+r^2\left(d\theta^2+\sin^2\theta\, d\varphi^2\right)\,,
\end{equation}
to `visualize' this metric we will embed it in a 3-dimensional Euclidian space written in cylindrical coordinates:
\begin{equation}\label{cyl-3d}
    ds^2=dz^2+dr^2+r^2d\varphi^2\,,
\end{equation}
if we consider the embedded surface to have an equation $z=z(r)$, we obtain for the line element~(\ref{cyl-3d}):
\begin{equation}\label{cyl-3d2}
    ds^2=\left[1+\left(\frac{dz}{dr}\right)^2\right]dr^2+r^2d\varphi^2\,.
\end{equation}
Confronting,~(\ref{fixed-t}), with~(\ref{cyl-3d2}), we easily conclude that:
\begin{equation}\label{emb-surf}
        \frac{dz}{dr}=\pm\left(\frac{r}{b(r)}-1\right)^{-1/2}\,.
\end{equation}
For the surface to be vertical, ie. $dz/dr\rightarrow\infty$, we must have a minimum radius at $r=b(r)=r_0$, the wormhole throat. Also to have asymptotic flatness, we need $dz/dr\rightarrow 0$ as, $r\rightarrow \infty$.

To have a wormhole solution the embedded surface must `flare out', that is the inverse of the imbedding function $r=r(z)$, must satisfy, $d^2r/dz^2>0$ near the throat $r_0$. From~(\ref{emb-surf}), we find:
\begin{equation}\label{flare-out}
        \frac{d^2r}{dz^2}=\frac{b-b'r}{2b^2}>0\,,
\end{equation}
 so that at the throat, $b'(r_0)<1$.
 There are also a set of conditions, aimed at insuring that a traveler could actually use a wormhole. These are called \emph{traversability conditions}, see \cite{Lobo-PhD, mor-thor}, for a detailed account. We refer explicitly, two of these conditions: one is linked with the gravitational acceleration felt by an observer at the initial and final points of his journey, this is, $g=-(1-b/r)^{-1/2}\Phi'\simeq -\Phi'$, and should be less the or equal to earth's acceleration so that the condition $|\Phi'|\leq g_\oplus$, must be met; the other, is related to the redshift of a signal sent from the initial or final point, towards infinity, this is, $\Delta \lambda/ \lambda = e^{-\Phi}-1 \approx-\Phi$ so that, we must have $|\Phi|\ll 1$. Given the first condition, a usual choice is to have the redshift function constant $\Phi'=0$.

 This section is just a very short presentation of this subject, for the original paper that set this subject in motion see \cite{mor-thor}, and more detailed discussions can be found at \cite{Lobo-PhD, Visser}.

 %%%%%%%%%%%%%%%%%%%%%%%%%%%%%%%%%%%%%%%%%%%%%%%%%%%%%%%%%%%%%%%%%%%%%%%%%%%%%%%%%%%%%%%%%%%%
% --------------------------------------chain--------------------------
%\include{referenc}
%\end{document}
% ----------------------------------------------------------------\left(

    \part{Applications}

    % ----------------------------------------------------------------
% Chapter on f(R) Theories of Gravity ************************************************
% **** -----------------------------------------------------------
%\documentclass[12pt]{report}
%\usepackage{dsfont}
%\usepackage{amssymb}
%\usepackage{graphicx}
% ----------------------------------------------------------------
%\vfuzz2pt % Don't report over-full v-boxes if over-edge is small
%\hfuzz2pt % Don't report over-full h-boxes if over-edge is small

% ----------------------------------------------------------------
%\begin{document}

% ----------------------------------------------------------------
\chapter{$f(R)$ gravity and wormholes}
Recent developments in observational cosmology brought about the need for two different phases of accelerated expansion of the universe. The first is inflation, that supposedly occurred in the early stages of the universe, and was succeeded by a radiation dominated expansion era. The second phase is the late-time cosmic acceleration, that is occurring in our present era. For the reasons we stated above, in section, \ref{cosmology}, (equation \ref{acceleq}), there is an intrinsic difficulty in the description of accelerated expansion since the condition for it's occurrence in FLRW models is $\rho+3p<0$, and therefore if we use a fluid with an equation of state $\omega=p/\rho$, this condition becomes $\omega<-1/3$, which in turn implies the use of a negative pressure fluid.

There is a `natural' choice for the source of this acceleration. The cosmological constant $\Lambda$, corresponding to a $\omega_\Lambda=-1$. The introduction of this constant, leads to a period of accelerated expansion. In fact, upon variation with respect to the metric, the action,
\begin{equation}\label{L-action}
    S=\int d^{\,4}x \sqrt{-g}(R-2\Lambda)\,,
\end{equation}
produces the following generalization of the Einstein Field Equations:
\begin{equation}
  R_{\mu\nu} - {1\over 2}Rg_{\mu\nu}
  + \Lambda g_{\mu\nu}
  = 8\pi GT_{\mu\nu}\,.
  \label{einsteinl}
\end{equation}
This interpretation however, suffers from serious drawbacks since this cosmological constant can not be easily (if at all), interpreted as a vacuum density in the context of field theories. The $\Lambda$ term, in equation (\ref{einsteinl}), is the source of another problem: once the possibility of a non-null cosmological constant has been introduced, setting this term to zero needs to be justified, just as setting any other term in any other equation to null. One is thus led to the so called `cosmological constant problem' \cite{Carroll-L}.

Another possibility, for the source of these accelerated expansion periods, comes from scalar fields $\phi$, with slowly varying potentials. These have been extensively studied and many variations of this theme exist --- Quintessence, K-essence, Tachyon Fields, models using the Chaplygin gas, to name but a few \cite{copeland}. However, no particular choice of field and potential, seems to generate a model in perfect accord with the experimental data. A particularly thorny problem appears to be, the possibility that the dark energy equation of state ``crosses the phantom divide'' (see \cite{copeland}, sec. V-D), that is, it may be in the region $\omega_\phi <-1$.

 We can not help but to feel some discomfort, when confronted with all these `material components', purposely added to the right-hand-side of Einstein's Field Equations, so that they may drive the two periods of accelerated expansion. Consequently, we are led to consider the possibility that gravity alone might, once conveniently modified account for the accelerated expansion, or at least, for the observations and theoretical difficulties that motivated the acceleration in the first place.

\section{The $f(R)$ modified theories of gravity}
%-----------------------------------------------

A modification of the Einstein-Hilbert gravitational Lagrangian
density involving an arbitrary function of the scalar invariant,
$f(R)$, was considered in \cite{Bu70}, and further developed in other works, for an up to date review see
\cite{fr-review}.

In this context, a renaissance of $f(R)$ modified theories of
gravity has been verified in an attempt to explain the late-time
accelerated expansion of the Universe (see Refs.
\cite{sot-faraoni} for a review). Earlier interest in $f(R)$
theories was motivated by inflationary scenarios as for instance,
in the Starobinsky model, where $f(R)=R-\Lambda + \alpha R^2$ was
considered \cite{Starobinsky:1980te}, and mentioned in the Introduction. In fact, it was shown that
the late-time cosmic acceleration can be indeed explained within
the context of $f(R)$ gravity \cite{Carroll:2003wy}. Furthermore,
the conditions of viable cosmological models have been derived
\cite{fr-review,viablemodels}, and an explicit coupling of an arbitrary
function of $R$ with the matter Lagrangian density has also been
explored \cite{coupling1}. Relative to the Solar System regime,
severe weak field constraints seem to rule out most of the models
proposed so far \cite{solartests,Olmo07}, although viable models
do exist \cite{solartests2}. In the context of dark matter, the
possibility that the galactic dynamics of massive test particles
may be understood without the need for dark matter was also
considered in the framework of $f(R)$ gravity models
\cite{darkmatter}.

We remark here, that many phenomenologically interesting terms can be accounted for, in the context of $f(R)$ theories, if we consider a series expansion of $f$,
\begin{equation}
f(R)=\cdots \frac{\alpha_i}{R^i} \cdots+\frac{\alpha_2}{R^2}+\frac{\alpha_1}{R}-2\Lambda+R+\beta_2 R^2+\beta_3 R^3\cdots\beta_jR^j\cdots,
\label{eq:fr}
\end{equation}
with, the $\alpha_i$ and $\beta_j$ having suitable dimensions. In the minimal case $\alpha_i=\beta_j=0$, we recover the Einstein-Hilbert action with a cosmological constant $\Lambda$ eq. (\ref{L-action}).

The metric formalism is usually considered in the literature,
this consists in varying the action with respect to $g^{\mu\nu}$.
However, other alternative approaches have been considered in the
literature, namely, the Palatini
formalism~\cite{Palatini,Sotiriou:2006qn}, already mentioned in chapter 2, where the metric and
the connections are treated as separate variables; and the
metric-affine formalism, where the matter part of the action now
depends and is varied with respect to the
connection~\cite{Sotiriou:2006qn}. The action for $f(R)$ modified
theories of gravity is given by
\begin{equation}
S=\frac{1}{2\kappa}\int d^4x\sqrt{-g}\;f(R)+S_M(g^{\mu\nu},\psi)
\,,
\end{equation}
where $\kappa =8\pi G$; throughout this chapter we consider
$\kappa=1$ for notational simplicity. Recall that $S_M(g^{\mu\nu},\psi)$ is
the matter action, defined as $S_M=\int d^4x\sqrt{-g}\;{\cal
L}_m(g_{\mu\nu},\psi)$, where ${\cal L}_m$ is the matter
Lagrangian density, in which matter is minimally coupled to the
metric $g_{\mu\nu}$ and $\psi$ collectively denotes the matter
fields.

Now, using the metric approach, by varying the action with respect
to $g^{\mu\nu}$, provides the following field equation
\begin{equation}
FR_{\mu\nu}-\frac{1}{2}f\,g_{\mu\nu}-\nabla_\mu \nabla_\nu
F+g_{\mu\nu}\Box F=\,T^m_{\mu\nu} \,,
    \label{field:eq}
\end{equation}
where $F\equiv df/dR$. Considering the contraction of Eq.
(\ref{field:eq}), provides the following relationship
\begin{equation}
FR-2f+3\,\Box F=\,T \,,
 \label{trace}
\end{equation}
which shows that the Ricci scalar is a fully dynamical degree of
freedom, and $T=T^{\mu}{}_{\mu}$ is the trace of the stress-energy
tensor.

In this Chapter, we extend the analysis of static and spherically
symmetric spacetimes considered in the literature (for instance,
see \cite{Multamaki}), and analyze traversable wormhole
geometries in $f(R)$ modified theories of gravity. Wormholes --- as mentioned in section \ref{wormholes} --- are
hypothetical tunnels in spacetime, possibly through which
observers may freely traverse. However, it is important to
emphasize that these solutions are primarily useful as
``gedanken-experiments'' and as a theoretician's probe of the
foundations of general relativity. In classical general
relativity, wormholes are supported by exotic matter, which
involves a stress-energy tensor that violates the null energy
condition (NEC) \cite{mor-thor,Visser}. Note that the NEC is
given by $T_{\mu\nu}k^\mu k^\nu \geq 0$, (Eq. \ref{NEC-gen}), where $k^\mu$ is {\it
any} null vector. Thus, it is an important and intriguing
challenge in wormhole physics to find a realistic matter source
that will support these exotic spacetimes. Several candidates have
been proposed in the literature, amongst which we refer to
solutions in higher dimensions, for instance in
Einstein-Gauss-Bonnet theory \cite{EGB1,EGB2}, wormholes on the
brane \cite{braneWH1}; solutions in Brans-Dicke theory
\cite{Nandi:1997en}; wormhole solutions in semi-classical gravity
(see Ref. \cite{Garattini:2007ff} and references therein); exact
wormhole solutions using a more systematic geometric approach were
found \cite{Boehmer:2007rm}; geometries supported by equations of
state responsible for the cosmic acceleration \cite{phantomWH},
solutions in conformal Weyl gravity were found \cite{Weylgrav},
and thin accretion disk observational signatures were also
explored \cite{Harko:2008vy}, etc (see Refs.
\cite{Lemos:2003jb,Lobo:2007zb} for more details and
 for a recent review).

Thus, we explore the possibility that wormholes be supported by
$f(R)$ modified theories of gravity. It is an effective stress
energy, which may be interpreted as a gravitational fluid, that is
responsible for the null energy condition violation, thus
supporting these non-standard wormhole geometries, fundamentally
different from their counterparts in general relativity. We also
impose that the matter threading the wormhole satisfies the energy
conditions.

This Chapter is organized in the following manner: In Sec.
\ref{Sec:3-II}, the spacetime metric, the effective field equations
and the energy condition violations in the context of $f(R)$
modified theories of gravity are analyzed in detail. In Sec.
\ref{Sec:III}, specific solutions are explored, and we conclude in
Sec. \ref{Sec:conclusion}.

\section{Wormhole geometries in $f(R)$ gravity}\label{Sec:3-II}

\subsection{Spacetime metric and gravitational field equations}

Consider the wormhole geometry given by the following static and
spherically symmetric metric
\begin{equation}
ds^2=-e^{2\Phi(r)}dt^2+\frac{dr^2}{1-b(r)/r}+r^2\,(d\theta^2 +\sin
^2{\theta} \, d\phi ^2) \,,
    \label{metric}
\end{equation}
where $\Phi(r)$ and $b(r)$ are arbitrary functions of the radial
coordinate, $r$, denoted as the redshift function, and the shape
function, respectively \cite{mor-thor}. The radial coordinate
$r$ is non-monotonic (as was already mentioned in Chapter 2) in that it decreases from infinity to a
minimum value $r_0$, representing the location of the throat of
the wormhole, where $b(r_0)=r_0$, and then it increases from $r_0$
back to infinity.

A fundamental property of a wormhole is --- as was stated earlier Sec.~\ref{wormholes}~---  that a flaring out
condition of the throat, given by $(b-b^{\prime}r)/b^{2}>0$, is
imposed \cite{mor-thor}, and at the throat
$b(r_{0})=r=r_{0}$, the condition $b^{\prime}(r_{0})<1$ is imposed
to have wormhole solutions. It is precisely these restrictions
that impose the NEC violation in classical general relativity.
Another condition that needs to be satisfied is $1-b(r)/r>0$. For
the wormhole to be traversable, one must demand that there are no
horizons present, which are identified as the surfaces with
$e^{2\Phi}\rightarrow0$, so that $\Phi(r)$ must be finite
everywhere. In the analysis outlined below, we consider that the
redshift function is constant, $\Phi'=0$, which simplifies the
calculations considerably, and provide interesting exact wormhole
solutions (If $\Phi'\neq 0$, the field equations become forth
order differential equations, and become quite intractable).

The trace equation (\ref{trace}) can be used to simplify the field
equations and then can be kept as a constraint equation. Thus,
substituting the trace equation into Eq. (\ref{field:eq}), and
re-organizing the terms we end up with the following gravitational
field equation
\begin{equation}
G_{\mu\nu}\equiv R_{\mu\nu}-\frac{1}{2}R\,g_{\mu\nu}= T^{{\rm
eff}}_{\mu\nu} \,,
    \label{field:eq2}
\end{equation}
where the effective stress-energy tensor is given by $T^{{\rm
eff}}_{\mu\nu}= T^{(c)}_{\mu\nu}+\tilde{T}^{(m)}_{\mu\nu}$. The
term $\tilde{T}^{(m)}_{\mu\nu}$ is given by
\begin{equation}
\tilde{T}^{(m)}_{\mu\nu}=T^{(m)}_{\mu\nu}/F \,,
\end{equation}
and the curvature stress-energy tensor, $T^{(c)}_{\mu\nu}$, is
defined as
\begin{eqnarray}
T^{(c)}_{\mu\nu}=\frac{1}{F}\left[\nabla_\mu \nabla_\nu F
-\frac{1}{4}g_{\mu\nu}\left(RF+\Box F+T\right) \right]    \,.
    \label{gravfluid}
\end{eqnarray}

It is also interesting to consider the conservation law for the
above curvature stress-energy tensor. Taking into account the
Bianchi identities, $\nabla^\mu G_{\mu\nu}=0$, and the
diffeomorphism invariance of the matter part of the action, which
yields $\nabla^\mu T^{(m)}_{\mu\nu}=0$, we verify that the
effective Einstein field equation provides the following
conservation law
\begin{equation}\label{conserv-law}
\nabla^\mu T^{(c)}_{\mu\nu}=\frac{1}{F^2}
T^{(m)}_{\mu\nu}\nabla^\mu F  \,.
\end{equation}

Relative to the matter content of the wormhole, we impose that the
stress-energy tensor that threads the wormhole satisfies the
energy conditions, and is given by the following anisotropic
distribution of matter
\begin{equation}
T_{\mu\nu}=(\rho+p_t)U_\mu \, U_\nu+p_t\,
g_{\mu\nu}+(p_r-p_t)\chi_\mu \chi_\nu \,,
\end{equation}
where $U^\mu$ is the four-velocity, $\chi^\mu$ is the unit
spacelike vector in the radial direction, i.e.,
$\chi^\mu=\sqrt{1-b(r)/r}\,\delta^\mu{}_r$. $\rho(r)$ is the
energy density, $p_r(r)$ is the radial pressure measured in the
direction of $\chi^\mu$, and $p_t(r)$ is the transverse pressure
measured in the orthogonal direction to $\chi^\mu$. Taking into
account the above considerations, the stress-energy tensor is
given by the following profile: $T^{\mu}{}_{\nu}={\rm
diag}[-\rho(r),p_r(r),p_t(r),p_t(r)]$.

Thus, the effective field equation (\ref{field:eq2}) provides the
following relationships
\begin{eqnarray}
\frac{b'}{r^2}&=&\frac{\rho}{F}+\frac{H}{F}
  \,,    \label{fieldtt}
     \\
-\frac{b}{r^3}&=&\frac{p_r}{F}+\frac{1}{F}\Bigg\{\left(1-\frac{b}{r}\right)
\times
   \nonumber    \\
&&\times\left[F'' -F'\frac{b'r-b}{2r^2(1-b/r)}\right] -H\Bigg\}
  \,,  \label{fieldrr} \\
-\frac{b'r-b}{2r^3}
     &=&\frac{p_t}{F}+\frac{1}{F}\left[\left(1-\frac{b}{r}\right)
     \frac{F'}{r}
     -H\right]
     \label{fieldthetatheta}  \,,
\end{eqnarray}
where the prime denotes a derivative with respect to the radial
coordinate, $r$. The term $H=H(r)$ is defined as
\begin{equation}
H(r)=\frac{1}{4}\left(FR+\Box F +T\right) \,,
\end{equation}
for notational simplicity. The curvature scalar, $R$, is given by
\begin{eqnarray}
R&=& \frac{2b'}{r^2}
    \,,
    \label{Ricciscalar}
\end{eqnarray}
and $\Box F$ is provided by the following expression
\begin{equation}
\Box F=\left(1-\frac{b}{r}\right)\left[F''
-\frac{b'r-b}{2r^2(1-b/r)}\,F'+\frac{2F'}{r}\right] \,.
\end{equation}

Note that the gravitational field equations
(\ref{fieldtt})-(\ref{fieldthetatheta}), can be reorganized to
yield the following relationships:
\begin{eqnarray}
\label{generic1} \rho&=&\frac{Fb'}{r^2}\,,
       \\
\label{generic2}
p_r&=&-\frac{bF}{r^3}+\frac{F'}{2r^2}(b'r-b)-F''\left(1-\frac{b}{r}\right)
     \,,   \\
\label{generic3}
p_t&=&-\frac{F'}{r}\left(1-\frac{b}{r}\right)+\frac{F}{2r^3}(b-b'r)\,,
\end{eqnarray}
which are the generic expressions of the matter threading the
wormhole, as a function of the shape function and the specific
form of $F(r)$. Thus, by specifying the above functions, one
deduces the matter content of the wormhole.

One may now adopt several strategies to solve the field equations.
For instance, if $b(r)$ is specified, and using a specific
equation of state $p_r=p_r(\rho)$ or $p_t=p_t(\rho)$ one can
obtain $F(r)$ from the gravitational field equations and the
curvature scalar in a parametric form, $R(r)$, from its definition
via the metric. Then, once $T=T^{\mu}{}_{\mu}$ is known as a
function of $r$, one may in principle obtain $f(R)$ as a function
of $R$ from Eq. (\ref{trace}).

\subsection{Energy condition violations}

A fundamental point in wormhole physics is the energy condition
violations, as mentioned above. However, a subtle issue needs to
be pointed out in modified theories of gravity, where the
gravitational field equations differ from the classical
relativistic Einstein equations. More specifically, we emphasize
that the energy conditions arise when one refers back to the
Raychaudhuri equation for the expansion where a term
$R_{\mu\nu}k^\mu k^\nu$ appears, with $k^\mu$ any null vector. The
positivity of this quantity ensures that geodesic congruences
focus within a finite value of the parameter labeling points on
the geodesics. However, in general relativity, through the
Einstein field equation one can write the above condition in terms
of the stress-energy tensor given by $T_{\mu\nu}k^\mu k^\nu \ge
0$. In any other theory of gravity, one would require to know how
one can replace $R_{\mu\nu}$ using the corresponding field
equations and hence using matter stresses. In particular, in a
theory where we still have an Einstein-Hilbert term, the task of
evaluating $R_{\mu\nu}k^\mu k^\nu$ is trivial. However, in $f(R)$
modified theories of gravity under consideration, things are not
so straightforward.

Now the positivity condition, $R_{\mu\nu}k^\mu k^\nu \geq 0$, in
the Raychaudhuri equation provides the following form for the null
energy condition $T^{{\rm eff}}_{\mu\nu} k^\mu k^\nu\geq 0$,
through the modified gravitational field equation
(\ref{field:eq2}), and it this relationship that will be used
throughout this work. For this case, in principle, one may impose
that the matter stress-energy tensor satisfies the energy
conditions and the respective violations arise from the higher
derivative curvature terms $T^{(c)}_{\mu\nu}$. Another approach to
the energy conditions considers in taking the condition
$T_{\mu\nu} k^\mu k^\nu\ge 0$ at face value. Note that this is
useful as using local Lorentz transformations it is possible to
show that the above condition implies that the energy density is
positive in all local frames of reference. However, if the theory
of gravity is chosen to be non-Einsteinian, then the assumption of
the above condition does not necessarily imply focusing of
geodesics. The focusing criterion is different and will follow
from the nature of $R_{\mu\nu} k^\mu k^\nu$.

Thus, considering a radial null vector, the violation of the NEC,
i.e., $T_{\mu\nu}^{{\rm eff}}\,k^\mu k^\nu < 0$ takes the
following form
\begin{equation}
\rho^{{\rm eff}}+p_r^{{\rm
eff}}=\frac{\rho+p_r}{F}+\frac{1}{F}\left(1-\frac{b}{r}\right)
\left[F''-F'\frac{b'r-b}{2r^2(1-b/r)}\right]\,.
     \label{NECeff}
\end{equation}
Using the gravitational field equations, inequality (\ref{NECeff})
takes the familiar form
\begin{equation}
\rho^{{\rm eff}}+p_r^{{\rm eff}}=\frac{b'r-b}{r^3}\,,
     \label{NECeff2}
\end{equation}
which is negative by taking into account the flaring out
condition, i.e., $(b'r-b)/b^2<0$, considered above.

At the throat, one has the following relationship
\begin{equation}
\rho^{{\rm eff}}+p_r^{{\rm eff}}|_{r_0}=
\frac{\rho+p_r}{F}\Big|_{r_0}+
\frac{1-b'(r_0)}{2r_0}\frac{F'}{F}\Big|_{r_0}<0 \,.
     \label{NECeffb}
\end{equation}
It is now possible to find the following generic relationships for
$F$ and $F'$ at the throat: $F'_0<-2r_0(\rho+p_r)|_{r_0}/(1-b')$
if $F>0$; and $F'_0>-2r_0(\rho+p_r)|_{r_0}/(1-b')$ if $F<0$.

Consider that the matter threading the wormhole obeys the energy
conditions. To this effect, imposing the weak energy condition
(WEC), given by $\rho \geq 0$ and $\rho + p_r \geq 0$, then Eqs.
(\ref{generic1})-(\ref{generic2}) yield the following
inequalities:
\begin{eqnarray}
\frac{Fb'}{r^2}\geq 0 \,, \label{WEC1}
  \\
\frac{(2F+rF')(b'r-b)}{2r^2}-F''\left(1-\frac{b}{r}\right)\geq 0
\,, \label{WEC2}
\end{eqnarray}
respectively.

Thus, if one imposes that the matter threading the wormhole
satisfies the energy conditions, we emphasize that it is the
higher derivative curvature terms that sustain the wormhole
geometries. Thus, in finding wormhole solutions it is fundamental
that the the functions $f(R)$ obey inequalities (\ref{NECeff}) and
(\ref{WEC1})-(\ref{WEC2}).

\section{Specific solutions}\label{Sec:III}

In this section, we are mainly interested in adopting the strategy
of specifying the shape function $b(r)$, which yields the
curvature scalar in a parametric form, $R(r)$, from its definition
via the metric, given by Eq. (\ref{Ricciscalar}). Then, using a
specific equation of state $p_r=p_r(\rho)$ or $p_t=p_t(\rho)$, one
may in principle obtain $F(r)$ from the gravitational field
equations. Finally, once $T=T^{\mu}{}_{\mu}$ is known as a
function of $r$, one may in principle obtain $f(R)$ as a function
of $R$ from Eq. (\ref{trace}).

\subsection{Traceless stress-energy tensor}

An interesting equation of state is that of the traceless
stress-energy tensor, which is usually associated to the Casimir
effect, with a massless field. Note that the Casimir effect is
sometimes theoretically invoked to provide exotic matter to the
system considered at hand. Thus, taking into account the traceless
stress-energy tensor, $T=-\rho+p_r+2p_t=0$, provides the following
differential equation
\begin{equation}
F''\left(1-\frac{b}{r}\right)-\frac{b'r+b-2r}{2r^2}F'-
\frac{b'r-b}{2r^3}F=0\,. \label{diffeqT}
\end{equation}
In principle, one may deduce $F(r)$ by imposing a specific shape
function, and inverting Eq. (\ref{Ricciscalar}), i.e., $R(r)$, to
find $r(R)$, the specific form of $f(R)$ may be found from the
trace equation (\ref{trace}).

For instance, consider the specific shape function given by
$b(r)=r_0^2/r$. Thus, Eq. (\ref{diffeqT}) provides the following
solution
\begin{eqnarray}
F(r)&=&C_1
\sinh\left[\sqrt{2}\,\arctan\left(\frac{r_0}{\sqrt{r^2-r_0^2}}
\right)\right]
   \nonumber   \\
&+&C_2\cosh\left[\sqrt{2}\,\arctan\left(\frac{r_0}{\sqrt{r^2-r_0^2}}
\right)\right]\,.
\end{eqnarray}

The stress-energy tensor profile threading the wormhole is given
by the following relationships

\begin{eqnarray}
\rho(r)&=&- \frac{r_0^2}{r^4}\Bigg\{C_1
\sinh\left[\sqrt{2}\,\arctan\left(\frac{r_0}{\sqrt{r^2-r_0^2}}
\right)\right]\nonumber \\
   &+&C_2\cosh\left[\sqrt{2}\,\arctan\left(\frac{r_0}{\sqrt{r^2-r_0^2}}
\right)\right]\Bigg\}\,,\\
p_r(r)&=&- \frac{r_0}{r^4}\Bigg\{\left(2C_2\sqrt{2(r^2-r_0^2)}+
3r_0C_1\right)
\sinh\left[\sqrt{2}\,\arctan\left(\frac{r_0}{\sqrt{r^2-r_0^2}}
\right)\right]
 \nonumber   \\
 &+&\left(2C_1\sqrt{2(r^2-r_0^2)}+
3r_0C_2\right)\cosh\left[\sqrt{2}\arctan\left(\frac{r_0}{\sqrt{r^2-r_0^2}}
\right)\right]\Bigg\},\\
p_t(r)&=& \frac{r_0}{r^4}\Bigg\{\left(C_2\sqrt{2(r^2-r_0^2)}+
r_0C_1\right)
\sinh\left[\sqrt{2}\,\arctan\left(\frac{r_0}{\sqrt{r^2-r_0^2}}
\right)\right]
 \nonumber   \\
 &+&\left(C_1\sqrt{2(r^2-r_0^2)}+
r_0C_2\right)\cosh\left[\sqrt{2}\,\arctan\left(\frac{r_0}{\sqrt{r^2-r_0^2}}
\right)\right]\Bigg\}\,.
\end{eqnarray}

\begin{figure}[h]
\centering
  \includegraphics[width=2.8in]{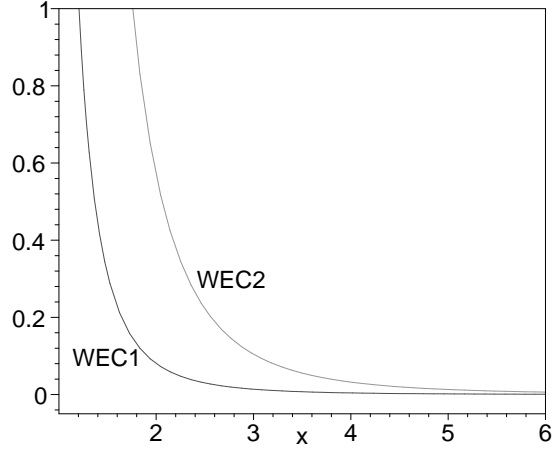}
  \caption{The stress-energy tensor satisfying the WEC, for the
  specific case of the traceless stress-energy tensor equation of
  state, and for the values $C_1=0$ and $C_2=-1$.
  We have considered the dimensionless quantities
  WEC1=$r_0^2\rho$, WEC2=$r_0^2(\rho+p_r)$ and $x=r/r_0$.}
 \label{Fig:WECT=0}
\end{figure}

One may now impose that the above stress-energy tensor satisfies
the WEC, which is depicted in Fig. \ref{Fig:WECT=0}, by
considering the values $C_1=0$ and $C_2=-1$.

\begin{figure}[h]
\centering
  \includegraphics[width=3.1in]{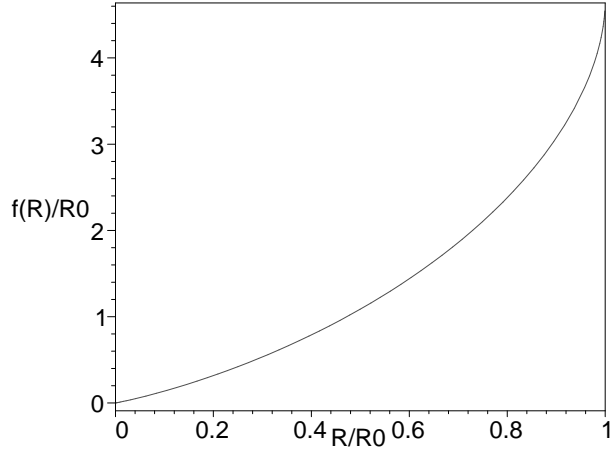}
  \caption{The specific form of $f(R)$, for the specific case of
  the traceless stress-energy tensor equation of state, by imposing
  the values $C_1=0$ and $C_2=-1$. The range is given by
  $0\leq R/R_0\leq 1$.}
 \label{Fig:WECT=0fR}
\end{figure}

For the specific shape function considered above, the Ricci
scalar, Eq. (\ref{Ricciscalar}), provides $R=-2r_0^2/r^4$ and is
now readily inverted to give $r=(-2r_0^2/R)^{1/4}$. It is also
convenient to define the Ricci scalar at the throat, and its
inverse provides $r_0=(-2/R_0)^{1/2}$. Substituting these
relationships into the consistency equation (\ref{trace}), the
specific form $f(R)$ is given by

\begin{eqnarray}
f(R)&=&-R\Bigg\{C_1\sinh\left[\sqrt{2}\arctan\left( \frac{1}
{\sqrt{\left(\frac{R_0}{R}\right)^{1/2}-1}} \right)\right]
    \nonumber   \\
&&\hspace{-0.5cm}+C_2\cosh\left[\sqrt{2}\arctan\left( \frac{1}
{\sqrt{\left(\frac{R_0}{R}\right)^{1/2}-1}}
\right)\right]\Bigg\}\,,
\end{eqnarray}

which is depicted in the Fig. \ref{Fig:WECT=0fR}, by imposing the
values $C_1=0$ and $C_2=-1$.

\subsection{Specific equation of state: $p_t=\alpha \rho$}

Many of the equations of state considered in the literature
involving the radial pressure and the energy density, such as the
linear equation of state $p_r=\alpha \rho$, provide very complex
differential equations, so that it is extremely difficult to find
exact solutions. This is due to the presence of the term $F''$ in
$p_r$. Indeed, even considering isotropic pressures does not
provide an exact solution. Now, things are simplified if one
considers an equation of state relating the tangential pressure
and the energy density, so that the radial pressure is determined
through Eq. (\ref{generic2}). For instance, consider the equation
of state $p_t=\alpha \rho$, which provides the following
differential equation:
\begin{equation}
F'\left(1-\frac{b}{r}\right)-\frac{F}{2r^2}
\bigg[b-b'r(1+2\alpha)\bigg]=0\,.
\label{diffeq1}
\end{equation}
In principle, as mentioned above one may deduce $F(r)$ by imposing
a specific shape function, and inverting Eq. (\ref{Ricciscalar}),
i.e., $R(r)$, to find $r(R)$, the specific form of $f(R)$ may be
found from the trace equation (\ref{trace}). In the following
analysis we consider several interesting shape functions usually
applied in the literature.

\subsubsection{1. Specific shape function:
$b(r)=r_0^2/r$}

First, we consider the case of $b(r)=r_0^2/r$, so that Eq.
(\ref{diffeq1}) yields the following solution
\begin{equation}
F(r)=C_1\left(1-\frac{r_0^2}{r^2}\right)^{\frac{1}{2}+\frac{\alpha}{2}}
\,.
\end{equation}

\begin{figure}[h]
\centering
  \includegraphics[width=2.8in]{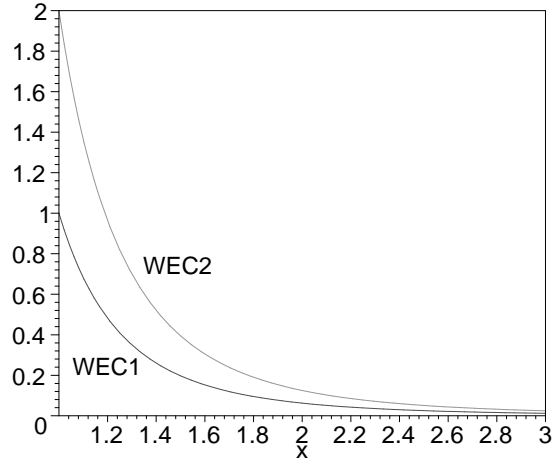}
  \caption{The stress-energy tensor satisfies the WEC,
  for the specific case of the equation of state $p_t=\alpha \rho$
  and considering the form function $b(r)=r_0^2/r$. We have
  imposed the values $C_1=-1$ and $\alpha=-1$, and
  considered the dimensionless quantities
  WEC1=$r_0^2\rho$, WEC2=$r_0^2(\rho+p_r)$ and $x=r/r_0$.}
 \label{Fig:WECptA}
\end{figure}

The gravitational field equations,
(\ref{generic1})-(\ref{generic3}), provide the stress-energy
tensor threading the wormhole, given by the following
relationships
\begin{eqnarray}
p_r(r)&=&\frac{C_1r_0^2}{r^6}\left(1-\frac{r_0^2}{r^2}\right)
^{-\frac{1}{2}+\frac{\alpha}{2}} \times
    \nonumber   \\
&&\times\left[2(r^2-r_0^2)+3\alpha
r^2-4r_0^2\alpha-r_0^2\alpha^2\right]\,,
   \\
p_t(r)&=&\alpha\rho(r)=-\frac{C_1r_0^2\alpha}{r^4}
\left(1-\frac{r_0^2}{r^2}\right)
^{\frac{1}{2}+\frac{\alpha}{2}} \,.
\end{eqnarray}
One may now impose that the above stress-energy tensor satisfies
the WEC, which is depicted in Fig. \ref{Fig:WECptA}, by imposing
the values $C_1=-1$ and $\alpha=-1$.

As in the previous case of the traceless stress-energy tensor, the
Ricci scalar, Eq. (\ref{Ricciscalar}), is given by $R=-2r_0^2/r^4$
and its inverse provides $r=(-2r_0^2/R)^{1/4}$. The inverse of the
Ricci scalar evaluated at the throat inverse is given by
$r_0=(-2/R_0)^{1/2}$. Substituting these relationships into the
consistency relationship (\ref{trace}), provides the specific form
of $f(R)$, which is given by
\begin{eqnarray}
f(R)&=&C_1R\left(1
-\sqrt{\frac{R}{R_0}}\right)^{\frac{\alpha}{2}-\frac{1}{2}}
\times
\nonumber   \\
\hspace{-1cm}&\times&
\left[\sqrt{\frac{R}{R_0}}
(\alpha^2+2\alpha+2)+(\alpha+2) \right]
 \,.
\end{eqnarray}

\begin{figure}[t]
\centering
  \includegraphics[width=3.1in]{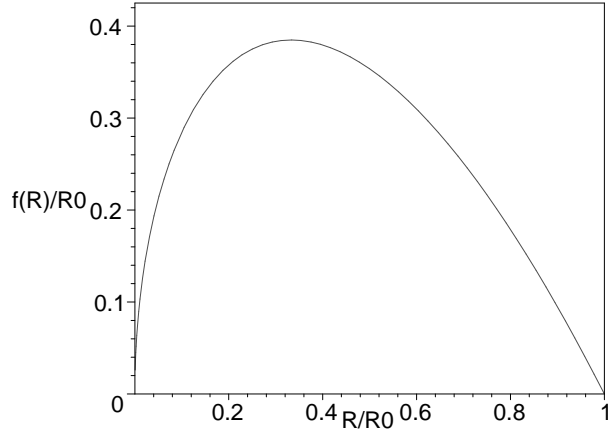}
  \caption{The profile of $f(R)$ is depicted for the
  specific case of the equation of state $p_t=\alpha \rho$
  and considering the form function $b(r)=r_0^2/r$. The values
  $C_1=-1$ and $\alpha=-1$ have been imposed,
  with the range given by $0\leq R/R_0\leq 1$.}
 \label{Fig:ptAfR}
\end{figure}

This function is depicted in Fig. \ref{Fig:ptAfR} as $f(R)/R_0$ as
a function as $R/R_0$, for the values $C_1=-1$ and $\alpha=-1$.

\subsubsection{2. Specific shape function: $b=\sqrt{r_0r}$}

Consider now the case of $b=\sqrt{r_0r}$, so that Eq.
(\ref{diffeq1}) yields the following solution
\begin{equation}
F(r)=C_1\left(1-\sqrt{\frac{r_0}{r}}\,\right)^{\frac{1}{2}-\alpha}
\,.
\end{equation}

The stress-energy tensor profile threading the wormhole is given
by the following relationships
\begin{eqnarray}
    p_t(r)&=&\alpha\rho(r)=\frac{C_1 \alpha}{2r^2}
    \frac{\left(1-\sqrt{\frac{r_0}{r}}\right)^{\frac{1}{2}
    -\alpha}}{\sqrt{\frac{r_0}{r}}}\,, \\
%\end{eqnarray}
%\begin{eqnarray}
    p_r(r)&=&-\frac{C_1r_0}{16r^3}
    \left(1-\sqrt{\frac{r_0}{r}}\right)^{-\frac{3}{2}-\alpha}\times
    \nonumber\\
    &&\hspace{-1.75cm}\times\bigg[10\sqrt{\frac{r}{r_0}}
    +\sqrt{\frac{r_0}{r}}
    \left(14\alpha+10\right)+
    \left(4\alpha^2-26\alpha+5\right)\bigg] \,.
\end{eqnarray}
Rather than consider plots of the WEC as before, we note that it
is possible to impose various specific values of $C_1$ and
$\alpha$ that do indeed satisfy the WEC.

Following the recipe prescribed above, the Ricci scalar is given
by $R=\sqrt{r_0}/r^{5/2}$ and is readily inverted to provide
$r=(\sqrt{r_0}/R)^{2/5}$. The inverse of the Ricci scalar at the
throat provides $r_0=1/\sqrt{R_0}$. Substituting these
relationships into the consistency relationship (\ref{trace}), the
specific form $f(R)$ is finally given by
\begin{eqnarray}
f(R)&=&-\frac{1}{8}\frac{C_1}{ R^\frac{2}{5}
-2\left(RR_0\right)^\frac{1}{5}+R_0^\frac{2}{5}} \Bigg\{
\left(R_0^\frac{1}{5}-R^\frac{1}{5} \right)^{\frac{1}{2}-\alpha}
R^\frac{3-2\alpha}{10} R_0^\frac{-21+10\alpha}{40}\times
\nonumber\\
&&\times \left[-8R R_0^\frac{2}{5}+\right(11+10\alpha
\left)R^\frac{4}{5}R_0^\frac{3}{5}+ \right(2-22\alpha
+4\alpha^2\left)R^\frac{3}{5}R_0^\frac{4}{5}+\right.\nonumber
\\
&& \left.+\left(-5+12\alpha-4\alpha^2\right)R^\frac{2}{5}R_0\right]\Bigg\}\,.
\end{eqnarray}
%\end{widetext}

\subsubsection{3. Specific shape function:
$b(r)=r_0+\gamma^2r_0(1-r_0/r)$}

Finally, it is also of interest to consider the specific shape
function given by $b(r)=r_0+\gamma^2r_0(1-r_0/r)$, with
$0<\gamma<1$, so that Eq. (\ref{diffeq1}) provides the following
solution
%\begin{widetext}
\begin{eqnarray}
F(r)=C_1\left(r-\gamma^2r_0\right)^{\frac{1}{2}
\frac{\gamma^2-2\alpha-1}{\gamma^2-1}} r^{-(\alpha+1)}\left(r-r_0\right)^{\frac{1}{2}
\frac{\gamma^2(1+2\alpha)-1}{\gamma^2-1}}
\end{eqnarray}
It is useful to write the last equation in the form
$F(r)=C_1X^ur^{-(\alpha+1)}Y^v$, where $X,\,Y,\,u,\, v$ are
defined as
\begin{eqnarray*}
&&X=r-\gamma^2r_0 \,, \qquad  Y=r-r_0\,,\\
&&u=\frac{\gamma^2-2\alpha-1}{2(\gamma^2-1)} \,, \qquad
v=\frac{\gamma^2(1+2\alpha)-1}{2(\gamma^2-1)}  \,.
\end{eqnarray*}

Thus, the stress energy tensor profile threading the wormhole is
given by the following expressions:
\begin{eqnarray}\nonumber
  p_r(r)&=& \frac{C_1}{2r^3}\Bigg\{X^uY^v\bigg[r^{-\alpha}
  \left(2\alpha^2+6\alpha+4\right) +
  \\ \nonumber
  &&  +r^{-(1+\alpha)}r_0
  \left(-7\alpha+2\alpha^2\gamma^2-3\gamma^2-7\alpha\gamma^2-2\alpha^2
 -3\right)+
 \\ \nonumber
 &&+r^{-(2+\alpha)}r_0^2\gamma^2\left(10\alpha^2+4\right)\bigg]+
\\ \nonumber
 &&+X^uY^{v-1}
 \bigg[r^{-\alpha}r_0v\left(-\gamma^2(5+4\alpha)-\alpha-5\right)
 +4r^{1-\alpha}v\left(1+\alpha\right)
 \\
 \nonumber
 &&+r^{-(1+\alpha)}r_0^2\gamma^2v\left(4\alpha+6\right)\bigg]+
 \\ \nonumber
 &&+X^uY^{v-2}\bigg[2r^{-\alpha}r_0\gamma^2v\left(v-\alpha \right)
 +2r^{1-\alpha}r_0v\left(-v+\gamma^2+1\right)
 \\
 &&+2r^{2-\alpha}v\left(v+1\right)\bigg]+
 \\ \nonumber
 &&X^{u-1}Y^v
 \bigg[r^{-\alpha}r_0u\left(4\alpha+5\right)\left(\gamma^2+1\right)
 -4r^{1-\alpha}\left(u-\alpha\right)
 \\ \nonumber
 && r^{-(1+\alpha)}r_0^2\gamma^2u\left(4\alpha-6\right)\bigg] +
 \\ \nonumber
 &&+X^{u-2}Y^v\big[2r^{-\alpha}r_0^2\gamma^2u\left(u-1\right)
 +2r^{1-\alpha}r_0u\left(1-u\right)
 +2r^{2-\alpha}u\left(u-1\right)\bigg] +
 \\ \nonumber
 &&+X^{u-1}Y^{v-1}\bigg[-4r^{-\alpha}r_0^2\gamma^2uv
 +4r^{1-\alpha}r_0uv\left(\gamma^2+1\right)
 -4r^{2-\alpha}uv\bigg]\Bigg\}\,,\\
  p_t(r)&=&\alpha\rho=C_1\alpha
  \gamma^2r_0^2X^ur^{-(5+\alpha)}Y^v\,.
\end{eqnarray}
As in the previous example, we will not depict the plot of the
functions, but simply note in passing that one may impose specific
values for the constants $\alpha$ and $C_1$ in order to satisfy
the WEC.

The Ricci scalar, Eq. (\ref{Ricciscalar}), provides
$R=2\gamma^2r_0^2/r^4$ and is now readily inverted to give
$r=(2\gamma^2r_0^2/R)^{1/4}$. The Ricci scalar at the throat is
given by $R_0=2\gamma^2/r_0^2$, and its inverse provides
$r_0=\gamma\sqrt{2/R_0}$. Substituting these relationships into
the consistency relationship (\ref{trace}), the specific form
$f(R)$ is given by
\newpage
\begin{eqnarray}
f(R)&=&\frac{C_1R}{2}\frac{\left(R_0
R\right)^{\frac{(\alpha+1)}{4}}}{
\gamma^2-\left(\frac{R_0}{R}\right)^\frac{1}{4}
(\gamma^2+1)+\left(\frac{R_0}{R}\right)^{\frac{1}{2}}}\times
\nonumber \\
%\nonumber
&&\times\left[\frac{\left(\frac{R_0}{R}\right)^\frac{1}{4}
-\gamma^2}{R_0^\frac{1}{2}}\right]^{\frac{1}{2}\frac{\gamma^2
-2\alpha-1}{\gamma^2-1}}\left[\frac{\left(\frac{R_0}{R}
\right)^\frac{1}{4}-1}{R_0^\frac{1}{2}}\right]^{\frac{1}{2}
\frac{\gamma^2(1-2\alpha)-1}{\gamma^2-1}}\times
\\\nonumber
&&\times\left[2\gamma^2(\alpha^2+2\alpha+2)
-\left(\frac{R_0}{R}\right)^\frac{1}{4}(3\alpha+4)
(\gamma^2+1)+\left(\frac{R_0}{R}\right)^\frac{1}{2}
(2\alpha+4)\right]\,.
\end{eqnarray}
%

%-----------------------------------------------
\section{Summary and Discussion}\label{Sec:conclusion}
%-----------------------------------------------

In general relativity, the NEC violation is a fundamental
ingredient of static traversable wormholes. Despite this fact, it
was shown that for time-dependent wormhole solutions the null
energy condition and the weak energy condition can be avoided in
certain regions and for specific intervals of time at the throat
\cite{dynamicWH}. Nevertheless, in certain alternative theories to
general relativity, taking into account the modified Einstein
field equation, one may impose in principle that the stress energy
tensor threading the wormhole satisfies the NEC. However, the
latter is necessarily violated by an effective total stress energy
tensor. This is the case, for instance, in braneworld wormhole
solutions, where the matter confined on the brane satisfies the
energy conditions, and it is the local high-energy bulk effects
and nonlocal corrections from the Weyl curvature in the bulk that
induce a NEC violating signature on the brane \cite{braneWH1}.
Another particularly interesting example is in the context of the
$D$-dimensional Einstein-Gauss-Bonnet theory of gravitation
\cite{EGB1}, where it was shown that the weak energy condition can
be satisfied depending on the parameters of the theory.

In this Chapter, we have explored the possibility that wormholes be
supported by $f(R)$ modified theories of gravity. We imposed that
the matter threading the wormhole satisfies the energy conditions,
and it is the higher order curvature derivative terms, that may be
interpreted as a gravitational fluid, that support these
non-standard wormhole geometries, fundamentally different from
their counterparts in general relativity. In the analysis outlined
above, we considered a constant redshift function, which
simplified the calculations considerably, yet provides interesting
enough exact solutions. One may also generalize the results of
this Chapter by considering $\Phi'\neq 0$, although the field
equations become forth order differential equations, and become
quite intractable. The strategy adopted to solve the field
equations was essentially to specify $b(r)$, and considering
specific equation of state, the function $F(r)$ was deduced from
the gravitational field equations, while the curvature scalar in a
parametric form, $R(r)$, was obtained from its definition via the
metric. Then, deducing $T=T^{\mu}{}_{\mu}$ as a function of $r$,
exact solutions of $f(R)$ as a function of $R$ from the trace
equation were found.

% ----------------------------------------------------------------
% ----------------------------------------------------------------

    % ----------------------------------------------------------------
% Chapter on Brams-Dicke Theories of Gravity ************************************************
% **** -----------------------------------------------------------
%\documentclass[12pt]{report}
%\usepackage{dsfont}
%\usepackage{amssymb}
%\usepackage{graphicx}
% ----------------------------------------------------------------
%\vfuzz2pt % Don't report over-full v-boxes if over-edge is small
%\hfuzz2pt % Don't report over-full h-boxes if over-edge is small

% ----------------------------------------------------------------
%\begin{document}

%\date{}%
%\dedicatory{}%
%\commby{}%
% ----------------------------------------------------------------

% ----------------------------------------------------------------
\chapter{Scalar-Tensor Theories of Gravity}

In this Chapter, we discus an apparent inconsistency in the wormhole solutions obtained for $f(R)$ modified theories of gravity. This is not intended as (not even nearly) a complete introduction to Scalar-Tensor theories of Gravity, the subject is not new and has had some time to mature, moreover, there are good references to this effect \cite{fujii}. Here, we simply present our analysis after a short motivation of the subject.

\section{Introduction}

%\subsection{Brief discussion: Formalism}

The scalar field, having very simple symmetry properties under coordinate transformations, has always been expected, one way or another, to play a role in our understanding of the gravitational interaction.  Indeed, as was mentioned above, Newton's gravitation was a scalar theory and, the first generalizations of SR wore scalar theories as well. However, one of the most surprising facts about gravity is, that it takes a rank two tensor field --- immensely more complex, when compared to a scalar field ---, to satisfactorily account for all the richness of gravitation.

This aesthetic desire to put the scalar field to good use, found a close companion, in the discussions (mainly philosophical), about the nature of space. On the one hand, we have the view according to which space exists on its own, and has absolute physical properties --- this found its apogee in the work of Isaac Newton ---, on the other, we have the notion that space is inextricably linked to the matter it contains and, that the only meaningful motion of a particle is motion relative to other matter in the universe. Although this last idea, never found a complete expression in a physical theory, it came to be known as \emph{Mach's principle}, and we can see immediately, that it implies that inertial forces, experienced in an accelerated laboratory are in fact gravitational effects of distant matter accelerated relative to the laboratory (see \cite{dinverno,BD1,BD2} for discussions). General Relativity, dos not fully embody these ideas. Indeed, for decades, the expectation was that one day some suitable boundary conditions to the field equations would be found, that could definitely bring the theory into accord with Mach's principle. Among other things, these boundary conditions would eliminate vacuum solutions.

One of the first attempts to include a scalar field is due to Jordan, he used a four dimensional curved manifold  embedded in a five-dimensional
flat space-time, proposing the following action:
\begin{equation}\label{Jordan-S}
    S_J=\frac{1}{16\pi G}\int d^4x\sqrt{-g}\left[\varphi_J^\gamma\left(R-\omega_j\frac{1}{\varphi_J^2}g^{\mu \nu}\partial_\mu\varphi_J\partial_\nu\varphi_J\right)\right]+S_{matter}(\varphi_J,\Psi)\,,
\end{equation}
$\varphi_J(x)$ is Jordan's scalar field, while $\gamma$ and $\omega_J$ are constants; we stress once more that a scalar field --- to in fact describe gravity, and not simply be considered a matter field ---, must be non minimally coupled to the geometry. This is represented here by the term $\varphi_J^\gamma R$.  We remark also, that coupling $L_{matter}$ to $\varphi_J$ leads to violations of the equivalence principle.

Later, a more general class of modifications based on the use of scalar fields wore proposed:
\begin{equation}\label{ST-lag}
    S_{ST}=\frac{1}{16\pi G}\int d^4x\sqrt{-g}\left(\varphi R-\omega(\varphi)\frac{1}{\varphi}g^{\mu \nu}\partial_\mu\varphi\partial_\nu\varphi-V(\varphi)\right)+S_{matter}(\Psi)
\end{equation}
$\omega(\varphi)$ is a function of $\varphi$.

The quantity, $\omega(\varphi)$ can be constrained using solar system tests:
\begin{equation}
    |\omega(\varphi_0)|> 40\;000.
\end{equation}
Where $\varphi_0$ is the present value of the scalar. For this to be applicable the mass of the scalar should be low (the field should be `light'), that is $\frac{\partial^2 V}{\partial \varphi^2´}$ evaluated at $\varphi_0)$ must be small.

Dimensionless coupling parameters, are expected to be of order unity, therefore, the above constraint is very unappealing. Scalar-Tensor theories, and the restricted version which we will study, (Brans-Dicke theories) are no longer considered a viable alternative to GR. They are, however, a very good example of a model theory.

%%%%%%%%%%%%%%%%%%%%%%%%%%%%%%%%%%%%%%%%%%%%%%%%%%%%%%%%%%%%%%%%%%%%%%%%%%%%%%%%%%%%%%%%%%%%

\section{Brans-Dicke wormholes}

In the last Chapter, traversable wormhole geometries in the context of $f(R)$
modified theories of gravity were constructed
\cite{mod-grav}. The matter threading the wormhole was imposed
to satisfy the energy conditions, so that it is the effective
stress-energy tensor containing higher order curvature derivatives
that is responsible for the NEC violation. Thus, the higher order
curvature terms, interpreted as a gravitational fluid, sustain
these non-standard wormhole geometries, fundamentally different
from their counterparts in general relativity. Furthermore, we
note that $f(R)$ modified theories of gravity are equivalent to a
Brans-Dicke theory with a coupling parameter $\omega=0$, and a
specific potential related to the function $f(R)$ and its
derivative. However, the value $\omega=0$ is apparently excluded
from the interval, $-3/2<\omega<-4/3$, of the coupling parameter,
extensively considered in the literature of static wormhole
solutions in vacuum Brans-Dicke theory.

In Brans-Dicke theory, analytical wormhole solutions were
constructed \cite{Agnese:1995kd,Nandi:1997en}.
It was shown that static wormhole solutions in vacuum Brans-Dicke
theory only exist in a narrow interval of the coupling parameter
\cite{Nandi:1997en}, namely, $-3/2<w<-4/3$. However, this result
is only valid for vacuum solutions and for a specific choice of an
integration constant of the field equations given by
$C(w)=-1/(w+2)$. The latter relationship was derived on the basis
of a post-Newtonian weak field approximation, and it is important
to emphasize that there is no reason for it to hold in the
presence of compact objects with strong gravitational fields.

In this context, we construct a general class of vacuum
Brans-Dicke wormholes that include the value of $\omega=0$, and
thus constructing a consistent bridge with the wormhole solutions
in $f(R)$ gravity found in \cite{mod-grav}. Furthermore, we
present the general condition for the existence of Brans-Dicke
wormhole geometries based on the NEC violation, and show that the
presence of effective negative energy densities is a generic
feature of these vacuum solutions.

\medskip

\subsection{General class of Brans wormholes}\label{Sec:II}

The matter-free action in
Brans-Dicke theory is given by
\begin{equation}\label{Jordan-F}
S=\frac{1}{2}\int d^{4}x(-g)^{\frac{1}{2}}\left[ \varphi
\mathbf{R} -\varphi ^{-1}\omega (\varphi )g^{\mu \nu }\varphi
_{,\mu }\varphi _{,\nu } \right] \,,
\end{equation}
where $\mathbf{R}$ is the curvature scalar, $\omega $ is a
constant dimensionless coupling parameter, and $\varphi$ is the
Brans-Dicke scalar. We adopt the convention $8\pi G=c=1$
throughout this Chapter.

We note here that this is not the only form that the Brans-Dicke actions can take. An equivalent way of writing the action is obtained using a conformal transformation. The result is the following action
\begin{equation}\label{Einstein-F}
S=\frac{1}{2}\int d^{4}x(-g)^{\frac{1}{2}}\left[
\mathbf{\tilde{R}} -g^{\mu \nu }\varphi
_{,\mu }\varphi _{,\nu } \right] \,,
\end{equation}

These are two equivalent forms of writing the action. The first is called the \emph{Jordan frame}, while the second is termed the \emph{Einstein frame}\footnote{If the theory contains a potential, this also changes. The first equation (\ref{Jordan-F}) is:$$S=\frac{1}{2}\int d^{4}x(-g)^{\frac{1}{2}}\left[ \varphi
\mathbf{R} -\varphi ^{-1}\omega (\varphi )g^{\mu \nu }\varphi
_{,\mu }\varphi _{,\nu } -V(\varphi)\right] \,,$$ and the second (\ref{Einstein-F}) becomes:$$S=\frac{1}{2}\int d^{4}x(-g)^{\frac{1}{2}}\left[
\mathbf{\tilde{R}} -g^{\mu \nu }\varphi
_{,\mu }\varphi _{,\nu } -\tilde{V}(\varphi)\right] \,.$$}\footnote{We also note, that the covariant derivatives would also change, but in this case they are equivalent to partial derivatives, since the argument is a scalar field.}. For more information on the possible representations see \cite{sot-PhD}.

The above action (\ref{Jordan-F}) provides the following field equations:
\begin{eqnarray}
\mathbf{G}_{\mu \nu }&=&-\frac{\omega }{ \varphi ^{2}}\left(
\varphi _{,\mu }\varphi _{,\nu }-\frac{1}{2}g_{\mu \nu }\varphi
_{,\sigma }\varphi ^{,\sigma }\right)
   \nonumber  \\
&&  -\frac{1}{\varphi }\left( \varphi _{;\mu }\varphi _{;\nu
}-g_{\mu \nu }\square^{2}\varphi \right) \,,  \\
\square ^{2}\varphi &=&0\,,
\end{eqnarray}
where $\mathbf{G}_{\mu \nu }$ is the Einstein tensor and $\square
^{2}\equiv \varphi ^{;\rho }{}_{;\rho }$.

It is useful to work in isotropic coordinates, with the metric
given by
\begin{equation}
ds ^{2}=-e^{2\alpha (r)}dt^{2}+e^{2\beta (r)}dr^{2}+e^{2\nu
(r)}r^{2}(d\theta ^{2}+\sin ^{2}\theta d\psi ^{2}).
\label{isometric}
\end{equation}
Throughout this work, we consider the Brans class I solution,
which corresponds to setting the gauge $\beta -\nu =0$. Thus, the
field equations yield the following solutions
\begin{eqnarray}
e^{\alpha (r)}&=&e^{\alpha _{0}}\left(
\frac{1-B/r}{1+B/r}\right)^{\frac{1}{ \lambda }} \label{feqs1}
\,,\\
e^{\beta (r)}&=&e^{\beta _{0}}\left(1+B/r\right) ^{2}\left(
\frac{1-B/r}{1+B/r }\right)^{\frac{\lambda -C-1}{\lambda }}\,,
\label{feqs2}
\\
\varphi (r)&=&\varphi _{0}\left(\frac{1-B/r}{1+B/r}\right)
^{\frac{C}{\lambda }}, \label{feqs3}
\\
\lambda ^{2}&\equiv & (C+1)^{2}-C\left( 1-\frac{\omega
C}{2}\right)>0\,, \label{feqs4}
\end{eqnarray}
where $\alpha _{0}$, $\beta _{0}$, $B$, $C$, and $\varphi _{0}$
are constants. Note that the asymptotic flatness condition imposes
that $\alpha _{0}=$ $\beta _{0}=0$, as can be readily verified
from Eqs. (\ref{feqs1}) and (\ref{feqs2}).

In order to analyze traversable wormholes in vacuum Brans-Dicke
theory, it is convenient to express the spacetime metric in the
original Morris-Thorne canonical form \cite{mor-thor}:
\begin{equation}
ds^{2}=-e^{2\Phi (R)}dt^{2}+\frac{dR^{2}}{1-b(R)/R}+R^{2}(d\theta
^{2}+\sin ^{2}\theta d\psi ^{2})
\end{equation}
where $R$ is the radial coordinate, $\Phi(R)$ and $b(R)$ are the redshift and shape functions, respectively. To be a wormhole solution, several properties are
imposed (as we mentioned in Chapter 2) \cite{mor-thor}, namely: The throat is located at
$R=R_0$ and $b(R_0)=R_0$. A flaring out condition of the throat is
imposed, i.e., $[b(R)-Rb'(R)]/b^2(R)>0$, which reduces to
$b'(R_0)<1$ at the throat, where the prime here denotes a
derivative with respect to $R$. The condition $1-b(R)/R\geq 0$ is
also imposed. To be traversable, one must demand the absence of
event horizons, so that $\Phi(R)$ must be finite everywhere.

Confronting the Morris-Thorne metric with the isotropic metric
(\ref{isometric}), the radial coordinate $r\rightarrow R$ is
redefined as
\begin{equation}
R=re^{\beta _{0}}\left(1+B/r\right)^{2}\left(
\frac{1-B/r}{1+B/r}\right)^{\Omega }\,,\qquad\Omega
=1-\frac{C+1}{\lambda }\,, \label{defineR}
\end{equation}
so that $\Phi(R)$ and $b(R)$ are given by
\begin{equation}
\Phi (R)=\alpha _{0}+\frac{1}{\lambda }\left\{ \ln \left[
1-\frac{B}{r(R)} \right]-\ln \left[1+\frac{B}{r(R)}\right]
\right\} , \label{redshift}
\end{equation}
\begin{equation}
b(R)=R\left\{ 1-\left[ \frac{\lambda
[r^{2}(R)+B^{2}]-2r(R)B(C+1)}{\lambda
[r^{2}(R)+B^{2}]}\right]^{2}\right\} \,,\label{shape}
\end{equation}
respectively. The wormhole throat condition $b(R_{0})=R_{0}$
imposes the minimum allowed $r$-coordinate radii $r_{0}^{\pm }$
given by
\begin{equation}
r_{0}^{\pm }=\alpha ^{\pm }B\,, \qquad  \alpha ^{\pm }=(1-\Omega
)\pm \sqrt{\Omega (\Omega -2)}\,. \label{throatr0}
\end{equation}
The values $R_{0}^{\pm }$ can be obtained from Eq. (\ref{defineR})
using Eq. (\ref{throatr0}). Note that $R\rightarrow \infty $ as
$r\rightarrow \infty $, so that $b(R)/R\rightarrow 0$ as
$R\rightarrow \infty $. The condition $ b(R)/R\leq 1$ is also
verified for all $R\geq $ $R_{0}^{\pm }$. The redshift function
$\Phi (R)$ has a singularity at $r=r_{S}=B$, so that the minimum
allowed values of $r_{0}^{\pm }$ must necessarily exceed
$r_{S}=B$. It can also be verified from Eq. (\ref{defineR}) that $
r_{0}^{\pm }\geq B$ which implies $R_{0}^{\pm }\geq 0$.

\subsection{Energy condition violations}

The energy density and the radial pressure of the wormhole
material are given by \cite{Nandi:1997en}
\begin{eqnarray}
\rho&=&-\frac{4B^{2}r^{4}Z^{2}[(C+1)^{2}-\lambda^{2}]}{\lambda^{2}(r^{2}
-B^{2})^{4}}\,,\label{rhoR}  \\
p_{r}&=&-\frac{4Br^{3}Z^{2}}{\lambda^{2}(r^{2}-B^{2})^{4}}
[\lambda C(r^{2} +B^{2})
    \nonumber   \\
&&-Br(C^{2}-1+\lambda^{2})] \label{prR}\,,
\end{eqnarray}
respectively, where $Z$ is defined as
\begin{equation}
Z\equiv\left(  \frac{r-B}{r+B}\right)^{(C+1)/\lambda}.
\end{equation}

Adding Eqs. (\ref{rhoR}) and (\ref{prR}), one arrives at
\begin{equation}
\rho+p_{r}=-\frac{4Br^{3}Z^{2}}
{\lambda^{2}(r^{2}-B^{2})^{4}}[\lambda
C(r^{2}+B^{2})+2Br(C+1-\lambda^{2})]\,, \label{NECviolation}
\end{equation}
which will be analyzed in the NEC violation below.

In \cite{Nandi:1997en}, the authors considered negative energy
densities, which consequently violates the weak energy condition
(WEC). Now, Eq. (\ref{rhoR}) imposes the following condition:
\begin{equation}
\left[C(\omega )+1\right]^2>\lambda^2 (\omega )\,,
   \label{Cwlambda}
\end{equation}
which can be rephrased as
\begin{equation}
C(\omega )\left[ 1-\frac{\omega C(\omega )}{2}\right] >0\,,
   \label{Cwlambda2}
\end{equation}
by taking into account Eq. (\ref{feqs4}). Note that the function
$C(\omega)$ is still unspecified.

However, it is important to emphasize that negative energy
densities are not a necessary condition in wormhole physics. The
fundamental ingredient is the violation of the NEC, $\rho+p_r<0$,
which is imposed by the flaring out condition
\cite{mor-thor}. To find the general restriction for
$\rho+p_{r}<0$ at the throat $r_0$, amounts to analyzing the
factor in square brackets in Eq. (\ref{NECviolation}), namely, the
condition $\lambda C(r_0^{2}+B^{2})+2Br_0(C+1-\lambda^{2})>0$.
Using Eqs. (\ref{feqs4}) and (\ref{throatr0}), the latter
condition is expressed as:
\begin{eqnarray}
&&(-1)^{s+t+1}\left[ (-1)^s(C+1)+(-1)^t\sqrt{C\bigg(1-\frac{\omega
C}{2}\bigg)}\right]\times
  \nonumber   \\
&&\times\frac{C \left(1-\omega
C/2\right)}{\sqrt{(4+2\omega)C^2+4(C+1)}}>0 \,,\label{gen-cond}
\end{eqnarray}
where $s,t=0,1$. Note that a necessary condition imposed by the
term in the square root, in square brackets, is precisely
condition (\ref{Cwlambda2}). Thus, a necessary condition for
vacuum Brans-Dicke wormholes is the existence of negative
effective energy densities. However, we emphasize that it is
condition (\ref{gen-cond}), i.e., the violation of the NEC at the
throat, that generic vacuum Brans-Dicke wormholes should obey.

\subsection{Specific forms of $C(\omega)$}

A specific choice of $C(\omega)$ considered extensively in the
literature, is the Agnese-La Camera function \cite{Agnese:1995kd}
given by
\begin{equation}
C(\omega )=-\frac{1}{\omega +2}\,.\label{Cw}
\end{equation}
Using this function, it was shown that static wormhole solutions
in vacuum Brans-Dicke theory only exist in a narrow interval of
the coupling parameter \cite{Nandi:1997en}, namely,
$-3/2<\omega<-4/3$. However, we point out that this result is only
valid for vacuum solutions and for the specific choice of
$C(\omega)$ considered by Agnese and  La Camera
\cite{Agnese:1995kd}. As mentioned before,
relationship (\ref{Cw}) was derived on the basis of a
post-Newtonian weak field approximation, and it is important to
emphasize that there is no reason for it to hold in the presence
of compact objects with strong gravitational fields. The choice
given by (\ref{Cw}) is a tentative example and reflects how
crucially the wormhole range for $\omega$ depends on the form of
$C(\omega)$. Evidently, different forms for $C(\omega)$ different
from Eq. (\ref{Cw}) would lead to different intervals for
$\omega$.

Note that in \cite{Nandietal}, the negative values of the coupling
parameter $\omega$ were extended to arbitrary positive values of
omega, i.e., $\omega<\infty$, in the context of two-way
traversable wormhole Brans solutions (we refer the reader to Ref
\cite{Nandietal} for specific details). An interesting example was
provided in Ref. \cite{Matsuda:1972uk}, in the context of
gravitational collapse in the Brans-Dicke theory, where the choice
$C(\omega)\sim-\omega^{-1/2}$ was analyzed. More specifically, the
authors in \cite{Nandi:1997en} considered
$C(\omega)=-q\omega^{-1/2}$, with $q<0$ so that $C(\omega)>0$.
Thus, the constraint (\ref{Cwlambda2}) is satisfied only if
$\omega> 4/q^2$. However, we will be interested in solutions which
include the value $\omega=0$, in order to find an equivalence with
the $f(R)$ solutions found in \cite{mod-grav}. The specific
choices we consider below possess the following limits, $C(\omega
)\rightarrow 0$, $ \lambda (\omega )\rightarrow 1$ as $\omega
\rightarrow \infty $, in order to recover the Schwarzschild
exterior metric in standard coordinates.

Another issue that needs to be mentioned is that the
above-mentioned interval imposed on $\omega$ was also obtained by
considering negative energy densities. In principle, the violation
of the WEC combined with an adequate choice of $C(\omega)$ could
provide a different viability and less restrictive interval
(including the value $\omega=0$) from the case of
$-3/2<\omega<-4/3$ considered in \cite{Nandi:1997en}. In this
context, we consider below different forms of $C(\omega)$ that
allow the value $\omega=0$ in the permitted range. Thus, to
satisfy the constraint (\ref{Cwlambda2}), both factors $C(\omega)$
and $[1-\omega C(\omega )/2]$ should both be positive, or both
negative.

\begin{figure}[h]
\centering
  \includegraphics[width=2.65in]{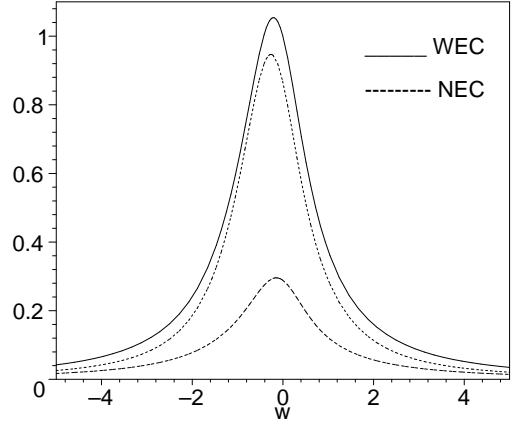}
  \caption{Plot of the energy conditions for
  $C(\omega)=(\omega^2+a^2)^{-1}$ for $a=1$.
  In particular, the WEC expressed by condition (\ref{Cwlambda2})
  is given by the solid line; and the NEC, expressed by the
  condition (\ref{gen-cond}), is given by the dashed curves. For the
latter, only the cases of $(s,t)=(0,1)$ and $(s,t)=(1,1)$ of
condition (\ref{gen-cond}) are allowed; and are depicted by the
small and large peaks, respectively.}
 \label{Fig:C-hat}
\end{figure}

Consider the following specific choice
\begin{equation}
C(\omega )=\frac{1}{\omega^2+a^2}\,,\label{Cw2}
\end{equation}
where $a$ is a real constant. The requirement that
$\lambda^{2}>0$, i.e., Eq. (\ref{feqs4}), is satisfied. The
function $C(\omega)$ is positive for all real $\omega$, and the
second term, in square brackets, of Eq. (\ref{Cwlambda2}), is
positive everywhere for $a^2>1/16$. Therefore, for this case,
condition (\ref{Cwlambda2}) is satisfied for all $\omega$. For
$a^2<1/16$, $[1-\omega C(\omega)/2]$ has two real roots, namely,
$\omega^0_\pm=(1\pm\sqrt{1-16a})/4$; the lesser value is positive
and thus both the second term and condition (\ref{Cwlambda2}) will
be positive at $\omega=0$. Thus, if $a^2<1/16$, the condition
(\ref{Cwlambda2}) is satisfied for $\omega\in \mathds{R}-
[\omega^0_-;\omega^0_+]$. Figure~\ref{Fig:C-hat} depicts condition
(\ref{Cwlambda2}) (depicted as a solid curve), i.e., negative
energy densities, and condition (\ref{gen-cond}) (depicted as the
dashed curves), i.e., the violation of the NEC, for $a=1$. For the
latter, only the cases of $(s,t)=(0,1)$ and $(s,t)=(1,1)$ of
condition (\ref{gen-cond}) are allowed; and are depicted in
Fig.~\ref{Fig:C-hat} by the small and large peaks, respectively.

In the limiting case, $C(\omega )\rightarrow 0$, $ \lambda (\omega
)\rightarrow 1$ as $\omega \rightarrow \infty $, one simply
recovers the Schwarzschild exterior metric in standard
coordinates. This can be verified from Eqs. (\ref{redshift}) and
(\ref{shape}), which impose $b(R)=2M$ and $b'|_{r_0}=0$. However,
in this limit, the inequality (\ref{gen-cond}) is violated, and
there are no traversable wormholes.

Consider a second specific choice given by
\begin{equation}
C(\omega)=A\exp\left(-\frac{\omega^2}{2}\right)\,.
\label{Cw3}
\end{equation}
The requirement that $\lambda^{2}>0$, i.e., Eq. (\ref{feqs4}), is
also satisfied. This function, for $A>0$, is positive for all
$\omega$. Therefore, in order to satisfy condition
(\ref{Cwlambda2}), the restriction $(1-\omega C(\omega)/2)>0$ is
imposed. We verify that if $0<A<2\exp(1/2)$, then $(1-\omega
C(\omega)/2)>0$ for all $\omega$, so that conditions
(\ref{Cwlambda2}) and (\ref{gen-cond}) are both satisfied. If
$A>2\exp(1/2)$, then the second term $(1-\omega C(\omega)/2)$ will
have two real positive roots, i.e., $\omega_{0,1}>0$. For this
choice of $A$, we have the following range of allowed $\omega$:
$\mathds{R}-]\omega_0,\omega_1[$. Moreover, since $\omega_0>0$,
the value $\omega=0$ will always be in the set of allowed values.

\begin{figure}[h]
\centering
  \includegraphics[width=2.65in]{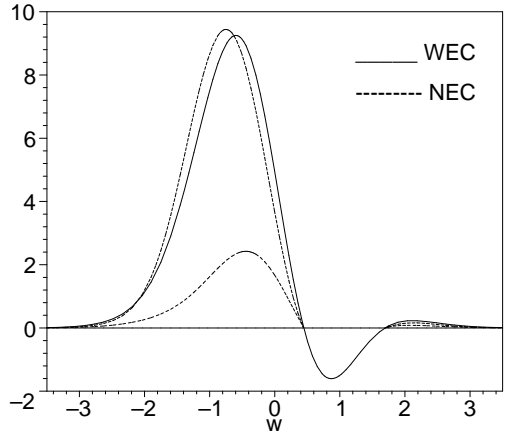}
  \caption{Plot of the energy conditions for
  $C(\omega)=A\exp(-\omega^2/2)$,
  with $A=3\exp(1/2)$.
  In particular, the WEC expressed by condition (\ref{Cwlambda2})
  is given by the solid line; and the NEC, expressed by the
  condition (\ref{gen-cond}), is given by the dashed curve.
  For the latter, only the cases of $(s,t)=(0,1)$ and $(s,t)=(1,1)$
  of condition (\ref{gen-cond}) are allowed; and are depicted by
  the smaller and larger peaks, respectively.}
 \label{Fig:2}
\end{figure}

Figure~\ref{Fig:2} depicts condition (\ref{Cwlambda2}) (depicted
as a solid curve), i.e., negative energy densities, and condition
(\ref{gen-cond}) (depicted as dashed curves), i.e., the violation
of the NEC for $A=3\exp(1/2)$. For the latter, only the cases of
$(s,t)=(0,1)$ and $(s,t)=(1,1)$ of condition (\ref{gen-cond}) are
allowed; and are depicted in Fig.~\ref{Fig:2} by the smaller and
larger peaks, respectively.

%-----------------------------------------------
%\subsection{Summary and Discussion}\label{Sec:conclusion}
%-----------------------------------------------

\section{Conclusion}
 Recently, in the context of $f(R)$ modified
theories of gravity, traversable wormhole geometries were
constructed. As $f(R)$ gravity is equivalent to a Brans-Dicke
theory with a coupling parameter $\omega=0$, one may be tempted to
find these solutions inconsistent with the permitted interval,
$-3/2<\omega<-4/3$, extensively considered in the literature of
static wormhole solutions in vacuum Brans-Dicke theory. Thus the
choice provided by Eq. (\ref{Cw}), in addition to the WEC and NEC
violation, reflects how crucially the range of $\omega$ depends on
the form of $C(\omega)$, and we have shown that adequate choices
of $C(\omega)$ provide different viability regions and less
restrictive intervals, that include $\omega=0$. In this context,
we have constructed a more general class of vacuum Brans-Dicke
wormholes that include the value of $\omega=0$, proving the
consistency of the solutions constructed in $f(R)$ gravity.
Furthermore, we deduced the general condition for the existence of
vacuum Brans-Dicke wormhole geometries, and have shown that the
presence of effective negative energy densities is a generic
feature of these vacuum solutions. It will also be interesting to
generalize this analysis in Brans-Dicke theory in the presence of
matter. Work along these lines is presently underway.

% ----------------------------------------------------------------

%\end{document}
% ----------------------------------------------------------------

    \part{Conclusions}

    %%%%%%%%%%%%%%%%%%
%%% Conclusions    %%%%%%%
%%%%%%%%%%%%%%%%%%%%%%%%%%%%%%%%

%\chapter{Theory of Gravitation Theories: a no Progress Report?}
\chapter{A `meta–-theory of gravitation'?}

An article by Sotiriou, Faraoni and Liberati\cite{meta},\footnote{This chapter's title was  adapted from there.} alerts us to the fact that, progress on the understanding of what could be called the `systematics of gravitation theories', --- and that consists largely on the axiomatization of gravitation --- has been scarce in the last decades. Such an axiomatization, could help us discriminate among the dizzying number of modified theories already existing and aid in the construction of new ones. They even enumerate some instances in which the existence of such a `meta–theory' would prove to be useful, among them are: the ever present problem of quantum gravity; a brief mention of `emergent gravity' scenarios; and an important experimental benefit, related to the fact that experimental tests seem to test principles more then theories.

This daunting effort, aiming to produce such a `theory of gravitation theories' is plagued by difficulties related to: (i) the definition and various versions of the Equivalence Principle, as well as it's relation to the Metric Postulates (of which we said something in Chapter 2); (ii) issues related to the definition of the Stress-Energy tensor and, which in turn can be traced to, (iii) the (notable) fact that, theories do not have a unique representation, leading to several conceptual biases.

Contrary to the authors of this article, however, we believe that the absence of noteworthy results, regarding this effort is not only to be expected, but also an indication that this programme may not be adequately suited to research in gravitation.  Although we stated in Chapter 2, that those traits of theories  related to the fact that they are theories (completeness, self-consistency, etc ), would be assumed,  this is perhaps an instance where these same features of physical theories probably come into play, and thus we are forced to mention them. Even if we knew all of the features a theory needs to have and, if we had identified all the principles involved, there is no determinism to assure us that attempting to incorporate  a certain principle, would result in a theory with a desired set of properties. There certainly is no superimposition in this case, asserting that the sum of the principles we use results in the sum of their respective properties. Just to give an example, there is an aesthetic choice to be made, when we attempt to modify gravity: we may `go along' with the particle physics trend, that views GR as nothing more than the field theory of the gravitational interaction, or we may  look upon it also as (and maybe more importantly) the dynamical and geometrical theory of space-time itself! Aesthetic choices such as these, cannot be easily incorporated into a supposed `meta--theory'. This is why we feel that this line of work, i.e., the attempt to produce a theory of gravitation theories, is not the next logical step to our `trial-and-error' approach (outlined in Chapter 1) despite the fact that it is often regarded as exactly this.

In summary, we began this thesis by briefly outlining the history of the emergence of both GR and modified theories in Chapter 1. In Chapter 2, we presented some of the principles that we find to be at the heart of Gravitation. These included the Dicke framework, the principle of equivalence, metric postulates, and the lagrangian formulation. We find, in retrospect, that although a rigorous and complete axiomatization is not possible (in the sense that we may not be able to find the smallest set of independent axioms), we do have a good idea of what principles come into play in gravitation theories. Also in this chapter, we explored some of the well known exact solutions: FLRW, Schwarzschild, and wormholes.

In chapter 3, we explored the possibility of the existence of wormholes in $f(R)$ modified theories of gravity. We found that it is possible to construct such solutions, and that the violation of the energy conditions they entail may be attributed to a curvature component of the effective stress-energy tensor. Since a correspondence between $f(R)$ and Brans-Dicke theories has long been established, and since wormholes in Brans-Dicke theories have been studied and severely constrained, an apparent contradiction arose. To deal with this apparent contradiction we critically reviewed the literature and proposed a new class of Brans-Dicke wormholes.

In regards to future perspectives of this work, we find that the exploration of more general forms of action, different variational formalisms and, more general modified theories, are some of  the routes to be pursued. Lately, interest arose, in actions involving the Gauss-Bonnet invariant. Theories have been explored where the action is a function of the Gauss-Bonnet invariant alone $f(\mathcal{G})$ \cite{F_G}, or of this variable and the Ricci scalar $f(R,\mathcal{G})$ \cite{sot-PhD,F_R_G}. These types of generalizations of the action increase the number of degrees of freedom of the theory, although this may also introduce some ghosts. Also, an interesting idea, is to have the Lagrangian be a function of $R$ and $\mathcal{L}_m$, the matter Lagrangian \cite{other_modified1}, or even a function of $T$ the Torsion Scalar \cite{other_modified2}. Non-minimal curvature-matter couplings have equally been explored in a wide range of circumstances~\cite{coupling1,coupling}. More recently, interest in the so-called Horava Gravity has emerged\cite{Horava}. Loop Quantum Gravity/Loop Quantum Cosmology (LQG/LQC) has also been a mainstream topic~\cite{LQG_C}, where one of the main attractions seems to be the possibility of constructing cosmological models, without an initial singularity, that is Big Bounce models. There has been some work on the use of the Palatini formalism in modified theories of gravity \cite{Lobo-pala}. As for wormholes, presently research continues in modified gravity in general, and in particular, Brans-Dicke theory \cite{wh-mg}.

It can be said, \emph{in nuce}, that the situation in modified theories of gravity, is complex and far from resolved. This is why we think, axiomatic formulations are very premature. In this subject, like in others, we must not be tempted to look for a panacea, lest we find nothing but a placebo!

%However, it can be argued that, it is dangerous to view this theory of scientific theories as a panacea, for it might turn out to be nothing more then a placebo.

%%%%%%% bachmatter %%%%%%%%%%%%%%%%%%%%%%
    \backmatter

    %%%%%%%%%%%%%%%%%%%%%%%%%%%%%%%%%%
%%%%%%%%%% References %%%%%%%%%%%
%%%%%%%%%%%%%%%%%%%%%%%%%%%%%%%%%


\addcontentsline{toc}{chapter}{Bibliography}


\begin {thebibliography}{99.}

%%%%%%%%% chapter 1 %%%%%%%%%%%%%5

\bibitem{dark_side} F.~S.~N.~Lobo, {`` The dark side of gravity: Modified theories of gravity''}, [arXiv:gr-qc/0807.1640v1].

\bibitem {how_far} R. P. Woodard, {``How Far Are We from the Quantum Theory of Gravity? ''},[arXiv:gr-qc/0907.4238].

\bibitem{sot-PhD} T.~P.~Sotiriou, {``Modified Actions for Gravity: Theory and Phenomenology''}, [arXiv:gr-qc/0710.4438v1].

\bibitem{mod-grav} Francisco~S.~N.~Lobo~and~Miguel~A.~Oliveira,{``Wormhole geometries in f(R) modified theories of gravity''}, Phys. Rev. D {\bf 80}, 104012 (2009).

\bibitem{brans-wh} Francisco~S.~N.~Lobo~and~Miguel~A.~Oliveira, {``General class of vacuum Brans-Dicke wormholes''},
Phys. Rev. D {\bf 81}, 067501 (2010).

\bibitem{Crawford} Paulo Crawford, {``Einstein's `Zurich Notebook'
and the Genesis of General Relativity''}, Boletim da SPM, Número especial Mira Fernandes, pp. 223–-245.

\bibitem{Pais} A.~Pais {\it ``Subtle is the Lord...'' The Science and the life of Albert Einstein}, (Cambridge University Press, 1982).

\bibitem{Weyl} H. Weyl, {\it Space Time Matter}, (Dover Publications, 1922).

\bibitem{Eddington} A. S. Eddington, {\it The Mathematical Theory of Relativity}, (Cambridge University Press, Cambridge, 1923).

\bibitem{Schrodinger} E. Schr\"{o}dinger, {\it Space-Time Structure}, (Cambridge University Press, Cambridge, 1963).

\bibitem{Bekenstein} J.~D.~Bekenstein, Phys. Rev. D \textbf{70}, 083509 (2004).

\bibitem{Milgrom} M.~Milgrom, Astrophys.\ J.\ \textbf{270}, 365 (1983).

\bibitem{Bu70} Buchdahl, H.A. 1970, Mon. Not. Roy. Astron. Soc., { 150}, 1.

%%%%%%%%%%%%%%% Chapter 2 %%%%%%%%%%%%%%%%%%%

\bibitem{Will} C. M. Will {\it Theory and Experiment in Gravitational Physics, revised edition}, (Cambridge University Press, Cambridge 1993).

\bibitem{wald} R.~M.~Wald, {\it General Relativity}, (University of Chicago Press, United States of America, 1984).

\bibitem{Pauli} W. Pauli, {\it Theory of Relativity}, (Pergamon Press, 1958).

\bibitem{Fock} V. A. Fock, {\it The Theory of Space Time and Gravitation}, (Pergmon Press, New York, 1964).

\bibitem{meta} T. P. Sotiriou, V. Faraoni, S. Liberati, ``Theory of Gravitation Theories: a no-progress report'', [arXiv:gr-qc/0707.2748v2].

\bibitem{dinverno} R. D'Inverno, {\it Introducing Einstein's Relativity}, (Cambridge University Press, Cambridge, 1998).

\bibitem{HawEll} S.~W.~Hawking \& G.~F.~R.~Ellis, {\it The Large Scale Structure of
Space-time}, (Cambridge University Press, Cambridge, 1973).

\bibitem{Gravitation} C.~W.~Misner, K.~S.~Thorne and  J.~A.~Wheeler, {\em Gravitation}, (W.~H.~Freeman and Co., San Francisco, 1973).

\bibitem{Schutz}  Bernard F. Schutz, {\it A First Course in General Relativity}, ( Cambridge University Press, Cambridge, 1985).

\bibitem{Goldstein} H.~Goldstein, {\it Classical Mechanics}, (Addison-Wesley, 1980).

\bibitem{ohanian} H. Ohanian, R. Ruffini, {\it Gravitation and Spacetime}, (W. W. Norton \& Company, New York, 1994).

\bibitem{Multamaki}
T. Multam\"{a}ki and I. Vilja,
{``Spherically symmetric solutions of modified field equations
in $f(R)$ theories of gravity,''}
Phys. Rev. D {\bf 74}, 064022 (2006).

\bibitem{Rindler} W.~Rindler, \ {\it Relativity: Special, General, and Cosmological}, \ Second Edition, \ (Oxford University Press, 2006).

\bibitem{Birkhoff} Valerio Faraoni, {``The Jebsen-Birkhoff theorem in alternative gravity''}, [arXiv:gr-qc/1001.2287].

\bibitem{CMB} WMAP~Collaboration (E. Komatsu (Texas U.) et al.), ``Seven-Year Wilkinson Microwave Anisotropy Probe (WMAP) Observations: Cosmological Interpretation'', Astrophys.J.Suppl. {\bf 192} (2011) 18, [arXiv:1001.4538].

\bibitem{observations}
A. Grant {\it et al},
%``The Farthest known supernova: Support for
%an accelerating Universe and a glimpse of the epoch of deceleration,''
Astrophys. J. {\bf 560} 49-71 (2001) [arXiv:astro-ph/0104455];
%
S. Perlmutter, M. S. Turner and M. White,
%``Constraining dark
%energy with SNe Ia and large-scale strucutre,''
Phys. Rev. Lett. {\bf 83} 670-673 (1999) [arXiv:astro-ph/9901052];
%
C. L. Bennett {\it et al},
%``First year {\it Wilkinson Microwave Anisotropy Probe}
%(WMAP) observations: Preliminary maps and basic results,''
Astrophys. J. Suppl. {\bf 148} 1 (2003) [arXiv:astro-ph/0302207];
%
G. Hinshaw {\it et al},
%``First year {\it Wilkinson Microwave Anisotropy Probe}
%(WMAP) observations: The angular power spectrum,''
[arXiv:astro-ph/0302217].

\bibitem{copeland} E.~J.~Copeland, M.~Sami, S.~Tsujikawa, {``Dynamics of dark energy''}, [arXiv:hep-th/0603057v3].

\bibitem{Lobo-PhD} F.~S.~N.~Lobo,
{\it Energy Conditions Violating Solutions in General Relativity:Travarseble Wormholes and ``Warp Drive'' Spacetimes}, (PhD Thesis, unpublished).

\bibitem{mor-thor} M.~S.~Morris~and~K.~S.~Thorne,\ {``Wormholes in spacetime and their use for interstellar travel: a toot for teaching General Relativity''},\  Am.~J.~Phys. {\bf 56,} 395 (1988).

\bibitem{Visser} M. Visser, {\it Lorentzian Wormholes: From Einstein to Hawking}, (American Institute of Physics,  August 1996).


%%%%%%%%%%%%%%%%%%%%% Chapter 3 %%%%%%%%%%%%%%%%%%%%%%%%%%%%%%%%%%%%%%%%%%%%%%%%%%%%%%%%%%%%%%%%%%%%%%%%%%%%%%%%%%%%%%%%%%%%%%%%%%%%%%%%%%%%%%%%%%%%%%%%%%%%%%%%%%%

\bibitem{Carroll-L} S.~Carroll, {``The Cosmological Constant''}, [arXiv:astro-ph/00004075].

\bibitem{fr-review} A.~De~Felice~and~S.~Tsujikawa,{``$f(R)$ theories''}, [arXiv:gr-qc/1002.4928v1].

\bibitem{sot-faraoni} T.~P.~Sotiriou~and~V.~Faraoni, {``f(R) theories of gravity''}, [arXiv:gr-qc/0805.1726v3].

\bibitem{Starobinsky:1980te}
A.~A.~Starobinsky, Phys.\ Lett.\  B {\bf 91}, 99 (1980).
%%CITATION = PHLTA,B91,99;%%

\bibitem{Carroll:2003wy}
  S.~M.~Carroll, V.~Duvvuri, M.~Trodden and M.~S.~Turner,
  ``Is cosmic speed-up due to new gravitational physics?''
  Phys.\ Rev.\  D {\bf 70}, 043528 (2004).
  %%CITATION = PHRVA,D70,043528;%%

\bibitem{viablemodels}
L.~Amendola, D.~Polarski and S.~Tsujikawa, Phys.\ Rev.\ Lett.\
\textbf{98}, 131302 (2007); S.~Capozziello, S.~Nojiri,
S.~D.~Odintsov and A.~Troisi, Phys.\ Lett.\ \textbf{B639}, 135
(2006); S.~Nojiri and S.~D.~Odintsov, Phys.\ Rev.\ \textbf{D74},
086005 (2006); M.~Amarzguioui, O.~Elgaroy, D.~F.~Mota and
T.~Multamaki, Astron.\ Astrophys.\ \textbf{454}, 707 (2006);
L.~Amendola, R.~Gannouji, D.~Polarski and S.~Tsujikawa, Phys.\
Rev.\ \textbf{D75}, 083504 (2007); T.~Koivisto, Phys.\ Rev.\  D
{\bf 76}, 043527 (2007); A.~A.~Starobinsky, JETP Lett.\  {\bf 86},
157 (2007); B.~Li, J.~D.~Barrow and D.~F.~Mota, Phys.\ Rev.\  D
{\bf 76}, 044027 (2007); S.~E.~Perez Bergliaffa, Phys.\ Lett.\
\textbf{B642}, 311 (2006); J.~Santos, J.~S.~Alcaniz,
M.~J.~Reboucas and F.~C.~Carvalho, Phys.\ Rev.\  D {\bf 76},
083513 (2007); G.~Cognola, E.~Elizalde, S.~Nojiri, S.~D.~Odintsov
and S.~Zerbini, JCAP \textbf{0502}, 010 (2005); V.~Faraoni, Phys.\
Rev.\ \textbf{D72}, 061501 (2005); V.~Faraoni, Phys.\ Rev.\
\textbf{D72}, 124005 (2005); L.~M.~Sokolowski, [gr-qc/0702097]
(2007); G.~Cognola, M.~Gastaldi and S.~Zerbini, [gr-qc/0701138]
(2007);
%
C.~G.~B\"ohmer, L.~Hollenstein and F.~S.~N.~Lobo, Phys.\ Rev.\  D
{\bf 76}, 084005 (2007);
%
S.~Carloni, P.~K.~S.~Dunsby and A.~Troisi, arXiv:0707.0106 [gr-qc]
(2007); K.~N.~Ananda, S.~Carloni and P.~K.~S.~Dunsby,
arXiv:0708.2258 [gr-qc] (2007); S.~Capozziello, R.~Cianci,
C.~Stornaiolo and S.~Vignolo, Class.\ Quant.\ Grav.\  {\bf 24},
6417 (2007); S.~Tsujikawa, Phys.\ Rev.\ D {\bf 77}, 023507 (2008);
S.~Nojiri, S.~D.~Odintsov and P.~V.~Tretyakov, Phys.\ Lett.\  B
{\bf 651}, 224 (2007); S.~Nojiri and S.~D.~Odintsov, Phys.\ Lett.\
B {\bf 652}, 343 (2007); G.~Cognola, E.~Elizalde, S.~Nojiri,
S.~D.~Odintsov, L.~Sebastiani and S.~Zerbini, Phys.\ Rev.\  D {\bf
77}, 046009 (2008).

\bibitem{coupling1}
  O.~Bertolami, C.~G.~Boehmer, T.~Harko and F.~S.~N.~Lobo,
  Phys.\ Rev.\  D {\bf 75}, 104016 (2007);
  %%CITATION = PHRVA,D75,104016;%%
%
%\bibitem{Bertolami:2007vu}
O. Bertolami and J. P\'aramos, Phys.\ Rev.\  D {\bf 77}, 084018
(2008);
%
O.~Bertolami, F.~S.~N.~Lobo and J.~Paramos,
  %``Non-minimum coupling of perfect fluids to curvature,''
  Phys.\ Rev.\  D {\bf 78}, 064036 (2008)
  [arXiv:0806.4434 [gr-qc]];
  %%CITATION = PHRVA,D78,064036;%%
%
O.~Bertolami, J.~Paramos, T.~Harko and F.~S.~N.~Lobo,
  %``Non-minimal curvature-matter couplings in modified gravity,''
  arXiv:0811.2876 [gr-qc];
  %%CITATION = ARXIV:0811.2876;%%
%
T.~Harko,
  %``Modified gravity with arbitrary coupling between matter and geometry,''
  Phys.\ Lett.\  B {\bf 669}, 376 (2008)
  [arXiv:0810.0742 [gr-qc]];
  %%CITATION = PHLTA,B669,376;%%
%
T.~P.~Sotiriou and V.~Faraoni,
  %``Modified gravity with R-matter couplings and (non-)geodesic motion,''
  Class.\ Quant.\ Grav.\  {\bf 25}, 205002 (2008);
  %[arXiv:0805.1249 [gr-qc]].
  %%CITATION = CQGRD,25,205002;%%
%
V.~Faraoni,
  %``A viability criterion for modified gravity with an extra force,''
  Phys.\ Rev.\  D {\bf 76}, 127501 (2007).
  %[arXiv:0710.1291 [gr-qc]].
  %%CITATION = PHRVA,D76,127501;%%

\bibitem{solartests}
T.~Chiba, Phys.\ Lett. \textbf{B575}, 1 (2003); A.~L.~Erickcek,
T.~L.~Smith and M.~Kamionkowski, Phys.\ Rev.\ \textbf{D74}, 121501
(2006); T.~Chiba, T.~L.~Smith and A.~L.~Erickcek, Phys.\ Rev.\
\textbf{D75}, 124014 (2007). S.~Nojiri and S.~D.~Odintsov, Phys.\
Lett.\  B {\bf 659}, 821 (2008); S.~Capozziello, A.~Stabile and
A.~Troisi, Phys.\ Rev.\  D {\bf 76}, 104019 (2007);
S.~Capozziello, A.~Stabile and A.~Troisi, Class.\ Quant.\ Grav.\
{\bf 25}, 085004 (2008).

\bibitem{Olmo07}
G.~J.~Olmo, Phys.\ Rev.\ D \textbf{75}, 023511 (2007).

\bibitem{solartests2}
W.~Hu and I.~Sawicki, Phys.\ Rev.\ \textbf{D76}, 064004 (2007);
S.~Nojiri and S.~D.~Odintsov, Phys.\ Rev.\ \textbf{D68}, 123512
(2003); V.~Faraoni, Phys.\ Rev.\ \textbf{D74}, 023529 (2006);
T.~Faulkner, M.~Tegmark, E.~F.~Bunn and Y.~Mao, Phys.\ Rev.\
\textbf{D76}, 063505 (2007); P.~J.~Zhang, Phys.\ Rev.\  D {\bf
76}, 024007 (2007); S.~Capozziello and S.~Tsujikawa, Phys.\ Rev.\
D {\bf 77}, 107501 (2008); I.~Sawicki and W.~Hu, Phys.\ Rev.\
\textbf{D75}, 127502 (2007); L.~Amendola and S.~Tsujikawa, Phys.\
Lett.\  B {\bf 660}, 125 (2008).

\bibitem{darkmatter}
S.~Capozziello, V.~F.~Cardone and A.~Troisi, JCAP \textbf{0608},
001 (2006); S. Capozziello, V. F. Cardone and A. Troisi, Mon. Not.
R. Astron. Soc. \textbf{375}, 1423 (2007); A.~Borowiec,
W.~Godlowski and M.~Szydlowski, Int. J. Geom. Meth. Mod. Phys.
\textbf{4} (2007) 183; C.~F.~Martins and P.~Salucci,
  Mon.\ Not.\ Roy.\ Astron.\ Soc.\  {\bf 381}, 1103 (2007);
%
  C.~G.~Boehmer, T.~Harko and F.~S.~N.~Lobo,
  %``Dark matter as a geometric effect in f(R) gravity,''
  arXiv:0709.0046 [gr-qc];
  %%CITATION = ARXIV:0709.0046;%%
%
 C.~G.~Boehmer, T.~Harko and F.~S.~N.~Lobo,
  %``Generalized virial theorem in f(R) gravity,''
  JCAP {\bf 0803}, 024 (2008);
  %%CITATION = JCAPA,0803,024;%%

\bibitem{Palatini}
M.~Ferraris, M.~Francaviglia and I.~Volovich, gr-qc/9303007
(1993); D.~N.~Vollick, Phys.\ Rev. D \textbf{68}, 063510 (2003);
E.~E.~Flanagan, Class.\ Quant.\ Grav.\ \textbf{21}, 417 (2003);
X.~H.~Meng and P.~Wang, Phys.\ Lett. \textbf{B584}, 1 (2004);
B.~Li and M.~C.~Chu, Phys.\ Rev.\ {\bf D74}, 104010 (2006);
N.~J.~Poplawski, Phys.\ Rev. \textbf{D74}, 084032 (2006); B.~Li,
K.~C.~Chan and M.~C.~Chu, Phys.\ Rev.\ {\bf D76}, 024002 (2007);
B.~Li, J.~D.~Barrow and D.~F.~Mota, Phys.\ Rev.\  D {\bf 76},
104047 (2007); A.~Iglesias, N.~Kaloper, A.~Padilla and M.~Park,
Phys.\ Rev.\  D {\bf 76}, 104001 (2007).

\bibitem{Sotiriou:2006qn}
T.~P.~Sotiriou and S.~Liberati, Annals Phys.\
\textbf{322}, 935 (2007).

\bibitem{EGB1}
B. Bhawal and S. Kar,
%``Lorentzian wormholes in Einstein-Gauss-Bonnet theory,''
Phys. Rev. D {\bf 46}, 2464-2468 (1992).

\bibitem{EGB2}
G. Dotti, J. Oliva, and R. Troncoso,
%``Static wormhole solution for higher-dimensional gravity in vacuum,''
Phys. Rev. D {\bf 75}, 024002 (2007)
%[arXiv:hep-th/0607062].
%%CITATION = HEP-TH 0607062;%%

\bibitem{braneWH1}
L. A. Anchordoqui and S. E. P Bergliaffa,
%``Wormhole surgery and cosmology on the brane: The world is not enough,''
Phys. Rev. D {\bf 62}, 067502 (2000);
%[arXiv:gr-qc/0001019];
%%CITATION = GR-GC 0001019;%%
%
K. A. Bronnikov and S.-W. Kim,
%``Possible wormholes in a brane world,''
Phys. Rev. D {\bf 67}, 064027 (2003);
%[arXiv:gr-qc/0212112];
%%CITATION = GR-GC 0212112;%%
%
M. La Camera,
%``Wormhole solutions in the Randall-Sundrum scenario,''
Phys. Lett. {\bf B573}, 27-32 (2003);
%[arXiv:gr-qc/0306017].
%%CITATION = GR-GC 0306017;%%
%
F.~S.~N.~Lobo,
  %``General class of braneworld wormholes,''
  Phys.\ Rev.\ {\bf D75}, 064027 (2007).
  %[arXiv:gr-qc/0701133].
  %%CITATION = PHRVA,D75,064027;%%

\bibitem{Nandi:1997en}
K.~K.~Nandi, B.~Bhattacharjee, S.~M.~K.~Alam and J.~Evans,
  %``Brans-Dicke wormholes in the Jordan and Einstein frames,''
  Phys.\ Rev.\  D {\bf 57}, 823 (1998).
  %%CITATION = PHRVA,D57,823;%%

\bibitem{Garattini:2007ff}
R.~Garattini and F.~S.~N.~Lobo,
  %``Self sustained phantom wormholes in semi-classical gravity,''
  Class.\ Quant.\ Grav.\  {\bf 24}, 2401 (2007);
  %[arXiv:gr-qc/0701020];
  %%CITATION = CQGRD,24,2401;%%
%
R.~Garattini and F.~S.~N.~Lobo,
  %``Self-sustained traversable wormholes in noncommutative geometry,''
  Phys.\ Lett.\  B {\bf 671}, 146 (2009).
  %[arXiv:0811.0919 [gr-qc]].
  %%CITATION = PHLTA,B671,146;%%

\bibitem{Boehmer:2007rm}
  C.~G.~Boehmer, T.~Harko and F.~S.~N.~Lobo,
  %``Conformally symmetric traversable wormholes,''
  Phys.\ Rev.\  D {\bf 76}, 084014 (2007);
  %[arXiv:0708.1537 [gr-qc]];
  %%CITATION = PHRVA,D76,084014;%%
%
C.~G.~Boehmer, T.~Harko and F.~S.~N.~Lobo,
  %``Wormhole geometries with conformal motions,''
  Class.\ Quant.\ Grav.\  {\bf 25}, 075016 (2008).
  %[arXiv:0711.2424 [gr-qc]].
  %%CITATION = CQGRD,25,075016;%%


\bibitem{phantomWH}
S.~Sushkov,
%``Wormholes supported by a phantom energy,''
Phys. Rev. D {\bf 71}, 043520 (2005);
%[arXiv:gr-qc/0502084];
%%CITATION = GR-QC 0502084;%%
%
F.~S.~N.~Lobo,
  %``Phantom energy traversable wormholes,''
  Phys.\ Rev.\ {\bf D71}, 084011 (2005);
  %[arXiv:gr-qc/0502099];
  %%CITATION = PHRVA,D71,084011;%%
%
  F.~S.~N.~Lobo,
  %``Stability of phantom wormholes,''
  Phys.\ Rev.\ {\bf D71}, 124022 (2005);
  %[arXiv:gr-qc/0506001].
  %%CITATION = PHRVA,D71,124022;%%
%
A.~DeBenedictis, R.~Garattini and F.~S.~N.~Lobo,
  %``Phantom stars and topology change,''
  Phys.\ Rev.\  D {\bf 78}, 104003 (2008);
  %[arXiv:0808.0839 [gr-qc]];
  %%CITATION = PHRVA,D78,104003;%%
%
\bibitem{Weylgrav}
F.~S.~N.~Lobo,
  %``General class of wormhole geometries in conformal Weyl gravity,''
  Class.\ Quant.\ Grav.\  {\bf 25}, 175006 (2008).
  %[arXiv:0801.4401 [gr-qc]].
  %%CITATION = CQGRD,25,175006;%%

\bibitem{Harko:2008vy}
  T.~Harko, Z.~Kovacs and F.~S.~N.~Lobo,
  %``Electromagnetic signatures of thin accretion disks in wormhole
  %geometries,''
  Phys.\ Rev.\  D {\bf 78}, 084005 (2008);
  %[arXiv:0808.3306 [gr-qc]];
%%CITATION = ARXIV:0808.3306;%%
%
T.~Harko, Z.~Kovacs and F.~S.~N.~Lobo,
  %``Thin accretion disks in stationary axisymmetric wormhole spacetimes,''
  Phys.\ Rev.\  D {\bf 79}, 064001 (2009).
  %[arXiv:0901.3926 [gr-qc]].
  %%CITATION = PHRVA,D79,064001;%%
\bibitem{Lemos:2003jb}
  J.~P.~S.~Lemos, F.~S.~N.~Lobo and S.~Quinet de Oliveira,
  %``Morris-Thorne wormholes with a cosmological constant,''
  Phys.\ Rev.\  D {\bf 68}, 064004 (2003).
  %[arXiv:gr-qc/0302049].
  %%CITATION = PHRVA,D68,064004;%%

\bibitem{Lobo:2007zb}
  F.~S.~N.~Lobo,
  ``Exotic solutions in General Relativity: Traversable wormholes and 'warp
  drive' spacetimes,''
  arXiv:0710.4474 [gr-qc].
  %%CITATION = ARXIV:0710.4474;%%

\bibitem{dynamicWH}
D.~Hochberg and M.~Visser,
  %``The null energy condition in dynamic wormholes,''
  Phys.\ Rev.\ Lett.\  {\bf 81}, 746 (1998);
  %[arXiv:gr-qc/9802048];
  %%CITATION = PRLTA,81,746;%%
%
  D.~Hochberg and M.~Visser,
  %``Dynamic wormholes, anti-trapped surfaces, and energy conditions,''
  Phys.\ Rev.\  D {\bf 58}, 044021 (1998);
  %[arXiv:gr-qc/9802046];
  %%CITATION = PHRVA,D58,044021;%%
%
S.~Kar,
  %``Evolving wormholes and the weak energy condition,''
  Phys.\ Rev.\  D {\bf 49}, 862 (1994);
  %%CITATION = PHRVA,D49,862;%%
%
 S.~Kar and D.~Sahdev,
  %``Evolving Lorentzian wormholes,''
  Phys.\ Rev.\  D {\bf 53}, 722 (1996);
  %[arXiv:gr-qc/9506094];
  %%CITATION = PHRVA,D53,722;%%
%
S.~W.~Kim,
  %``The Cosmological model with traversable wormhole,''
  Phys.\ Rev.\  D {\bf 53}, 6889 (1996);
  %%CITATION = PHRVA,D53,6889;%%
%
A.~V.~B.~Arellano and F.~S.~N.~Lobo,
  %``Evolving wormhole geometries within nonlinear electrodynamics,''
  Class.\ Quant.\ Grav.\  {\bf 23}, 5811 (2006).
  %[arXiv:gr-qc/0608003].
  %%CITATION = CQGRD,23,5811;%%

%%%%%%%%%%%%%%%%%%%%%%%%%%%% Chapter 4 %%%%%%%%%%%%%%%%%

\bibitem{fujii} Y.~Fujii~\&~K.~Maeda, {\it The Scalar-Tensor Theory of Gravitation}, (Cambridge University Press, 2004).

\bibitem{BD1} C. Brans~and~R. Dicke{``Mach's Principle and a Relativistic Theory of Gravitation''}, Phys. Rev. {\bf 124}, 925 (1961).

\bibitem{BD2} C. Brans {``Mach's Principle and a Relativistic Theory of Gravitation II''}, Phys. Rev. D {\bf 125}, 2163 (1962).

\bibitem{Agnese:1995kd}
  A.~G.~Agnese and M.~La Camera,
  %``Wormholes in the Brans-Dicke theory of gravitation,''
  Phys.\ Rev.\  D {\bf 51}, 2011 (1995).
  %%CITATION = PHRVA,D51,2011;%%

%\bibitem{Anchordoqui:1996jh}
%  L.~A.~Anchordoqui, S.~E.~Perez Bergliaffa and D.~F.~Torres,
  %``Brans-Dicke wormholes in non-vacuum spacetime,''
%  Phys.\ Rev.\  D {\bf 55}, 5226 (1997).
  %[arXiv:gr-qc/9610070].
 Boletim da SPM, Número especial Mira Fernandes, pp. 1–245 %%CITATION = PHRVA,D55,5226;%%

%\bibitem{Nandi:1997en}
%K.~K.~Nandi, B.~Bhattacharjee, S.~M.~K.~Alam and J.~Evans,
  %``Brans-Dicke wormholes in the Jordan and Einstein frames,''
 % Phys.\ Rev.\  D {\bf 57}, 823 (1998).
  %%CITATION = PHRVA,D57,823;%%

\bibitem{Nandietal}
K.~K.~Nandi, A.~Islam and J.~Evans,
  %``Brans wormholes,''
  Phys.\ Rev.\  D {\bf 55}, 2497 (1997);
  %[arXiv:0906.0436 [gr-qc]].
  %%CITATION = PHRVA,D55,2497;%%
%
A.~Bhattacharya, I.~Nigmatzyanov, R.~Izmailov and K.~K.~Nandi,
  %``Brans-Dicke Wormhole Revisited,''
  Class.\ Quant.\ Grav.\  {\bf 26} (2009) 235017.
  %[arXiv:0910.1109 [gr-qc]].
  %%CITATION = CQGRD,26,235017;%%

\bibitem{Matsuda:1972uk}
  T.~Matsuda,
  %``On the gravitational collapse in brans-dicke theory of gravity,''
  Prog.\ Theor.\ Phys.\  {\bf 47}, 738 (1972).
  %%CITATION = PTPKA,47,738;%%



%%%%%%%%%%%%%%%% conclusion %%%%%%

\bibitem{F_G} Antonio De Felice and Shinji Tsujikawa, {``Solar system constraints on f(G) gravity models''}, [arXiv:gr-qc/0907.1830]; Ratbay Myrzakulov, Diego Sáez-Gómez, Anca Tureanu, {``On the $\Lambda$CDM Universe in $f(G)$ gravity''}, [arXiv:gr-qc/1009.0902]; Shuang-Yong Zhou, Edmund J. Copeland, Paul M. Saffin, {``Cosmological Constraints on $f(G)$ Dark Energy Models''}, [arXiv:gr-qc/0903.4610].

\bibitem{F_R_G} Antonio De Felice, Jean-Marc Gerard, Teruaki Suyama, {``Cosmological perturbation in f(R,G) theories with a perfect fluid''}, [arXiv:gr-qc/1005.1958]; Antonio De Felice, Takahiro Tanaka, {``Inevitable ghost and the degrees of freedom in f(R,G) gravity''},[arXiv:gr-qc/1006.4399].

\bibitem{other_modified1} Tiberiu Harko, Francisco S. N. Lobo, {``$f\left(R,L_m\right)$ gravity''}, [arXiv:gr-qc/1008.4193].

\bibitem{other_modified2} Ratbay Myrzakulov, {``Accelerating universe from F(T) gravities''}, [arXiv:gr-qc/1006.1120].

\bibitem{coupling} Jay D. Tasson, {``Lorentz Symmetry and Matter-Gravity Couplings''}, [arXiv: hep-ph/1010.3990]; Sante Carloni, Emilio Elizalde and, Pedro J Silva, {``Matter couplings in Horava-Lifshitz and their cosmological applications''}. [arXiv: hep-ph/1009.5319].

\bibitem{Horava} Petr Horava, {``Quantum Gravity at a Lifshitz Point''}, [arXiv:hep-th/0901.3775].

\bibitem{LQG_C}Francesco Cianfrani, Giovanni Montani, {``Shortcomings of the Big Bounce derivation in Loop Quantum Cosmology''}, [arXiv:gr-qc/1006.1814]; Ramon Herrera, {``Warm inflationary model in loop quantum cosmology''}, [arXiv:gr-qc/1006.1299]; L.J. Garay, M. Martín-Benito, G.A. Mena Marugán {``Inhomogeneous Loop Quantum Cosmology: Hybrid Quantization of the Gowdy Model''}, [arXiv:gr-qc/1005.5654].

\bibitem{Lobo-pala} Tiberiu Harko and Francisco S. N. Lobo, {``Palatini formulation of modified gravity with a nonminimal curvature-matter coupling''}, [arXiv:gr-qc/1007.4415].

\bibitem{wh-mg} Ernesto F. Eiroa, Claudio Simeone, {``Brans-Dicke cylindrical wormholes''}, [arXiv:gr-qc/1008.0382]; S. Habib Mazharimousavi, M. Halilsoy, Z. Amirabi, {``Higher dimensional thin-shell wormholes in Einstein-Yang-Mills-Gauss-Bonnet gravity''}, [arXiv:gr-qc/1007.4627 ].

\end{thebibliography}
\end{document}